\documentclass[a4paper,aps,prd,10pt,preprintnumbers,showpacs,twocolumn,superscriptaddress,nofootinbib,amsmath,amssymb]{revtex4-1}
\usepackage[utf8]{inputenc}
\usepackage[T1]{fontenc}
\usepackage{cmap}
\usepackage{graphicx}
\usepackage{tabularx}
\usepackage{makecell} % 用于增加单元格高度
\usepackage{amsmath}
\usepackage{makecell}
\usepackage[utf8]{inputenc}
\usepackage[T1]{fontenc}
\usepackage{amsfonts}
\usepackage{amsmath}
\usepackage{amssymb}
\usepackage{amsthm}
\usepackage{mathrsfs}
\usepackage{euscript}
\usepackage{color}
\usepackage{braket}
\usepackage{tensor}
\usepackage{mathtools}
\usepackage{float}
\usepackage{graphicx} 
\usepackage{subcaption}

\begin{document}

\title{Proca-Maxwell System in an Infinite Tower of Higher-Derivative Gravity}
\author{Chen-Hao Hao}\affiliation{Department of Physics, Key Laboratory of Low Dimensional Quantum Structures and Quantum  Control of Ministry of Education, and Synergetic Innovation Center for Quantum Effects and Applications, Hunan Normal University, Changsha, Hunan 410081, P. R. China}
\affiliation{ Institute of Interdisciplinary Studies, Hunan Normal University, Changsha, Hunan 410081, P. R. China}
\author{Yong-Qiang Wang}\affiliation{Lanzhou Center for Theoretical Physics, Key Laboratory of Theoretical Physics of Gansu Province, School of Physical Science and Technology, Lanzhou University,\\ Lanzhou 730000, China}
\affiliation{Institute of Theoretical Physics $\&$ Research Center of Gravitation, Lanzhou University,\\ Lanzhou 730000, China}
\author{Jieci Wang}\email{jcwang@hunnu.edu.cn, corresponding author}
\affiliation{Department of Physics, Key Laboratory of Low Dimensional Quantum Structures and Quantum  Control of Ministry of Education, and Synergetic Innovation Center for Quantum Effects and Applications, Hunan Normal University, Changsha, Hunan 410081, P. R. China}
\affiliation{ Institute of Interdisciplinary Studies, Hunan Normal University, Changsha, Hunan 410081, P. R. China}

%============================
\begin{abstract}
	We numerically construct a five-dimensional Proca-Maxwell system coupled to an infinite tower of higher-derivative gravity, parameterized by the correction order and coupling constant. While the first-order correction case recovers standard Einstein gravity results, and the second-order correction (Gauss-Bonnet) case fails to resolve the central singularity in the vanishing frequency limit, we demonstrate that higher-order corrections effectively regularize the spacetime, yielding globally regular solutions. A key finding is the emergence of a ``frozen state'' in the supercritical regime: as the field frequency approaches zero, matter concentrates entirely within a critical radius, creating a regular core that externally mimics an extremal black hole. We further reveal that introducing the electric charge fundamentally alters this behavior; the electrostatic repulsion counteracts the gravitational collapse, effectively ``unfreezing'' the system and preventing the formation of the critical core. Significantly, unlike models relying on exotic matter, our solutions satisfy all standard energy conditions across the entire parameter space, establishing a physically viable pathway for constructing regular black hole mimickers.
\end{abstract}

	\maketitle
	\flushbottom
	
	\section{Introduction}\label{Sec1}
	
	General Relativity predicts that the final outcome of gravitational collapse of ordinary matter is a black hole, which possesses event horizon and conceals a singularity within it \cite{Hawking:1970zqf,Penrose:1964wq}. However, the singularity is not a physically acceptable result, as it represents infinite matter density and spacetime curvature, signifying a breakdown of causality, even though the existence of the event horizon ostensibly conceals this pathology \cite{Senovilla:1998oua,Penrose:1969pc}. It is widely believed that quantum gravity can resolve this issue, but a mature theory of quantum gravity remains elusive. Therefore, exploring alternative approaches to address this problem constitutes a significant topic in current gravitational theory research.
	
	Historically, a common approach was to directly modify the metric to replace the singularity with a regular core \cite{Duan:1954bms,Sakharov:1966aja,Bardeen1,Dymnikova:1992ux}, however, such proposals often lack deep physical motivation and are therefore unsatisfactory. Other conventional methods involve modified gravity theory \cite{Berej:2006cc,Junior:2023ixh,Aros:2019quj} or introducing exotic matter within classical gravity \cite{Hayward:2005gi,Ayon-Beato:1998hmi,Ayon-Beato:2000mjt,Bronnikov:2000vy,Bronnikov:2000yz,Dymnikova:2004zc,Fan:2016hvf,Lan:2023cvz,Wang:2025oek,Li:2025kou,Uktamov:2026hsw}. Yet, exotic matter sources, such as phantom fields or nonlinear electromagnetic fields, are frequently plagued by physical unnaturalness and dynamical instabilities \cite{Bronnikov:2012ch,Gonzalez:2008wd,Hao:2023kvf}. A more compelling avenue, well-motivated by string theory and effective field theory arguments, is to modify the gravitational sector itself through higher-curvature corrections. 
	
	Recently, a significant breakthrough was achieved in $D \geq 5$ dimensions, where it was shown that an infinite tower of higher-derivative terms can resolve the Schwarzschild singularity \cite{Bueno:2024dgm,Bueno:2024zsx,Hao:2025utc}. This framework falls within the class of ``quasi-topological gravity'' \cite{Oliva:2010eb,Myers:2010ru,Dehghani:2011vu,Ahmed:2017jod,Cisterna:2017umf,Frolov:2024hhe,Frolov:2025ddw} and provides a robust basis for an effective gravitational action \cite{Bueno:2019ltp}. The success of this pure gravity regularization scheme highlights its potential to address fundamental issues in black hole physics \cite{Konoplya:2024hfg,DiFilippo:2024mwm,Bueno:2025qjk,Aguayo:2025xfi,Tsuda:2026xjc,Borissova:2026dlz,Ling:2025ncw}.
	
	On the other hand, solutions describing matter fields coupled to gravity are often restricted to highly idealized models \cite{Oppenheimer:1939ue}, realistic scenarios necessitate numerical approaches, specifically for scalar or vector fields possessing internal $U(1)$ symmetries. Since the seminal work on scalar ``Boson stars'' \cite{Kaup:1968zz,Ruffini:1969qy}, the field has expanded to include massive spin-1 vector fields, known as ``Proca stars'' \cite{Brito:2015pxa,DiGiovanni:2018bvo,Liang:2025myf}. These macroscopic quantum objects resist gravitational collapse via the Heisenberg uncertainty principle and can achieve densities comparable to black holes \cite{Colpi:1986ye}. Astrophysical interest in these exotic compact objects (ECOs) has surged, as they are viable dark matter candidates \cite{Matos:1999et,Matos:2000ss,Hu:2000ke,Hui:2016ltb} and black hole mimickers \cite{Cardoso:2019rvt,Glampedakis:2017cgd,Herdeiro:2021lwl}. Notably, gravitational wave analyses suggest that events like GW190521 could be consistent with Proca star mergers \cite{LIGOScientific:2020iuh,CalderonBustillo:2020fyi}.
	
	Building on the premise that General Relativity may not be the ultimate theory of gravity, the study of Exotic Compact Objects (ECOs) has been extended to various modified gravity theories, such as scalar-tensor gravity \cite{Torres:1998xw,Whinnett:1999sc,Brihaye:2019puo}, $f(R)$ gravity \cite{Maso-Ferrando:2021ngp,Maso-Ferrando:2023wtz}, $f(T)$ gravity \cite{Ilijic:2020vzu} and Gauss-Bonnet gravity \cite{Baibhav:2016fot,Hartmann:2013tca,Henderson:2014dwa,Brihaye:2013zha}. Recently, Refs. \cite{Ma:2024olw,Chen:2025iuy} systematically investigated Boson stars and Proca stars in quasi-topological gravity theory featuring an infinite tower of higher-derivative gravity terms, and discovered their ``frozen states''. This phenomenon is characterized by the complex scalar field's frequency approaching zero, while the metric components $g_{tt}$ and $1/g_{rr}$ simultaneously tend towards zero at a certain critical radius $r_{c}$ (though never strictly vanishing) and matter concentrates within this radius. Further related studies can be found in \cite{Zhang:2025nem,Sun:2024mke,Huang:2024rbg,Yue:2023sep,Brihaye:2025dlq}.
	
\begin{figure}[!htbp]
		\begin{center}
			\begin{subfigure}[b]{0.46\textwidth}
				\includegraphics[width=\textwidth]{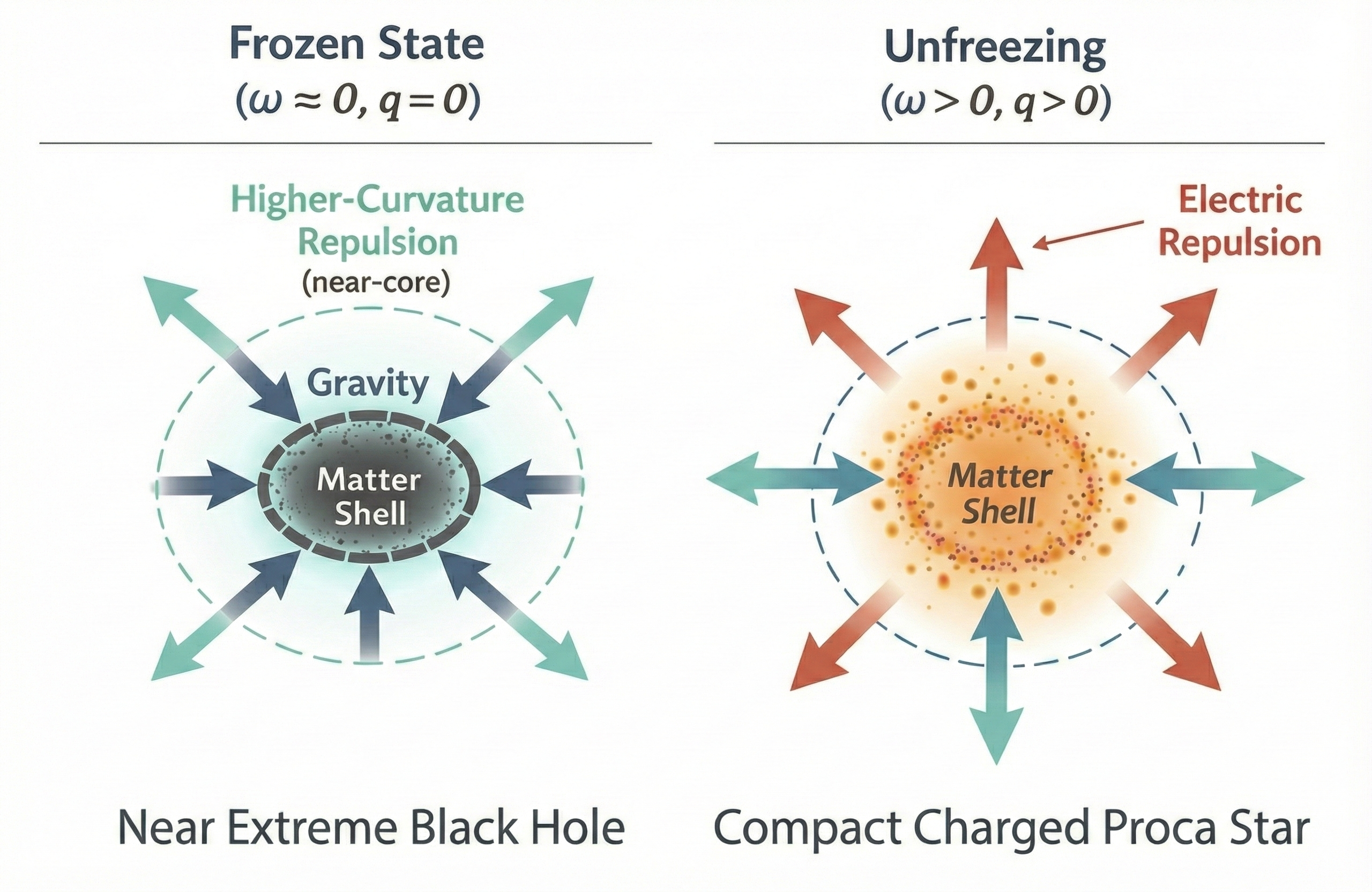}
			\end{subfigure}
		\end{center}
		\caption{Schematic representation of the equilibrium mechanism: gravitational attraction versus higher-order curvature and electromagnetic repulsive forces.}
		\label{p0}
	\end{figure}

	While previous works \cite{Bueno:2024eig} have shown that simplified matter configurations form regular black holes in this theory, bosonic fields instead collapse into peculiar ``frozen star'' under overwhelming gravity. To explore more realistic scenarios, we gauge the global $U(1)$ symmetry of the Proca field, naturally introducing a Maxwell field. This gauge interaction provides a crucial long-range Coulomb repulsion. To investigate whether this electrostatic repulsion can compete with gravitational attraction and higher-curvature effects—thereby preventing the system from completely ``freezing'', we numerically construct a spherically symmetric charged Proca-Maxwell model minimally coupled to an infinite tower of higher-derivative gravity. The numerical results indicate that for a correction order $n=1$ or $\alpha=0$, the model corresponds to the charged Proca star in Einstein gravity. When the correction order is $n=2$, the model represents the five-dimensional charged Proca star solution in Gauss-Bonnet gravity, whereas for $n \geq 3$ up to infinity, the solution exhibits a frozen phenomenon similar to that in \cite{Ma:2024olw,Chen:2025iuy}. However, our key finding is that the introduction of electric charge $q$ fundamentally alters this behavior. As the charge $q$ increases, the star gradually ``unfreezes'' and demonstrates several new properties. Whether it is a finite number of corrections or up to infinite order, these solutions have been obtained and discussed for the first time. This model is shaped by the delicate equilibrium between gravitational attraction, higher-order curvature repulsion, and long-range electromagnetic repulsion (Fig.~\ref{p0}), potentially offering new insights into black hole mimickers.
	
	The paper is organized as follows. In Sec.~\ref{sec2}, we construct a model of five-dimensional Proca-Maxwell system within an infinite tower of higher-derivative gravity theory. Sec.~\ref{sec3} is dedicated to the determination of the boundary conditions required for the numerical solving and to introduce the physical quantities of interest. In Sec.~\ref{sec4}, we present the numerical results obtained within $n=1,2,3,\infty$, followed by a discussion of these results. Finally, we summarize the obtained results in Sec.~\ref{sec5}.
	
	\section{The Model Setup}\label{sec2}
	Motivated by the effective field theory (EFT) arguments discussed in the Introduction, which necessitate the inclusion of higher-curvature corrections in the strong-gravity regime, we now establish the theoretical framework of our model. Specifically, we consider a Proca-Maxwell system minimally coupled to an infinite tower of higher-derivative gravity, whose action in a $D$-dimensional spacetime is formulated as
	\begin{equation}\label{equ1}
		S = \int \mathrm{d}^D x \sqrt{|g|} \left[ \frac{R}{16\pi G} + \sum_{n=2}^{n_{\text{max}}} \frac{\alpha_n \mathcal{Z}_n}{16\pi G} + \mathcal{L}_P + \mathcal{L}_M \right],
	\end{equation} 
	where $g$ is the determinant of the metric tensor $g_{ab}$, $R$ denotes the Ricci scalar, $\mathcal{Z}_n$ represents the Lagrangian density of the nth order, and $\alpha_n$ is the coupling constant corresponding to the order of correction. The detailed form of $\mathcal{Z}_n$ can be found in \cite{Bueno:2024dgm}. This specific form of the higher-order corrections is physically motivated by EFT arguments. It has been conjectured that, generic higher-order gravities can be mapped to generalized quasi-topological theories via field redefinitions, thereby capturing universal quantum gravity features \cite{Bueno:2019ltp}. However, absent a rigorous proof for infinite-order expansions, the specific series adopted here serves as a phenomenological ansatz designed to ensure mathematical tractability. We emphasize that, in this work, we do not consider more complicated non-minimal matter field couplings, aiming for results that are simple yet representative. The Lagrangian density for the Proca matter field and the Maxwell field, are denoted by 
	\begin{equation}\label{equ2}
		\mathcal{L}_{P}=-\frac{1}{4} \mathcal{W}_{\mu\nu} \overline{\mathcal{W}}^{\mu\nu}-\frac{1}{2} m^{2} \mathcal{B}_{\mu} \overline{\mathcal{B}}^{\mu},
	\end{equation}
	\begin{equation}
		\label{LM}
		\mathcal{L}_M =-\frac{1}{4}F_{\mu\nu}F^{\mu\nu}.
	\end{equation}
	
	Here, $m$ is the mass of the Proca field, $\mathcal{W}_{\mu\nu}$ is the Proca field tensor, $\mathcal{B}_{\mu}$ is the Proca potential 1-form, $F_{\mu\nu}$ is the Maxwell field tensor. The Proca field tensor $\mathcal{W}_{\mu\nu}$ is defined in terms of the Proca potential as $W_{\mu\nu} := \mathcal{D}_\mu \mathcal{B}_\nu - \mathcal{D}_\nu \mathcal{B}_\mu$, where $\mathcal{D}_\mu := \nabla_\mu - iqA_\mu$ denotes the gauge invariant derivative which includes the coupling to the Maxwell potential 1-form $A_{\mu}$ with $q$ the corresponding electric charge parameter of the Proca field. Meanwhile, the Maxwell field $F_{\mu\nu}$ is defined in terms of the potential 1-form $A_{\mu}$ in the usual way, $F_{\mu\nu} := \nabla_\mu A_\nu - \nabla_\nu A_\mu$. Next, we introduce the static spherically symmetric spacetime metric and the ansatz for the Proca field and Maxwell field
	\begin{equation}\label{equ3}
		\mathrm ds^2=-\sigma(r)^2N(r)\mathrm dt^2+\frac{\mathrm dr^2}{N(r)}+r^2\mathrm d\Omega_{D-2}^2,
	\end{equation}
	where $N(r)$ and $\sigma(r)$ are two undetermined functions depend only on the radial distance $r$, and the ansatz for the Proca field and the Maxwell field as follows:
	\begin{equation}\label{equ4}
		\mathcal{B}=[f(r)dt+ih(r)dr]e^{-i\omega t},
	\end{equation}
	\begin{equation}\label{equ5}
		A_\mu dx^\mu= V(r) dt.
	\end{equation}
	
	Since standard quasi-topological terms are trivial in four dimensions, we investigate spherically symmetric solutions in five-dimensional spacetime-a critical arena for probing non-perturbative higher-curvature gravity and holographic dualities. We set $G = 1/4\pi$ and adopt the specific coupling ansatz $\alpha_n = \alpha^{n-1}$. It is worth noting that alternative coupling choices introduce only minor quantitative variations without altering the system's fundamental properties. Crucially, recent works \cite{Bueno:2025zaj,Borissova:2026klg} have successfully extended non-polynomial quasi-topological frameworks to four dimensions. The subsequent discovery of `frozen' neutron stars therein \cite{Tan:2025hht} strongly suggests that our model admits a similar 4D generalization, paving the way for solutions with direct astrophysical significance.
	By substituting ansatz (\ref{equ3}), (\ref{equ4}) and (\ref{equ5}) into (\ref{equ1}) and varying the action, we obtain the equation of motion
	\begin{equation}\label{equ6}
		\begin{split}
			[r^4 \mathcal{H}(\psi)]' = &\frac{2r^3}{3N \sigma^2} \Big[ m^2 f^2 + N \big( h^2 (m^2 N \sigma^2 + (\omega+q V)^2) \\
			&- 2 h (\omega+q V) f' + f'^2 + V'^2 \big) \Big],
		\end{split}
	\end{equation}
	\begin{equation}\label{equ7}
		\sigma' \frac{d \mathcal{H}(\psi)}{d\psi} = \frac{2m^2 r^3}{3N^2 \sigma^2} \left( f^2 + h^2 N^2 \sigma^2 \right),
	\end{equation}
	\begin{equation}\label{equ8}
		m^2 h N \sigma^2 - h (\omega + qV)^2 + (\omega + qV) f' = 0,
	\end{equation}
	\begin{equation}\label{equ9}
		\begin{split}
			&r f (\omega + qV)^2 + N \sigma \Big[ (\omega + qV) \big( rN\sigma h' +  \\& h(3N\sigma + r\sigma N' + rN\sigma') \big)
			- qrhN\sigma V' \Big] = 0,
		\end{split}
	\end{equation}
	\begin{equation}\label{equ10}
		q r h^2 \sigma (\omega + qV) - q r h \sigma f' + r \sigma' V' - \sigma (3V' + rV'') = 0, 
	\end{equation}
	where 
	\begin{equation}
		\mathcal{H}(\psi) \equiv \psi + \sum_{n=2}^{n_{\text{max}}} \alpha^{n-1} \psi^n, \quad \psi \equiv \frac{1 - N(r)}{r^2}.
	\end{equation}
	These equations form a system of ODEs to be solved numerically. Before proceeding, it is worth noting that in the absence of a matter field, the equations of motion reduce to the results of regular black hole presented in \cite{Bueno:2024dgm}
	\begin{equation}\label{equ11}
		\frac{d\sigma}{dr}=0 ,\quad\frac{d}{dr}\left[r^{4}\mathcal{H}(\psi)\right]=0.
	\end{equation}
	By solving (\ref{equ11}), we can deduce that $\sigma(r)=1$(required by normalization of the time coordinate at infinity), we obtain
	\begin{equation}\label{equ12}
		\mathcal{H}(\psi)=\frac{\tilde{m}}{r^{4}},
	\end{equation}
	where $\tilde{m}$ is an integration constant which is proportional to the ADM mass $M$. It takes the form (we restored the gravitational constant $G$ when presenting the analytical solutions.)
	\begin{equation}\label{equ13}
		\tilde{m}=\frac{8GM}{3\pi}.
	\end{equation}
	
	For $n=2$, by solving (\ref{equ12}), $N(r)$ should be
	\begin{equation}\label{equ14}
		N(r)= 1-\frac{-r^2+\sqrt{\frac{32\alpha GM}{3\pi}+r^4}}{2\alpha},
	\end{equation}
	is classical 5D Gauss-Bonnet black holes solution. Similarly, for $n=3$ and $n=\infty$ (the expression for $n=4$ or more higher order is too complicated to present), we have the following expressions for $N(r)$ (for the detailed derivation process, please see  Appendix \ref{app:vacuum_ninf}):
	
	For $n=3$:
	\begin{equation}\label{equ15}
		\begin{split}
			N(r)=1-\frac{1}{6}\left(\frac{2^{2/3}\tilde{N}(r)}{\pi^{1/3}\:\alpha^{2}}
			\:-\frac{2\:r^{2}}{\alpha}\:-\:\frac{4\:(2\:\pi)^{1/3}\:r^{4}}{\tilde{N}(r)}\:\right)
		\end{split} 
	\end{equation}
	with
	\begin{equation}
		\begin{split}
			&\tilde{N}(r)=\left(7\:\pi\:r^{6}\:\alpha^{3}\:+\:72G\:M\:r^{2}\:\alpha^{4}\:\right.\\
			&\left.+\:3\:\sqrt{\:r^{4}\:\alpha^{6}\:\left(9\:\pi^{2}\:r^{8}\:+\:112G\:M\:\pi\:r^{4}\:\alpha\:+\:576\:G^{2}\:M^{2}\:\alpha^{2}\:\right)}\:\right)^{1/3},
		\end{split}
	\end{equation} 
	and for $n=\infty$:
	\begin{equation}\label{equ16}
		N(r)= 1-\frac{8GMr^2}{3\pi r^4+8GM\alpha}.
	\end{equation}
	
	Furthermore, in the electrovacuum limit (vanishing Proca field), the system reduces to the charged black hole solutions. For $n=2$, this recovers the well-known charged 5D Gauss-Bonnet black hole
	\begin{equation}\label{equn2q}
		N(r)= 1 + \frac{r^2}{2 \alpha} - \frac{\sqrt{r^6 - \frac{4 G \left( Q^2 - 8 M \pi^2 r^2 \right) \alpha}{3 \pi^3}}}{2 r \alpha},
	\end{equation}
	Omitting the explicit and lengthy expressions for $n = 3,4$, the form for $n=\infty$ is (for the detailed derivation process, see  Appendix \ref{app:electrovac_ninf}):
	\begin{equation}\label{equninq}
		N(r)= 1 + \frac{G r^2 \left( Q^2 - 8 M \pi^2 r^2 \right)}{3 \pi^3 r^6 - G Q^2 \alpha + 8 G M \pi^2 r^2 \alpha}.
	\end{equation}
	
	In (\ref{equn2q}) and (\ref{equninq}), $Q$ represents the electric charge and $M$ represents the ADM mass. It is easy to see that taking $Q$ as 0 reduces the result back to (\ref{equ14}) and (\ref{equ16}). Expand $N(r)$ at the origin, one obtains
	\begin{equation}\label{n2ex}
		\begin{split}
			N(r)_{n=2}=& 1- \frac{i G Q}{\sqrt{3} \pi^{3/2} \sqrt{G \alpha} r} + \frac{4 i G M \sqrt{\frac{\pi}{3}} r}{Q \sqrt{G \alpha}}  \\
			&+ \frac{r^2}{2 \alpha} + \frac{8 i G M^2 \pi^{5/2} r^3}{\sqrt{3} Q^3 \sqrt{G \alpha}} + O(r)^4,
		\end{split}
	\end{equation}
	\begin{equation}\label{ninfex}
		N(r)_{n=\infty}= 1 -  \frac{r^2}{ \alpha} + O(r)^4.
	\end{equation}
	
	For $n = 2$, the pathological behavior of the metric indicates that when $r$ is less than the curvature singularity $r_0$, the internal metric cannot be defined \cite{Wiltshire:1985us}. For $n=\infty$, the de Sitter core replaces the singularity at $r=0$. However, with the coupling parameter $\alpha_n$ chosen in this work, the black hole solution corresponding to Eq.~\eqref{equninq} develops a singularity near $r=0$. Achieving global regularity requires stricter constraints, and relevant studies on charged regular black holes in quasi-topological gravity can be found in \cite{Hao:2025utc,Hennigar:2025yqm,Xie:2025auj}.
	
	\section{Boundary Conditions and Physical Quantities}\label{sec3}
	
	To construct global solutions, we impose appropriate boundary conditions on the system of ODEs derived in Section \ref{sec2}. Regularity at the origin requires that the metric and matter functions satisfy
	\begin{equation}\label{equ17}
		\begin{split}
			N(0)=1,\quad \sigma(0)=\sigma_0,\quad h(0)=0, \\
			V(0)=V_0,\quad \partial_rf(0)=0, \quad \partial_rV(0)=0.
		\end{split}
	\end{equation}
	
	At infinity, we assume the spacetime is asymptotically flat, so the boundary conditions are
	\begin{equation}\label{equ18}
		N(\infty)=1,\quad \sigma(\infty)=1,\quad f(\infty)=0, \quad h(\infty)=0.
	\end{equation}
	
	The ADM mass $M$ is an important physical quantity of the model and an indicator for monitoring the reliability of numerical values. It can be extracted from the asymptotic subleading behavior of the metric component $g_{tt}$:
	\begin{equation}\label{equ19}
		g_{tt}=-\sigma(r)^2N(r)=-1+\frac{8GM}{3\pi r^2}+...\quad.
	\end{equation}
	
	Additionally, the matter fields $\mathcal{B}$ is invariant under a global $U(1)$ transformation. The corresponding total conserved particle number is
	\begin{equation}
		N_P=-\int j^t_P\left | g \right |^{1/2}d \Omega_3,
	\end{equation}
	with the conserved current of the Proca field
	\begin{equation}
		\ j^{\alpha}_P=\frac{i}{2}\left(\overline{\mathcal{W}}^{\mu \nu} \mathcal{B}_{\nu}-\mathcal{W}^{\mu \nu} \overline{\mathcal{B}}_{\nu}\right),
	\end{equation}
	and the total electric charge in spacetime is $Q = qN_P$. After having the ADM mass and the total number of particles, we can define the binding energy as follows
	\begin{equation}\label{equ20}
		E_B:= mN_P - M.
	\end{equation}
	This binding energy is the difference between the total mass-energy of the star \(M\) and the total rest mass of the bosonic particles. It thus reflects the net balance of kinetic and potential energy contributions. Generally, \(E_B > 0\) indicates a gravitationally bound state, which may be stable or unstable, while \(E_B < 0\) typically corresponds to an unstable state \cite{Kusmartsev:1990cr,Kusmartsev:1989nc,Tamaki:2010zz,Kleihaus:2011sx}.
	
	To characterize the matter distribution and spacetime curvature, we evaluate the components of the energy-momentum tensor and the Kretschmann scalar. The energy density $\rho \equiv -T^0_0$, the principal pressures $P_1 \equiv T^1_1$ (radial) and $P_2 \equiv T^2_2$ (tangential) and the Kretschmann scalar are given by
	\begin{equation}\label{rou}
		\begin{split}
			\rho = -T^0_0 = \frac{1}{2N\sigma^2} \Big[& m^2 f^2 + N \big( h^2 (m^2 N \sigma^2 + (\omega + qV)^2) \\
			&- 2h (\omega + qV) f' + f'^2 + V'^2 \big) \Big],
		\end{split}
	\end{equation}
	\begin{equation}\label{P1}
		\begin{split}
			P_1 = T^1_1 = \frac{1}{2N\sigma^2} \Big[& m^2 f^2 + N \big( -h^2 (-m^2 N \sigma^2 + (\omega + qV)^2) \\
			&+ 2h (\omega + qV) f' - f'^2 - V'^2 \big) \Big],
		\end{split}
	\end{equation}
	\begin{equation}\label{P2}
		\begin{split}
			P_2 = T^2_2 = \frac{1}{2N\sigma^2} \Big[& m^2 f^2 + N \big( h^2 (-m^2 N \sigma^2 + (\omega + qV)^2) \\
			&- 2h (\omega + qV) f' + f'^2 + V'^2 \big) \Big],
		\end{split}
	\end{equation}
	\begin{equation}
		\begin{split}
			K& = R^{\mu\nu\rho\sigma}R_{\mu\nu\rho\sigma} = \frac{1}{r^4 \sigma^2} \Big[  3 r^2 (4 N^2 + 3 r^2 N'^2) \sigma'^2 + \\
			&6 r^2 \sigma N' \sigma' (2 N + r^2 N'') + \sigma^2 (12 (-1+N)^2 + 6 r^2 N'^2 + r^4 N'') \\
			&+ 12 r^4 N N' \sigma' \sigma'' + 4 r^4 N \sigma N''\sigma'' + 4 r^4 N^2 \sigma''^2 \Big].
		\end{split}
	\end{equation}
	
	It is evident that $P_1 \neq P_2$, rendering the properties of this more complex and matter field model within the present gravitational framework distinct from isotropic perfect fluid in \cite{Bueno:2025tli}. Moreover, the energy conditions implied by linear combinations of these physical quantities also require numerical investigation.
	
	\section{Numerical Results}\label{sec4}
	To facilitate numerical computations, we set $4\pi G=1, m=1$, and employ the following scaling transformations to obtain the dimensionless variables:
	\begin{equation}
		r\to r/m,\quad \omega\to\omega m.
	\end{equation}
	Additionally, we introduce a new radial variable
	\begin{equation}
		x=\frac{r}{1+r}.
	\end{equation}
	Through this transformation, we can change the range of the radial coordinate from $\tilde{r}\in[0,\infty)$ to a finite interval $x\in[0,1]$, which is more suitable for numerical integration. The system of differential equations is solved using the finite element method. We discretize the integration domain $0\leq x\leq1$ into 10000 grid points and employ the Newton-Raphson method for iteration. To ensure high numerical precision, we enforce a relative error tolerance of less than $10^{-5}$. 
	
	For clarity, we adopt the following plotting conventions throughout this section: in all figures involving the ADM mass $M$ and particle number $N_P$, solid lines represent $M$, and dashed lines represent $N_P$. In the field profile plots, solid lines denote $f$, en-dash lines denote $h$, and dashed lines denote $V$.
	
	\subsection{$n\leq 2$: Einstein and Gauss-Bonnet gravity}
	
	The correction orders $n=1$ and $n=2$ correspond to five-dimensional charged Proca star solutions in Einstein gravity and Gauss-Bonnet gravity respectively. To our knowledge, while some neutral boson star solutions in higher dimensions can be found in \cite{Hartmann:2010pm,Blazquez-Salcedo:2019qrz,Brihaye:2013zha}, the charged cases remain uninvestigated. Since any order of our modified gravity reduces to Einstein gravity when the coupling parameter $\alpha=0$ , we will focus our discussion on the $n=2$ case to illustrate the effects of the coupling parameter $\alpha$.
	
	\begin{figure}[!htbp]
		\begin{center}
			\begin{subfigure}[b]{0.23\textwidth}
				\includegraphics[width=\textwidth]{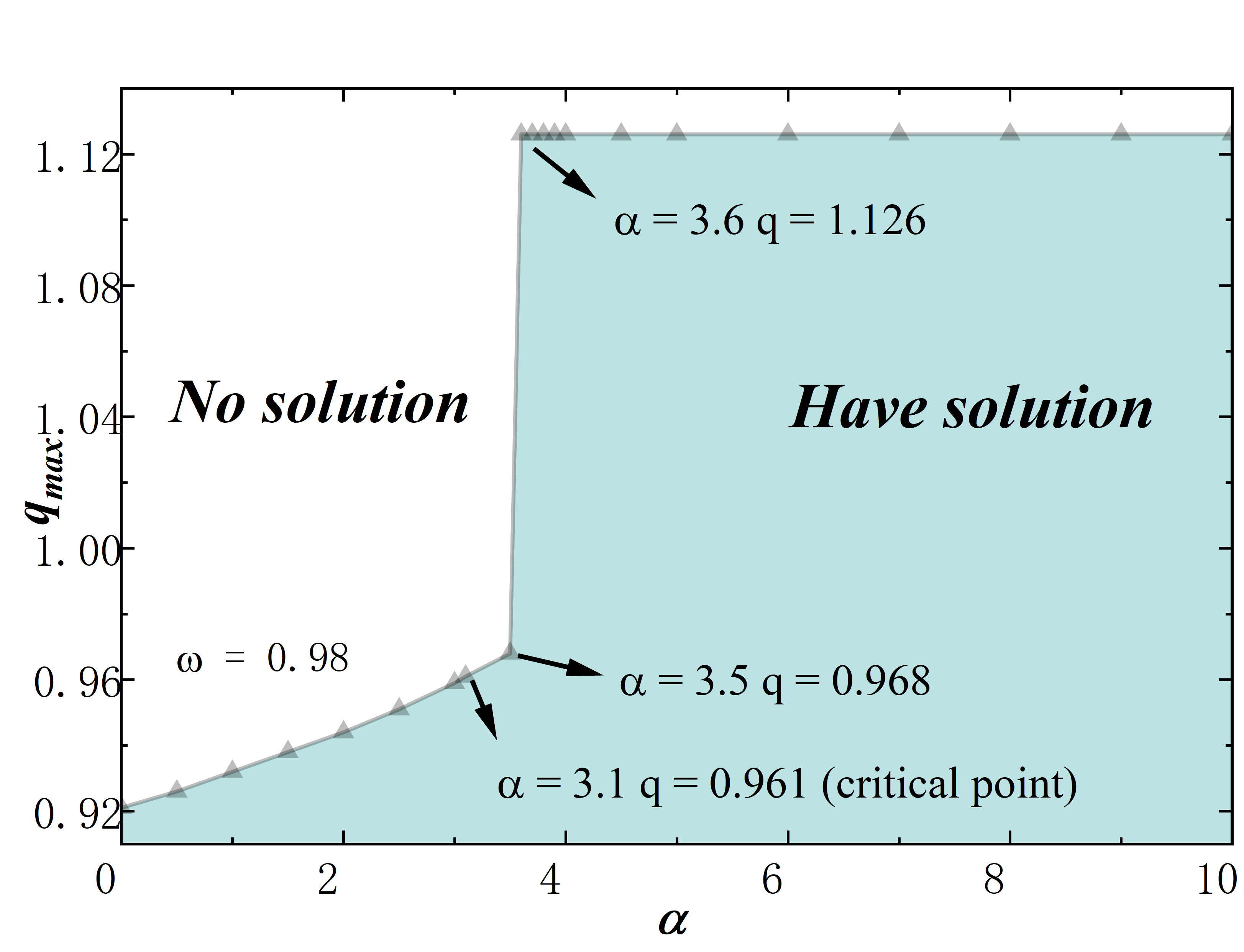}
				\caption{}
			\end{subfigure}
			\begin{subfigure}[b]{0.23\textwidth}
				\includegraphics[width=\textwidth]{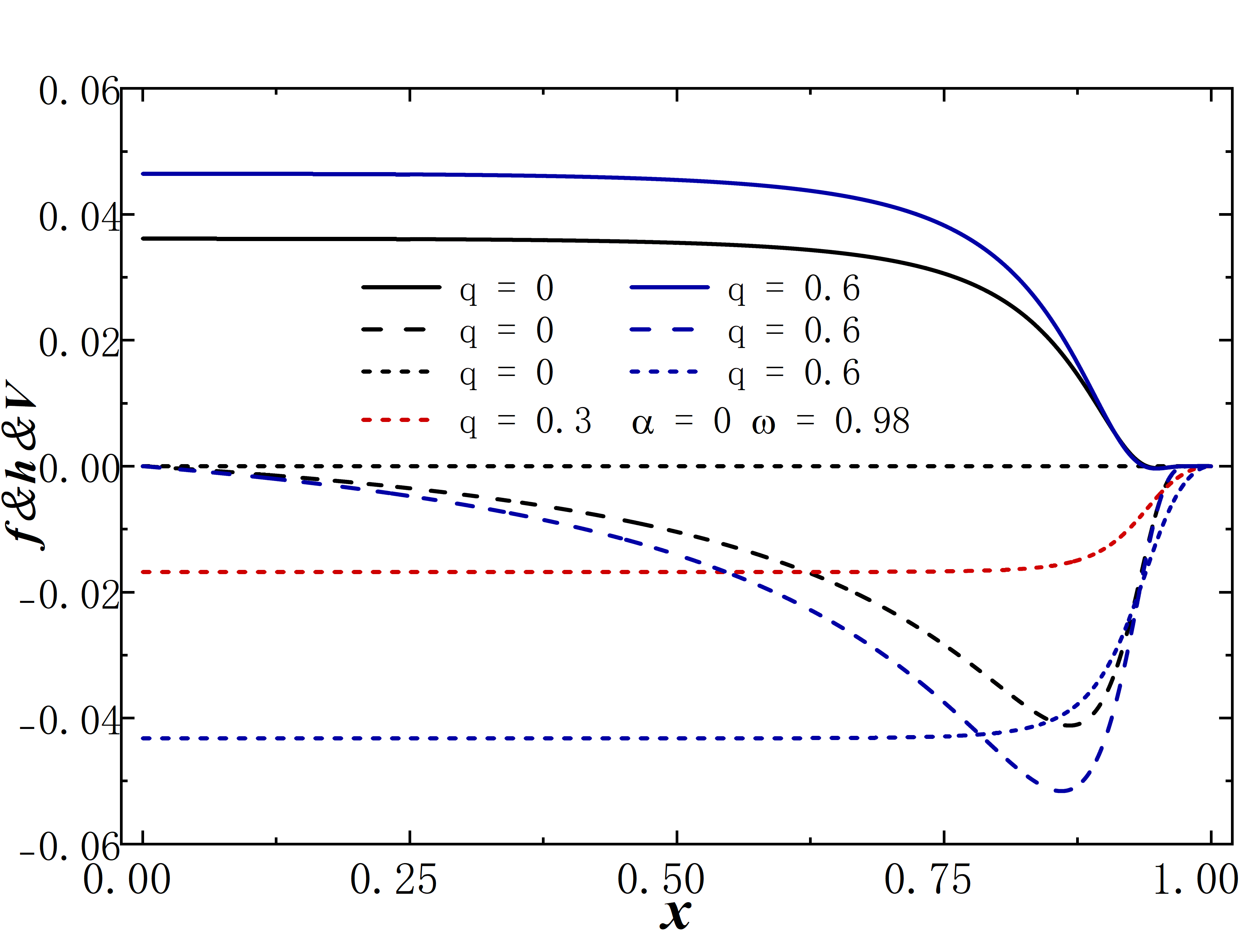}
				\caption{}
			\end{subfigure}
			\begin{subfigure}[b]{0.23\textwidth}
				\includegraphics[width=\textwidth]{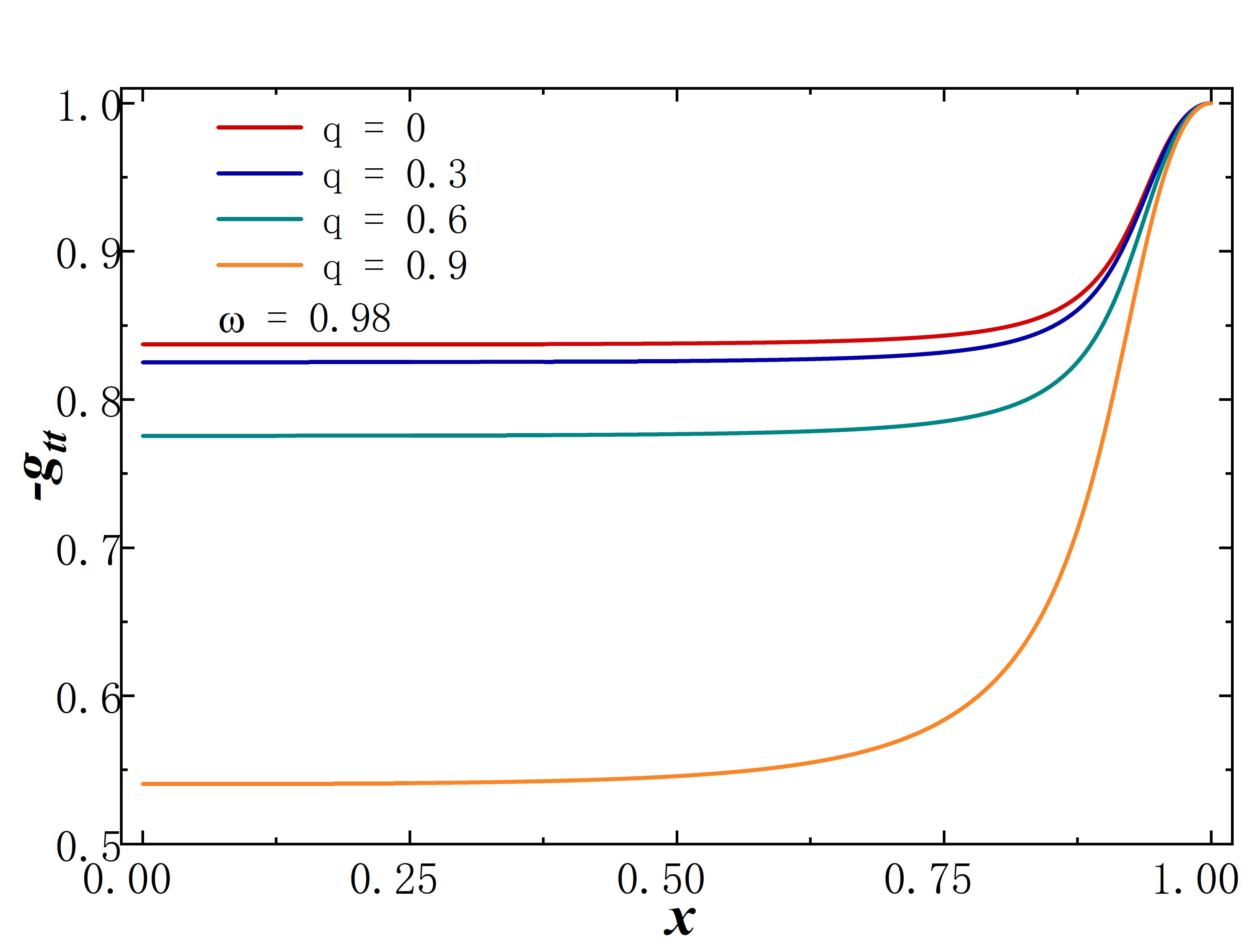}
				\caption{}
			\end{subfigure}
			\begin{subfigure}[b]{0.23\textwidth}
				\includegraphics[width=\textwidth]{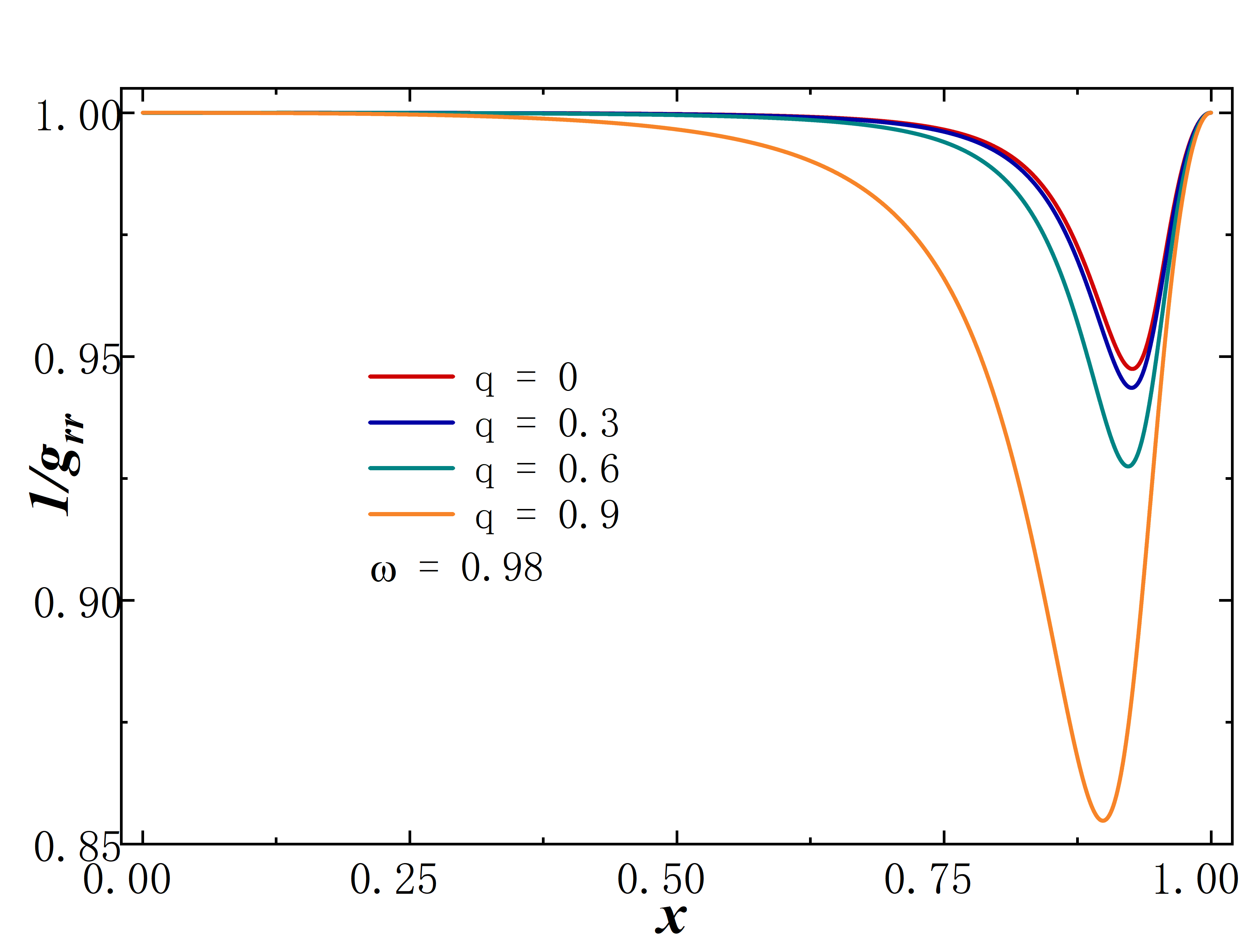}
				\caption{}
			\end{subfigure}
		\end{center}
		\caption{(a): Relationship between the maximum value of the charge parameter $q$ and the coupling parameter $\alpha$ when the fixed frequency $\omega$ is 0.98. (b): Field functions at different $q$ when the $\omega$ is 0.98 and $\alpha$ is 0. The solid lines denote $f$, en-dash lines denote $h$, and dashed lines denote $V$. (c) and (d): The metric component $-g_{tt}$ and $1/g_{rr}$ vs. the radial coordinate $x$ when the $\omega$ is 0.98 and $\alpha$ is 0.}
		\label{p1}
	\end{figure}
	
\begin{figure}[!htbp]
		\begin{center}
			\begin{subfigure}[b]{0.23\textwidth}
				\includegraphics[width=\textwidth]{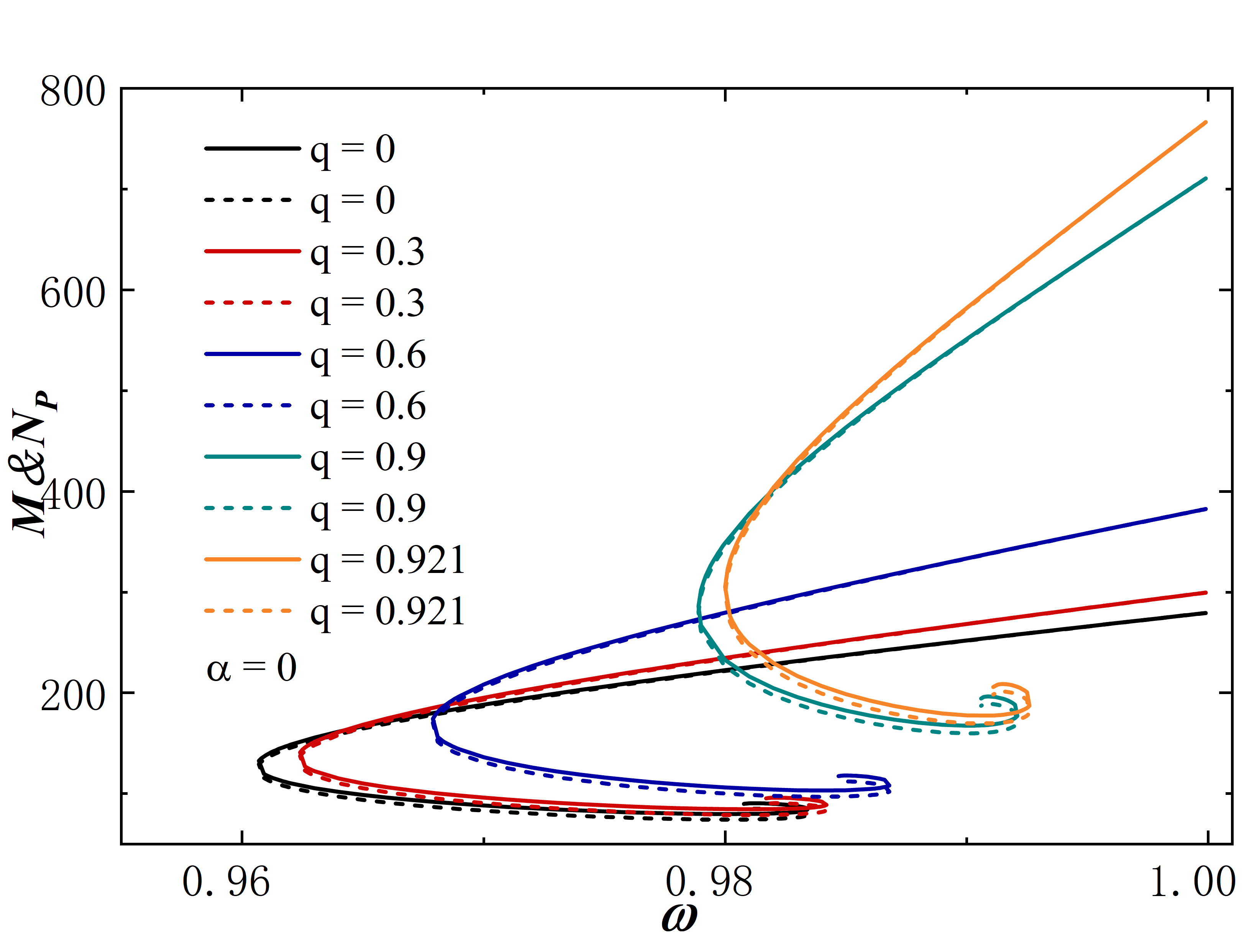}
				\caption{}
			\end{subfigure}
			\begin{subfigure}[b]{0.23\textwidth}
				\includegraphics[width=\textwidth]{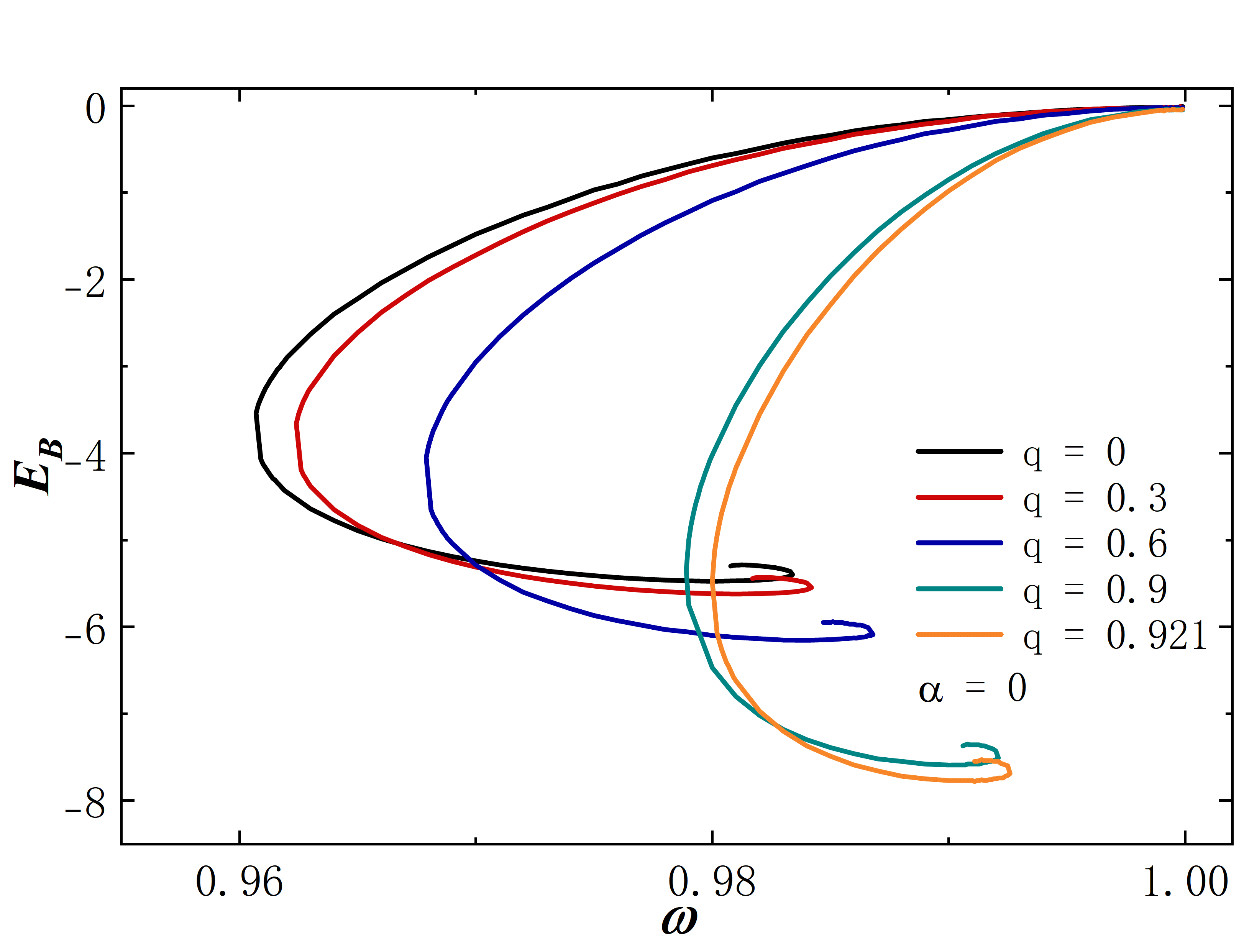}
				\caption{}
			\end{subfigure}
		\end{center}
		\caption{(a): The ADM mass and conserved particle number, as functions of the field frequency $\omega$ with different $q$ under $\alpha = 0$. The solid lines represent $M$, and dashed lines represent $N_P$. (b): The corresponding binding energy.}
		\label{p2}
	\end{figure}

	We first delineate the domain of existence in the $(\alpha, q)$ plane. Fig.~\ref{p1} (a) illustrates the maximum allowable charge $q_{\text{max}}$ for a fixed frequency $\omega=0.98$. The system exhibits a qualitative change in the branch structure: for weak coupling ($\alpha \lesssim 3.5$), $q_{\text{max}}$ grows linearly with $\alpha$, but undergoes a sharp non-linear saturation towards $q \approx 1.126$ in the strong coupling regime. Crucially, the threshold $\alpha \approx 3.1$ marks a topological change in the solution space, separating the standard ``multi-branch spiral structure'' solutions from the ``single non-spiraling branch'' characteristic of high-curvature gravity. To establish a baseline, we examine the Einstein gravity limit ($\alpha=0$). The field profiles (Fig.~\ref{p1} b) and metric functions (Fig.~\ref{p1} c-d) display standard regular behavior. As shown in Fig.~\ref{p2}, the ADM mass $M$ and particle number $N_P$ follow the characteristic spiral structure as a function of frequency $\omega$. 
	
	However, the Einstein-Proca-Maxwell system in five dimensions appears dynamically fragile. As the charge $q$ increases, the frequency domain supporting solutions narrows drastically, shrinking to a width of $\approx 0.02$ at the limiting charge $q=0.921$. Crucially, the binding energy remains negative ($E_B < 0$) throughout this regime, indicating that these states are gravitationally unbound and likely unstable against perturbations.
	
	\begin{figure}
		\begin{center}
			\begin{subfigure}[b]{0.23\textwidth}
				\includegraphics[width=\textwidth]{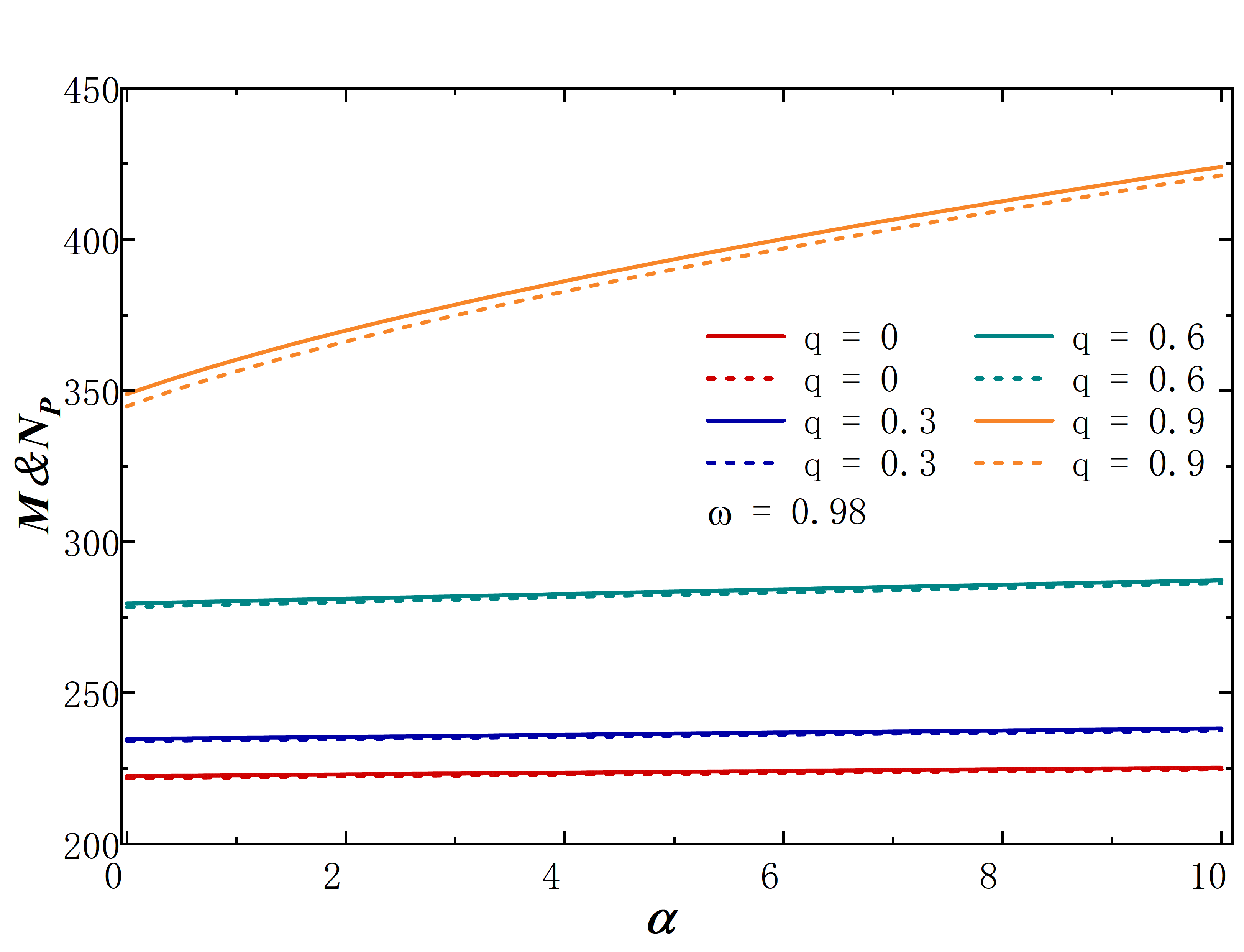}
				\caption{}
			\end{subfigure}
			\begin{subfigure}[b]{0.23\textwidth}
				\includegraphics[width=\textwidth]{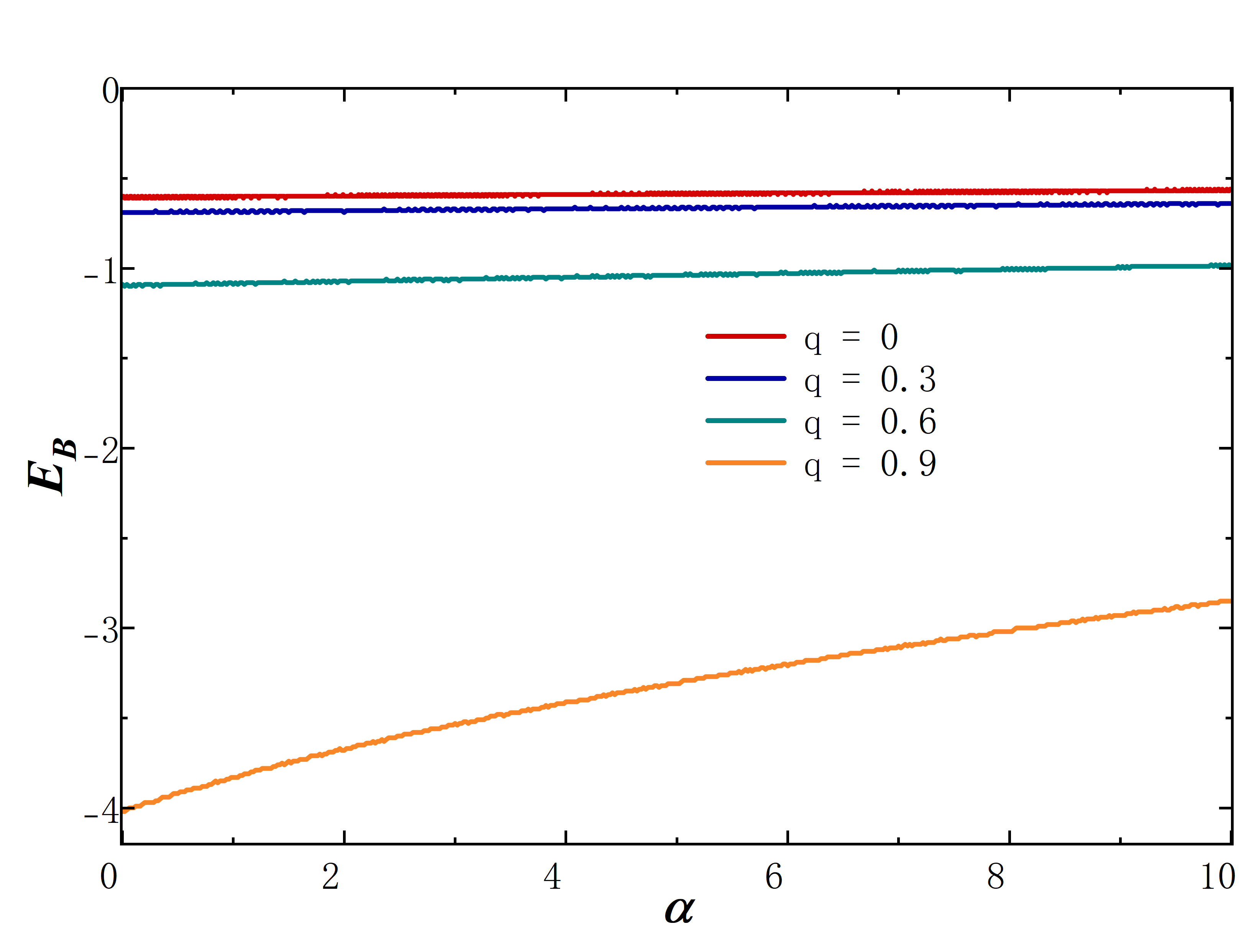}
				\caption{}
			\end{subfigure}
		\end{center}
		\caption{(a): The ADM mass and conserved particle number vs. $\alpha$ with different $q$ under $\omega = 0.98$. The solid lines represent $M$, and dashed lines represent $N_P$. (b): The corresponding binding energy.}
		\label{p3}
	\end{figure}
	
\begin{figure}[!htbp]
		\begin{center}
			\begin{subfigure}[b]{0.23\textwidth}
				\includegraphics[width=\textwidth]{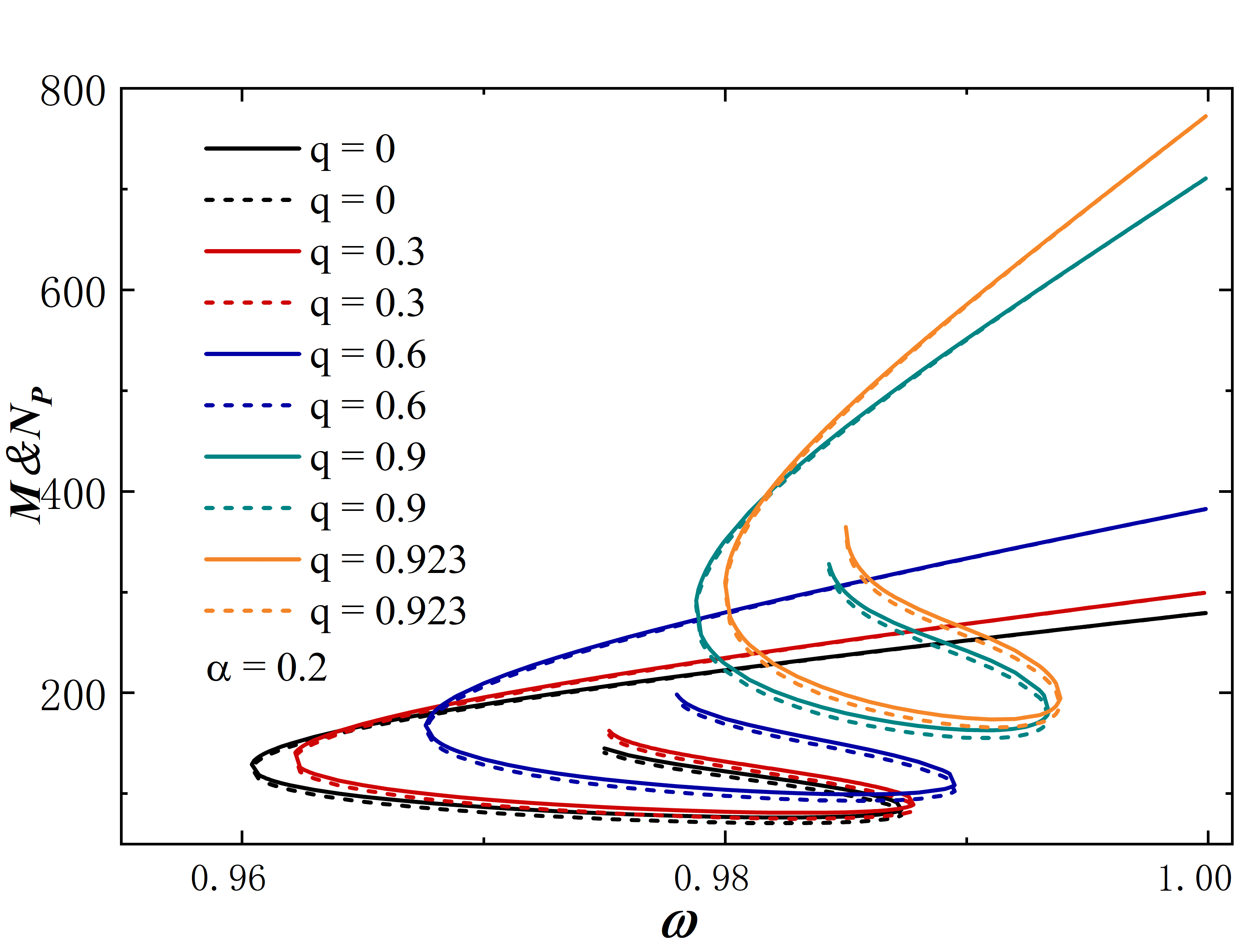}
				\caption{}
			\end{subfigure}
			\begin{subfigure}[b]{0.23\textwidth}
				\includegraphics[width=\textwidth]{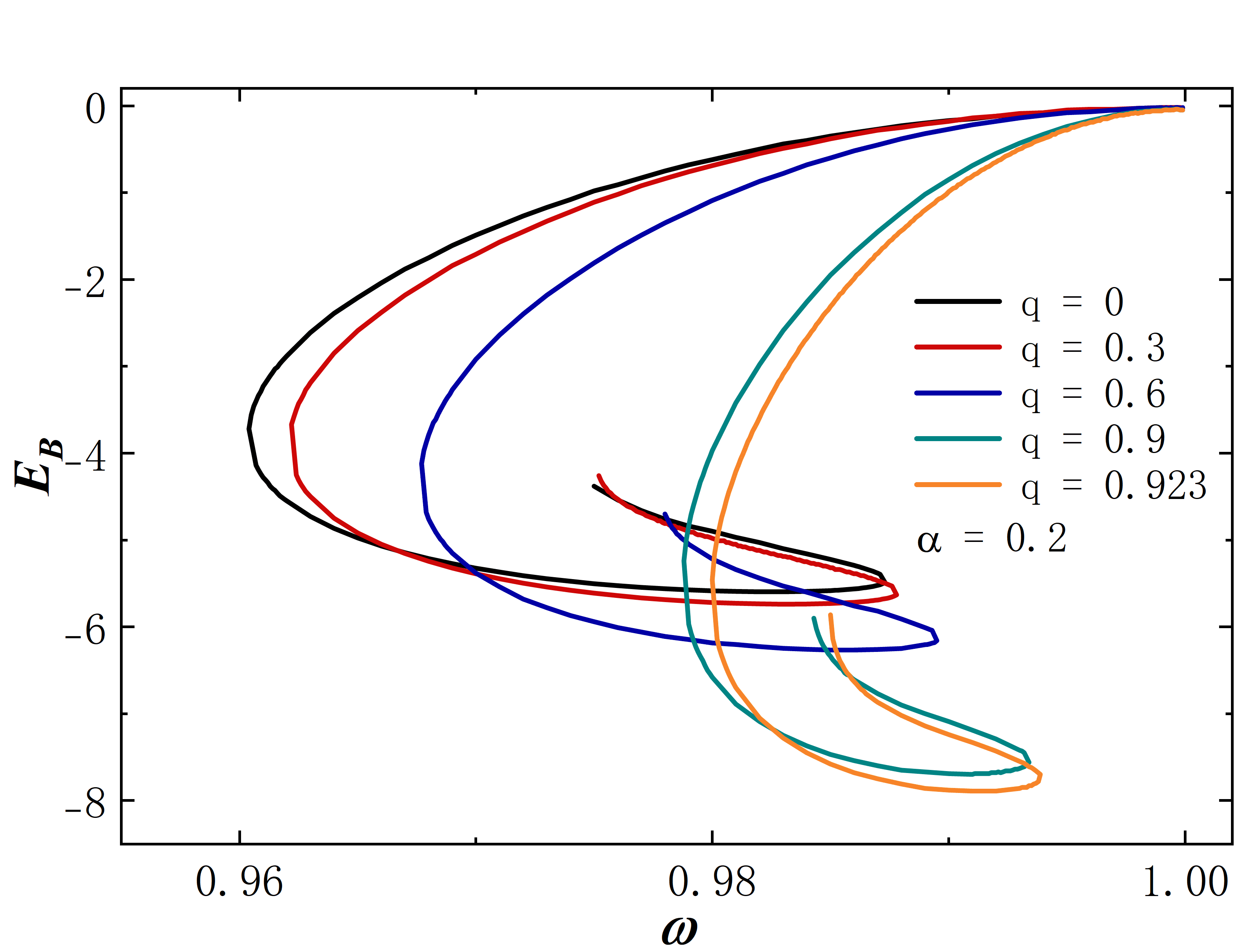}
				\caption{}
			\end{subfigure}
			\begin{subfigure}[b]{0.23\textwidth}
				\includegraphics[width=\textwidth]{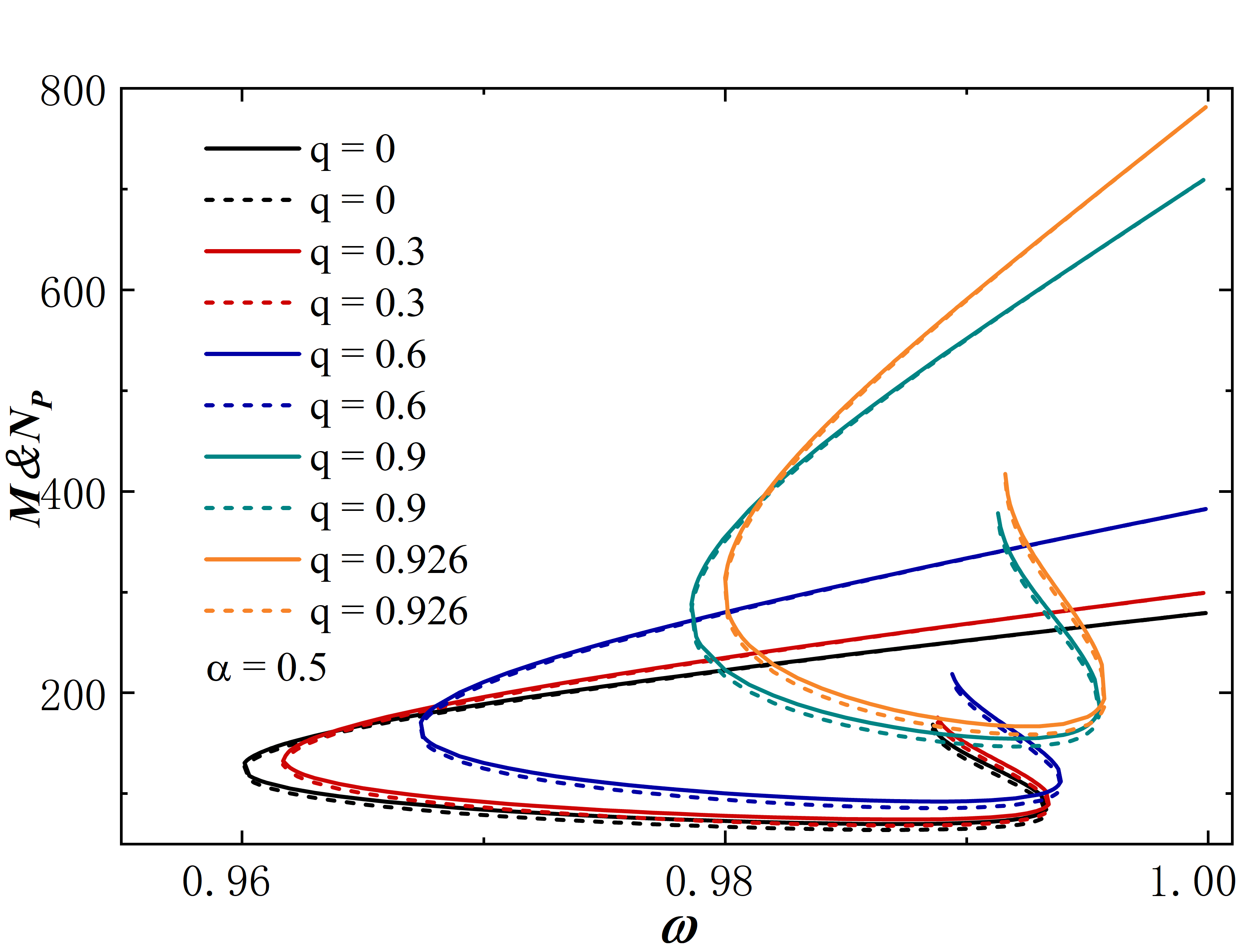}
				\caption{}
			\end{subfigure}
			\begin{subfigure}[b]{0.23\textwidth}
				\includegraphics[width=\textwidth]{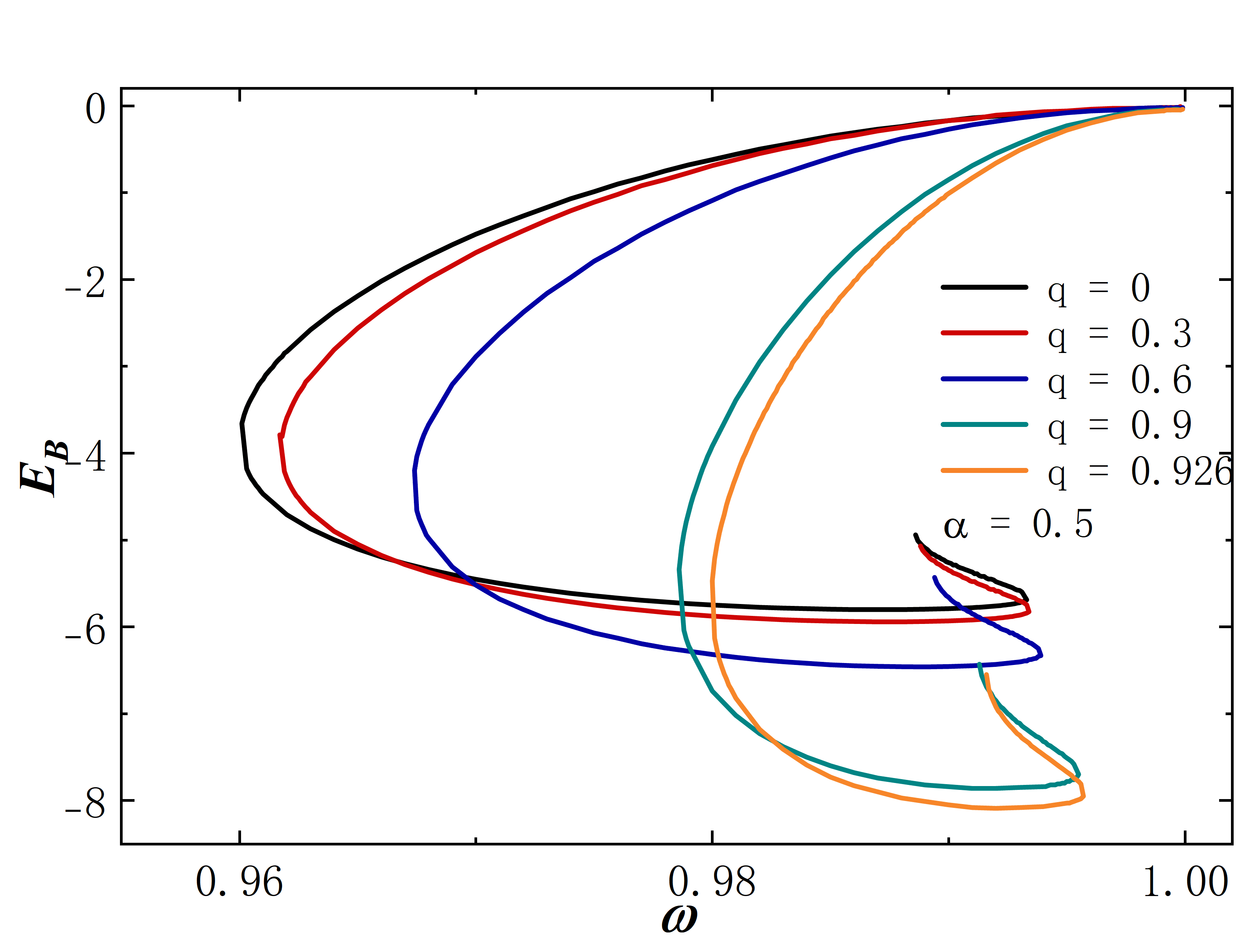}
				\caption{}
			\end{subfigure}
		\end{center}
		\caption{(a) and (b): The ADM mass and conserved particle number, as functions of the field frequency $\omega$ with different $q$ under $\alpha = 0.2$ and the corresponding binding energy. (c) and (d): The ADM mass and conserved particle number, as functions of the field frequency $\omega$ with different $q$ under $\alpha = 0.5$ and the corresponding binding energy. In (a) and (c), the solid lines represent $M$, and dashed lines represent $N_P$.}
		\label{p4}
	\end{figure}

	As shown in Fig.~\ref{p3}, increasing $\alpha$ leads to a monotonic increase in both the ADM mass $M$ and particle number $N_P$ for a fixed frequency. Similarly, increasing the charge $q$ further amplifies these quantities. However, despite the inclusion of higher-curvature terms, solutions in the weak coupling regime (e.g., $\alpha=0.2$ and $0.5$, shown in Fig.~\ref{p4}) retain a negative binding energy. This suggests that perturbative Gauss-Bonnet corrections alone are insufficient to stabilize charged Proca stars in this parameter regime.
	
	\begin{figure}[!htbp]
		\begin{center}
			\begin{subfigure}[b]{0.23\textwidth}
				\includegraphics[width=\textwidth]{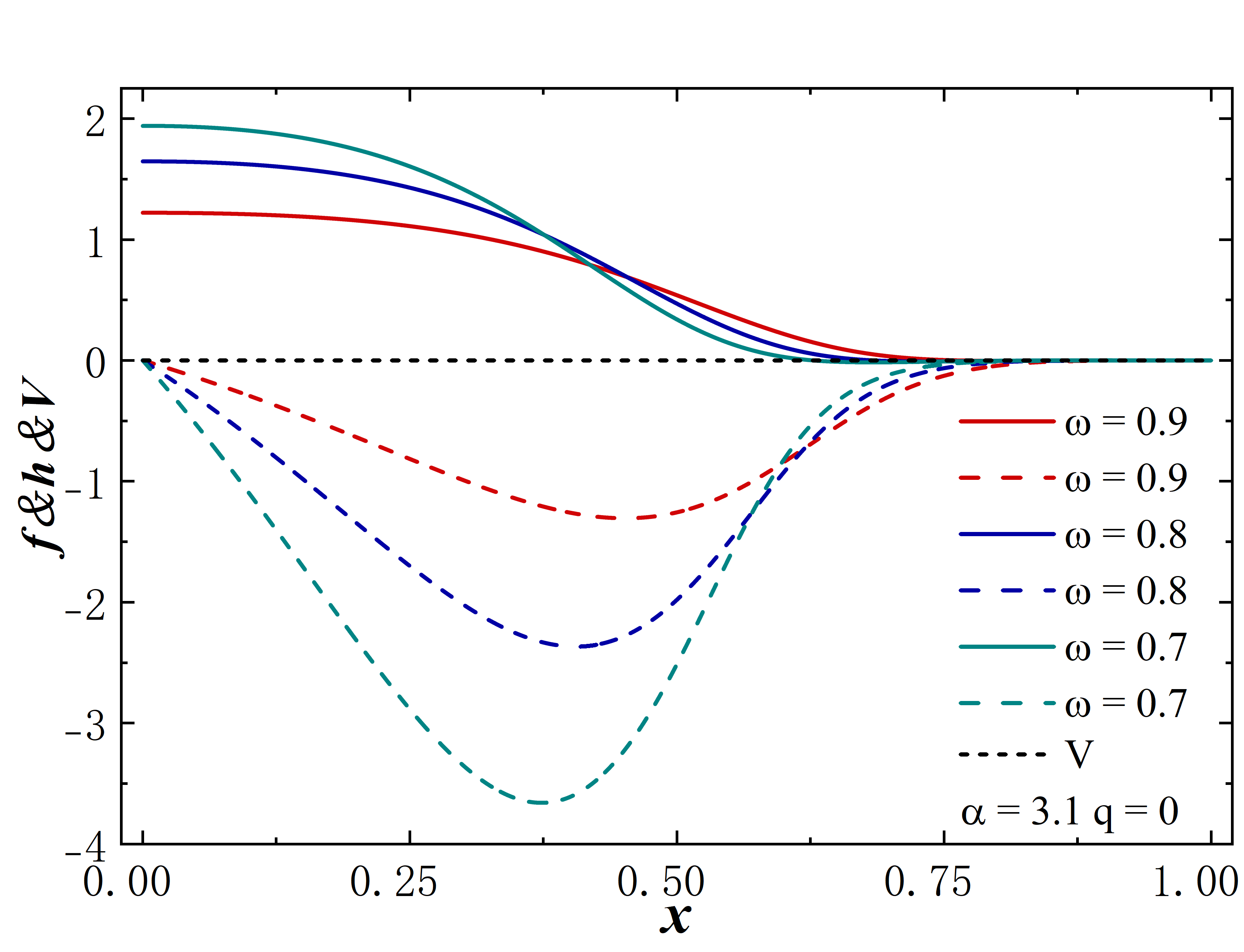}
				\caption{}
			\end{subfigure}
			\begin{subfigure}[b]{0.23\textwidth}
				\includegraphics[width=\textwidth]{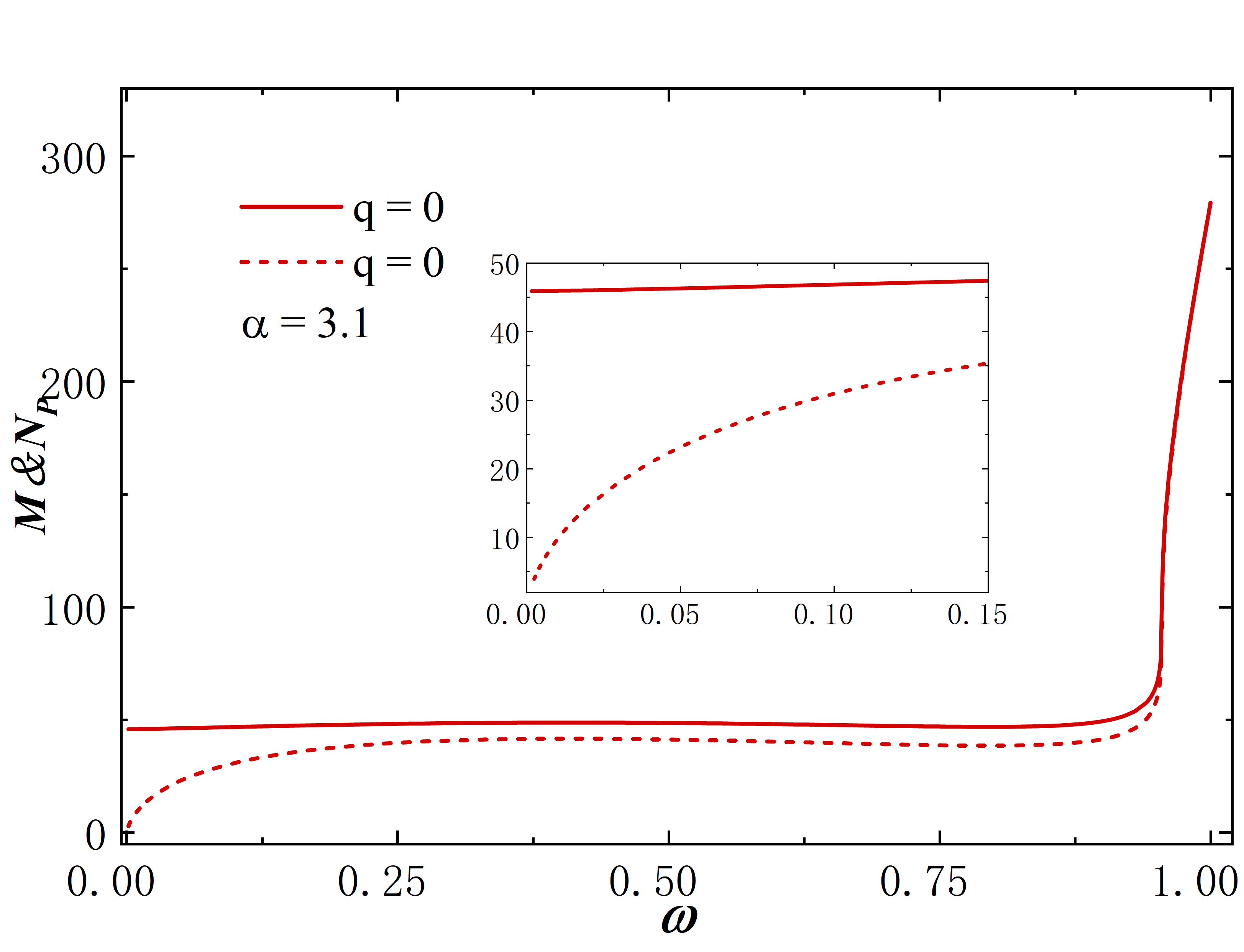}
				\caption{}
			\end{subfigure}
		\end{center}
		\caption{(a): Field functions at different $\omega$ when the $\alpha$ is 3.1 and $q$ is 0. The solid lines denote $f$, en-dash lines denote $h$, and dashed lines denote $V$. (b): The ADM mass and conserved particle number, as functions of the field frequency $\omega$ under $\alpha = 3.1$ and $q = 0$. The solid lines represent $M$, and dashed lines represent $N_P$.}
		\label{p5}
	\end{figure}
	
	We now proceed to the regime of strong coupling, specifically $\alpha \ge 3.1$. At this critical value, the solution landscape changes fundamentally. As illustrated in Fig.~\ref{p5}, the characteristic spiral structure observed in the $M$-$\omega$ and $N_P$-$\omega$ diagrams disappears. Instead, the solutions form a single, monotonic branch extending continuously from the vacuum limit $\omega \to 1$ down to the zero-frequency limit $\omega \to 0$. In this low-frequency limit, the ADM mass $M$ approaches a finite constant, while the particle number $N_P$ vanishes.
	
	In the supercritical regime ($\alpha \ge 3.1$), the system exhibits consistent physical behavior; without loss of generality, we analyze the case $\alpha=4$.
	
\begin{figure}
		\begin{center}
			\begin{subfigure}[b]{0.23\textwidth}
				\includegraphics[width=\textwidth]{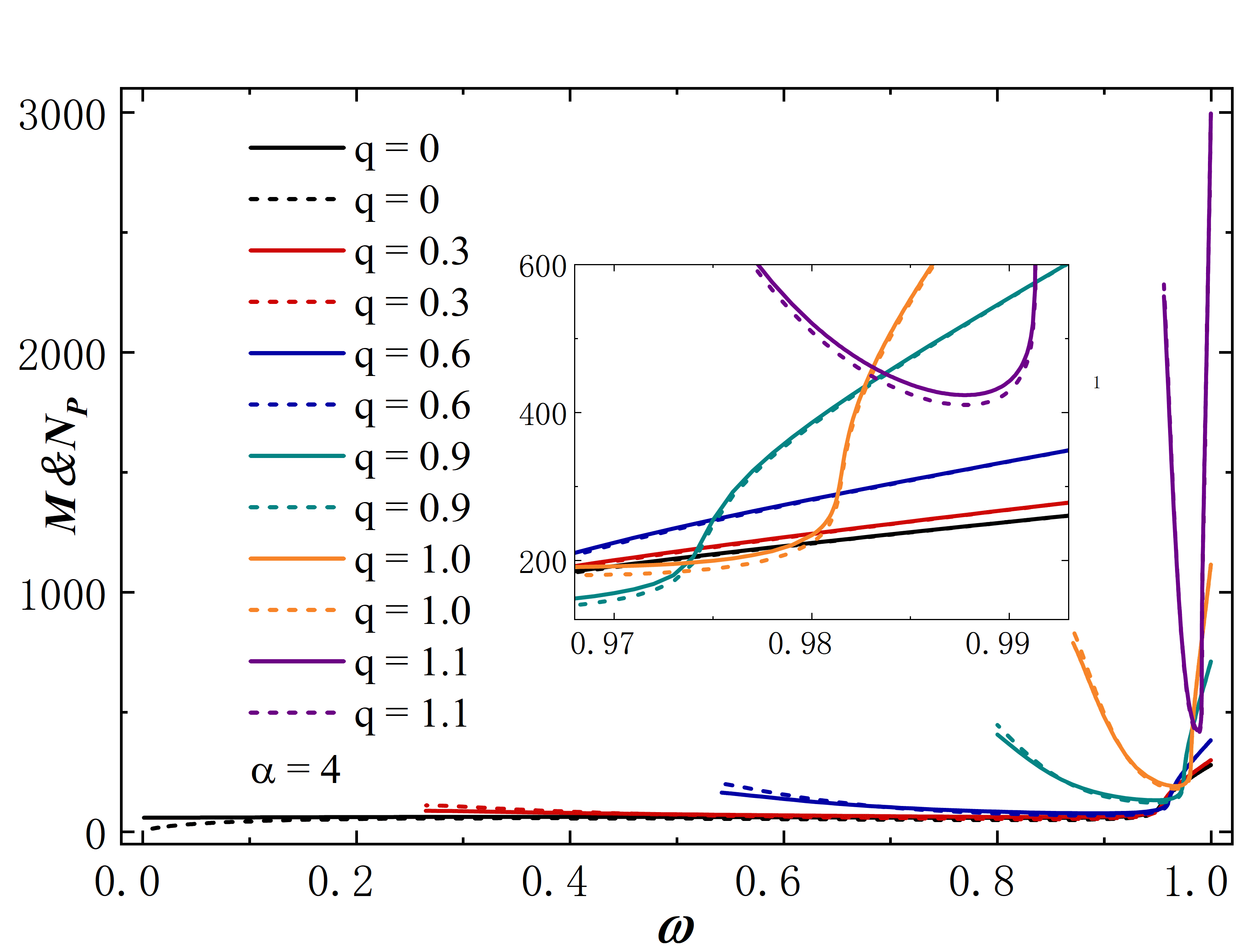}
				\caption{}
			\end{subfigure}
			\begin{subfigure}[b]{0.23\textwidth}
				\includegraphics[width=\textwidth]{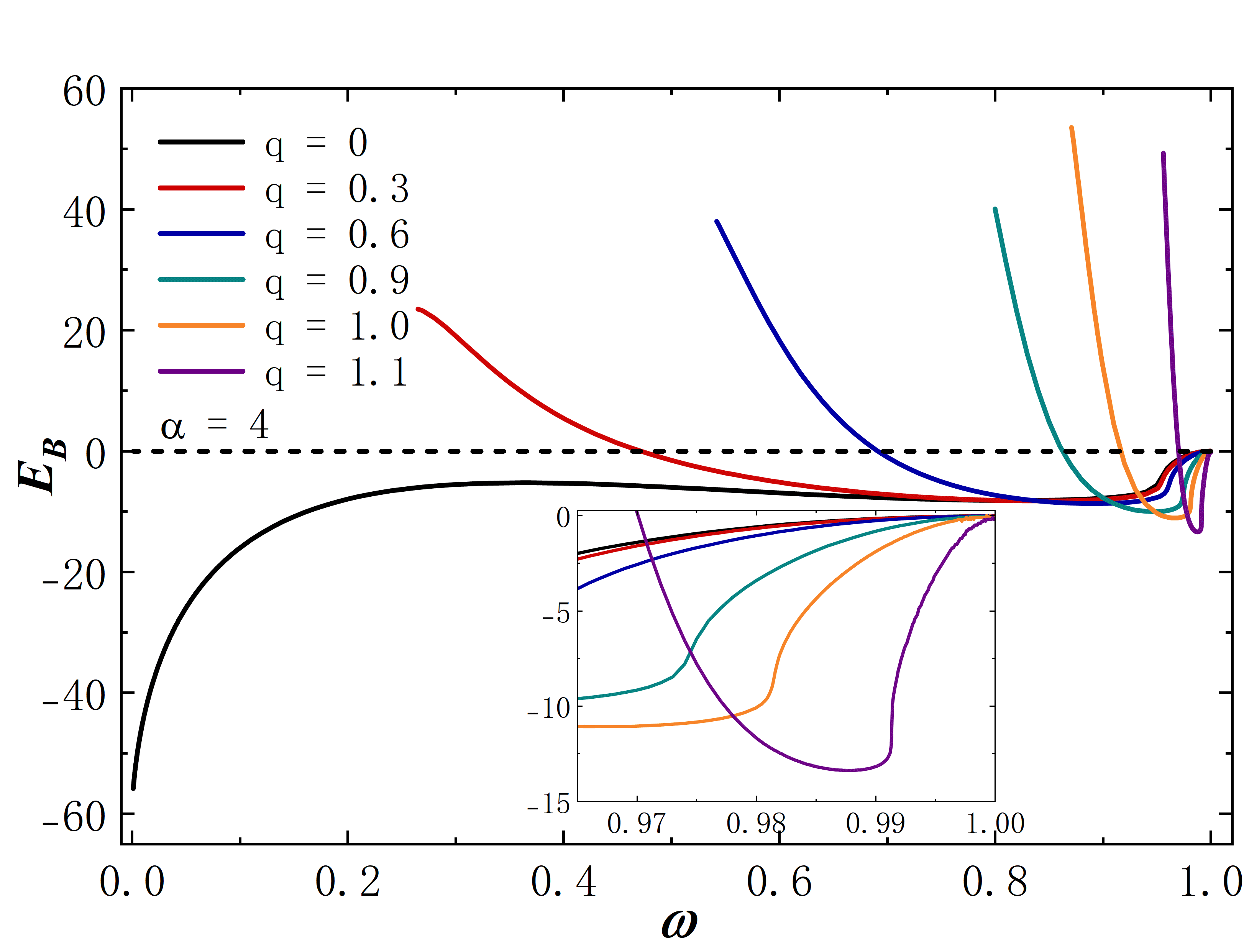}
				\caption{}
			\end{subfigure}
		\end{center}
		\caption{$M$, $N_P$ and the $E_B$ vs. the frequency $\omega$ with different $q$ under $\alpha = 4$. In (a), the solid lines represent $M$, and dashed lines represent $N_P$.}
		\label{p6}
	\end{figure}

	Fig.~\ref{p6} presents the mass, particle number, and binding energy for various charges. For the neutral case ($q=0$), solutions exist over the full frequency range $(0, 1)$. However, the presence of electric charge narrows this domain: while the upper limit remains $\omega \to 1$, the lower frequency bound $\omega_{\text{min}}$ increases monotonically with $q$.
	Most notably, unlike the Einstein and perturbative Gauss-Bonnet cases, the supercritical solutions with $q \neq 0$ exhibit a broad region of positive binding energy ($E_B > 0$). This indicates that the system has entered a gravitationally bound state, suggesting a potential stability. While a rigorous stability analysis requires time-dependent simulations, this positive binding energy is a strong indicator of physical viability.
	
	\begin{figure}
		\begin{center}
			\begin{subfigure}[b]{0.23\textwidth}
				\includegraphics[width=\textwidth]{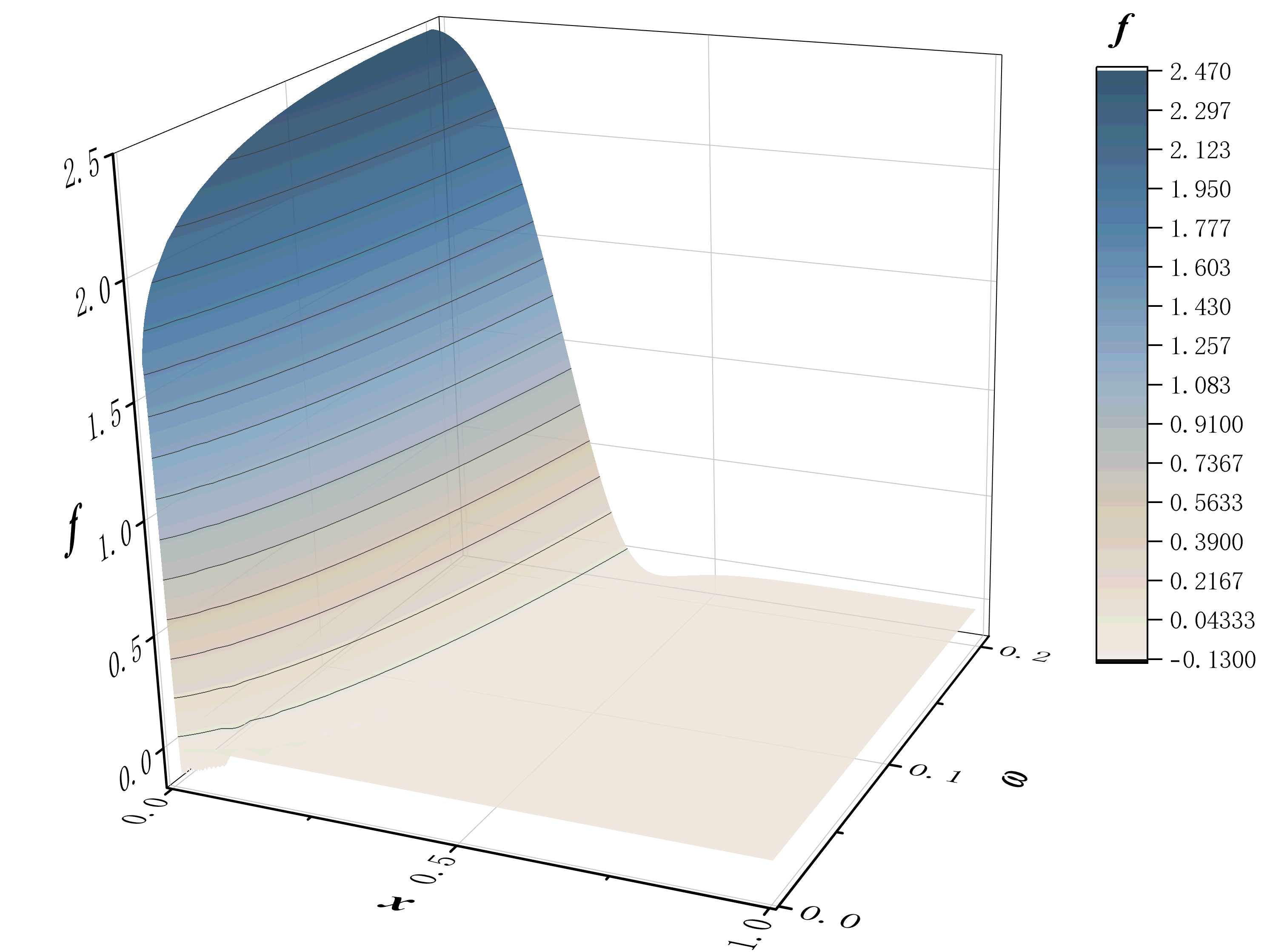}
				\caption{}
			\end{subfigure}
			\begin{subfigure}[b]{0.23\textwidth}
				\includegraphics[width=\textwidth]{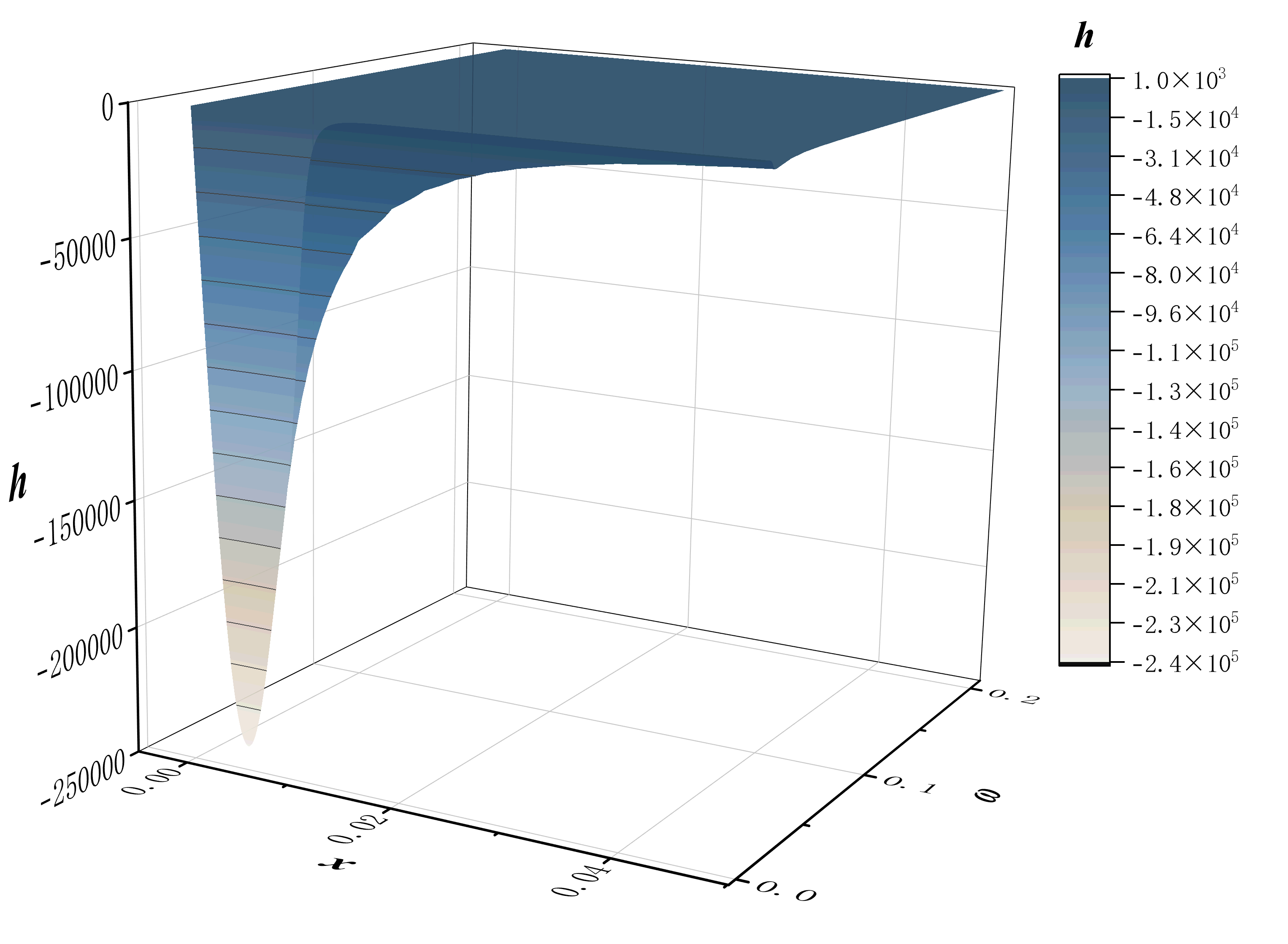}
				\caption{}
			\end{subfigure}
			\begin{subfigure}[b]{0.23\textwidth}
				\includegraphics[width=\textwidth]{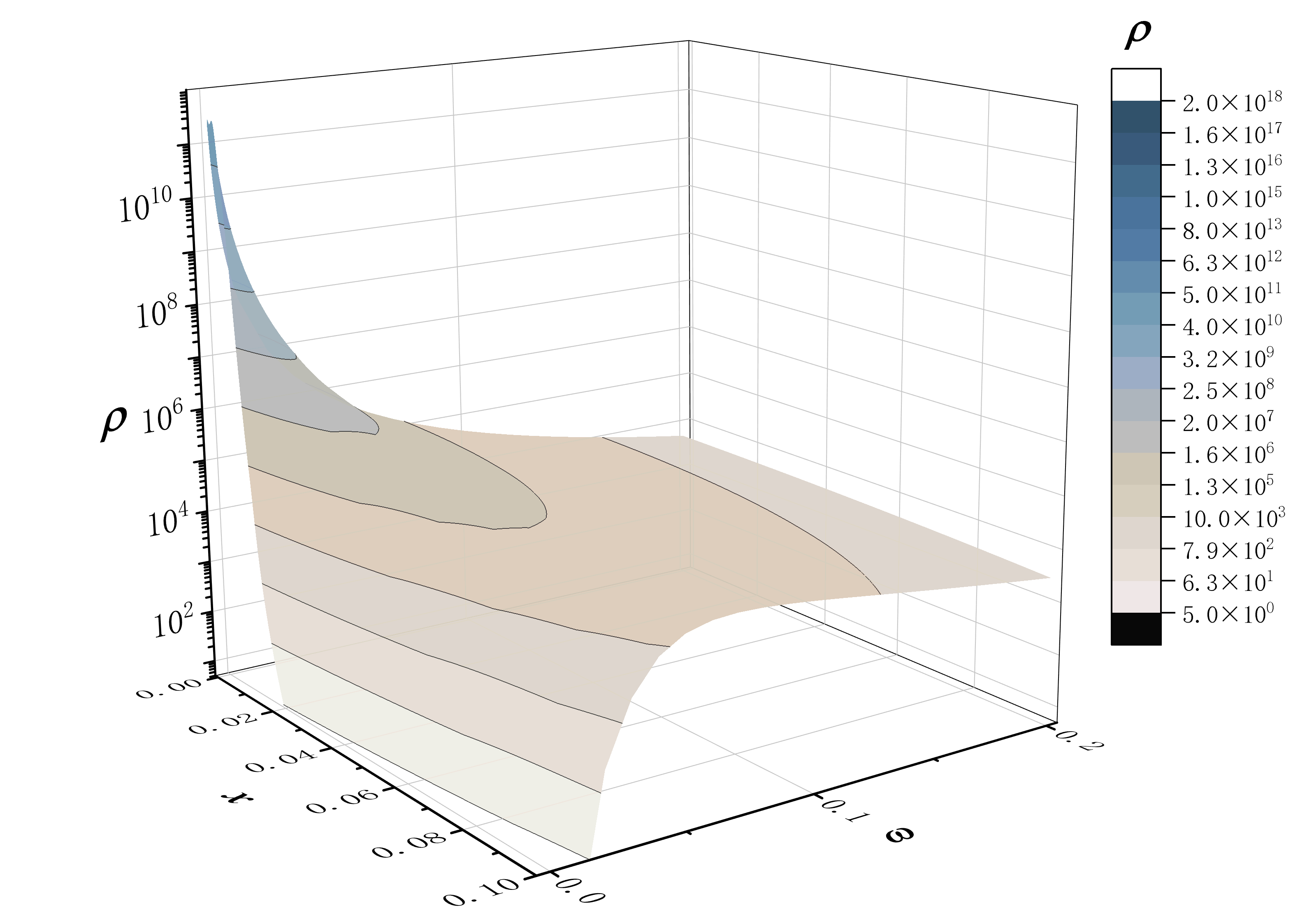}
				\caption{}
			\end{subfigure}
			\begin{subfigure}[b]{0.23\textwidth}
				\includegraphics[width=\textwidth]{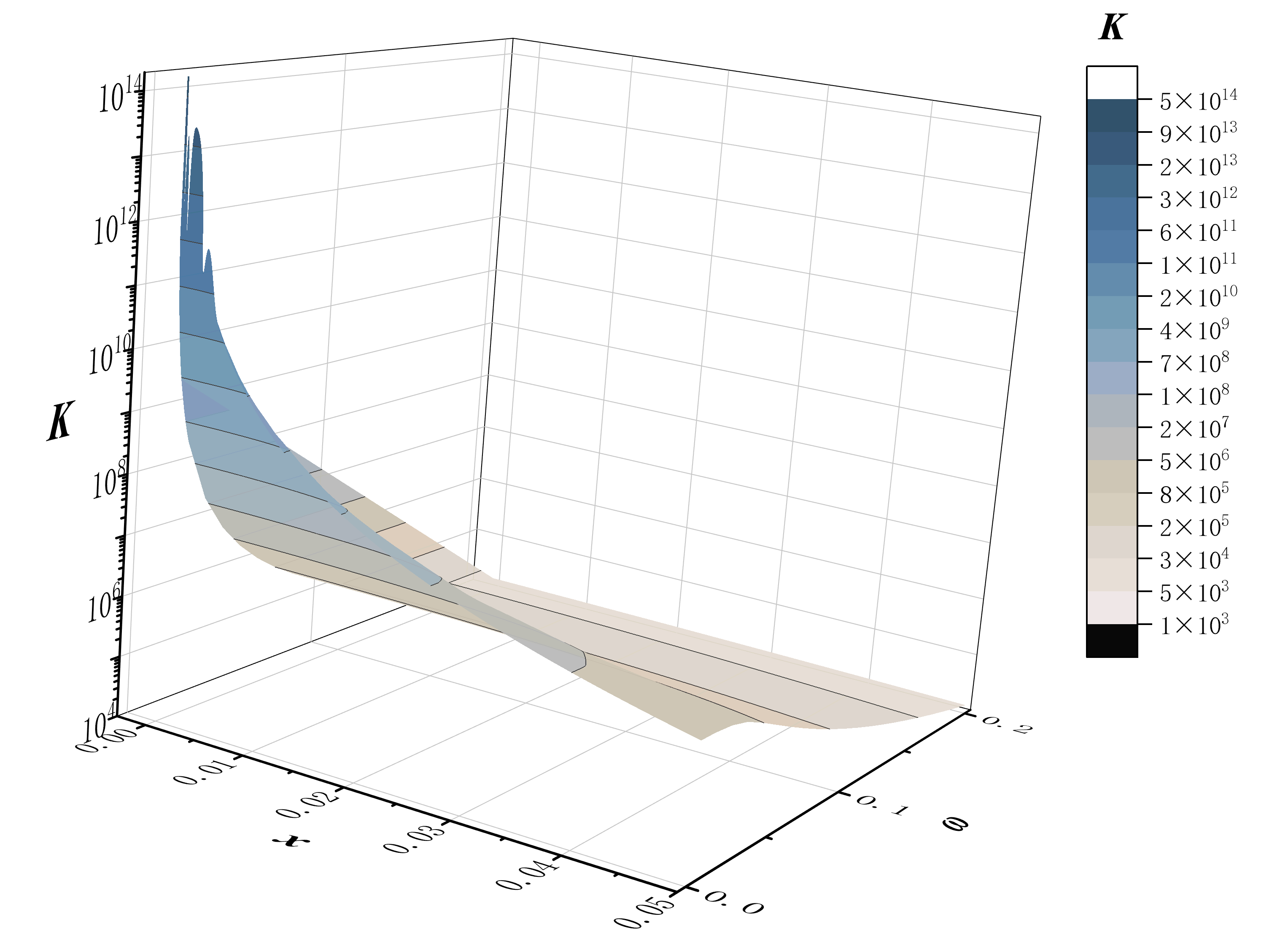}
				\caption{}
			\end{subfigure}
		\end{center}
		\caption{Three-dimensional surface plots of the Proca field components $f$ (a) and $h$ (b), the matter energy density $\rho$ (c), and the Kretschmann scalar $K$ (d) as functions of the radial coordinate $x$ and the field frequency $\omega$.}
		\label{p7}
	\end{figure}
	
\begin{figure}[!htbp]
		\begin{center}
			\begin{subfigure}[b]{0.23\textwidth}
				\includegraphics[width=\textwidth]{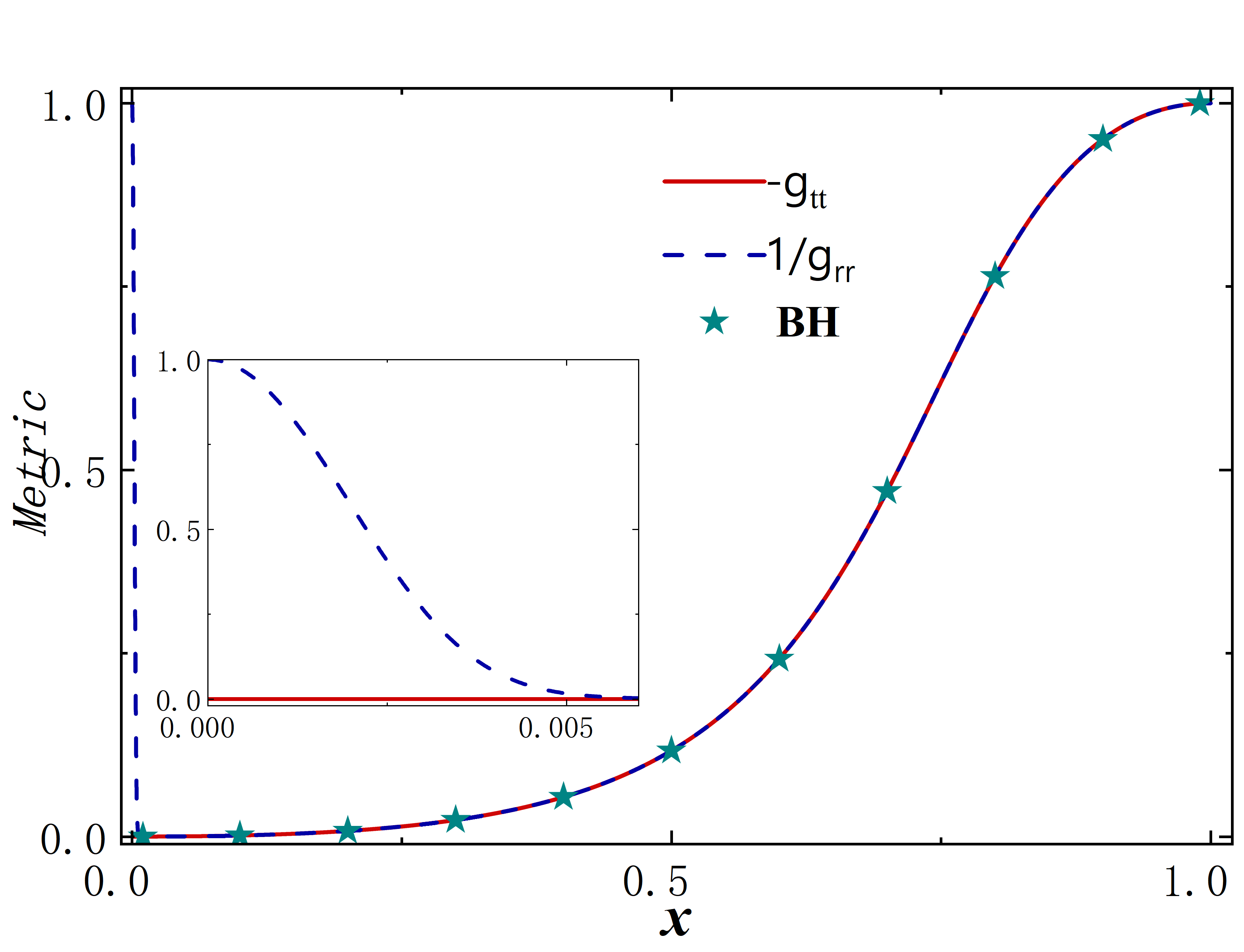}
				\caption{}
			\end{subfigure}
			\begin{subfigure}[b]{0.23\textwidth}
				\includegraphics[width=\textwidth]{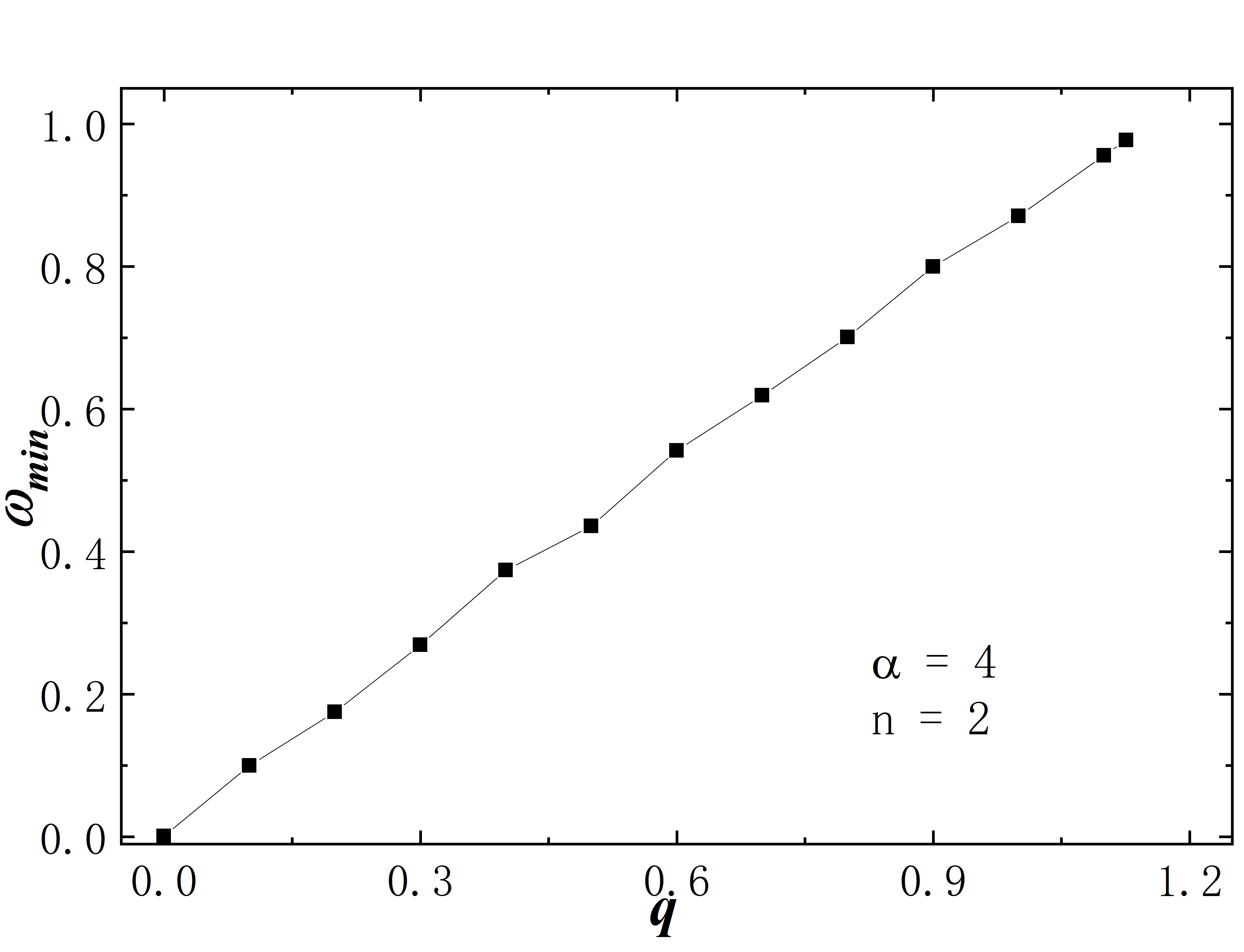}
				\caption{}
			\end{subfigure}
		\end{center}
		\caption{(a): The comparison of metrics between the charged Proca star with $\omega = 0.001$ and the 5D extreme Gauss-Bonnet BH. (b): Relationship between the minimum value of the $\omega$ and the electric parameter $q$.}
		\label{p8}
	\end{figure}

	Now, we turn our attention to the physical properties of the $q=0$ solution in the $\omega\rightarrow0$ limit. In this regime, the matter fields and energy density retract towards the origin (Fig.~\ref{p7}), leading to a divergence in the Kretschmann scalar $K$. This implies that for $n=2$, the ``frozen'' limit corresponds to a singular configuration rather than a regular soliton. Fig.~\ref{p8} (a) offers a geometric interpretation: externally, the spacetime is indistinguishable from an extremal Gauss-Bonnet black hole; internally, however, the diverging $1/g_{rr}$ component reveals that the Gauss-Bonnet term alone is insufficient to regularize the core.
	
	The introduction of electric charge ($q \neq 0$) acts as a regularizing barrier against this zero-frequency collapse. As illustrated in Fig.~\ref{p8} (b), the minimum accessible frequency $\omega_{\text{min}}$ rises monotonically with $q$. This ``unfreezing'' effect is driven by Coulomb repulsion, which forbids the extreme matter concentration required for the $\omega \to 0$ state.
	
\begin{figure}[!htbp]
		\begin{center}
			\begin{subfigure}[b]{0.23\textwidth}
				\includegraphics[width=\textwidth]{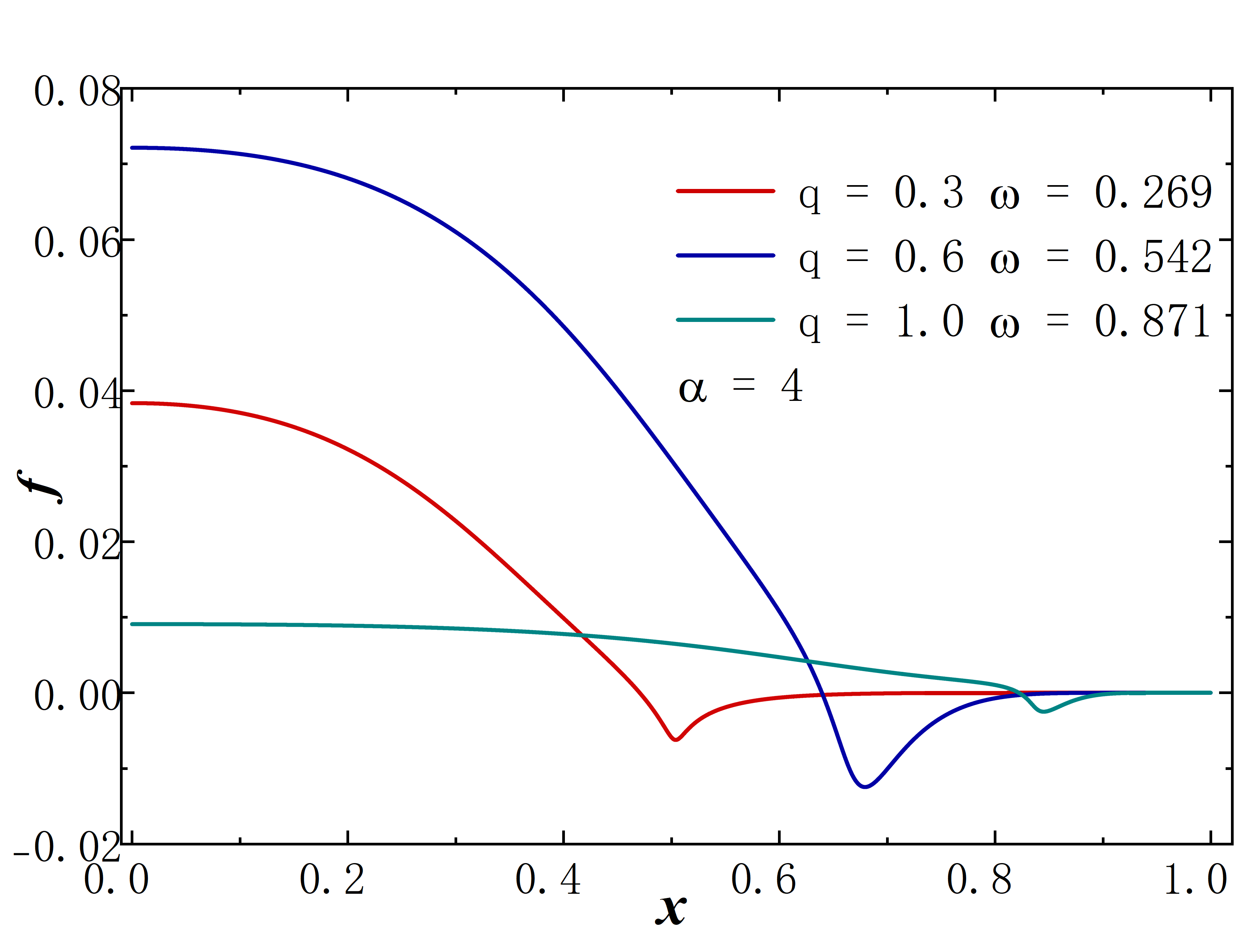}
				\caption{}
			\end{subfigure}
			\begin{subfigure}[b]{0.23\textwidth}
				\includegraphics[width=\textwidth]{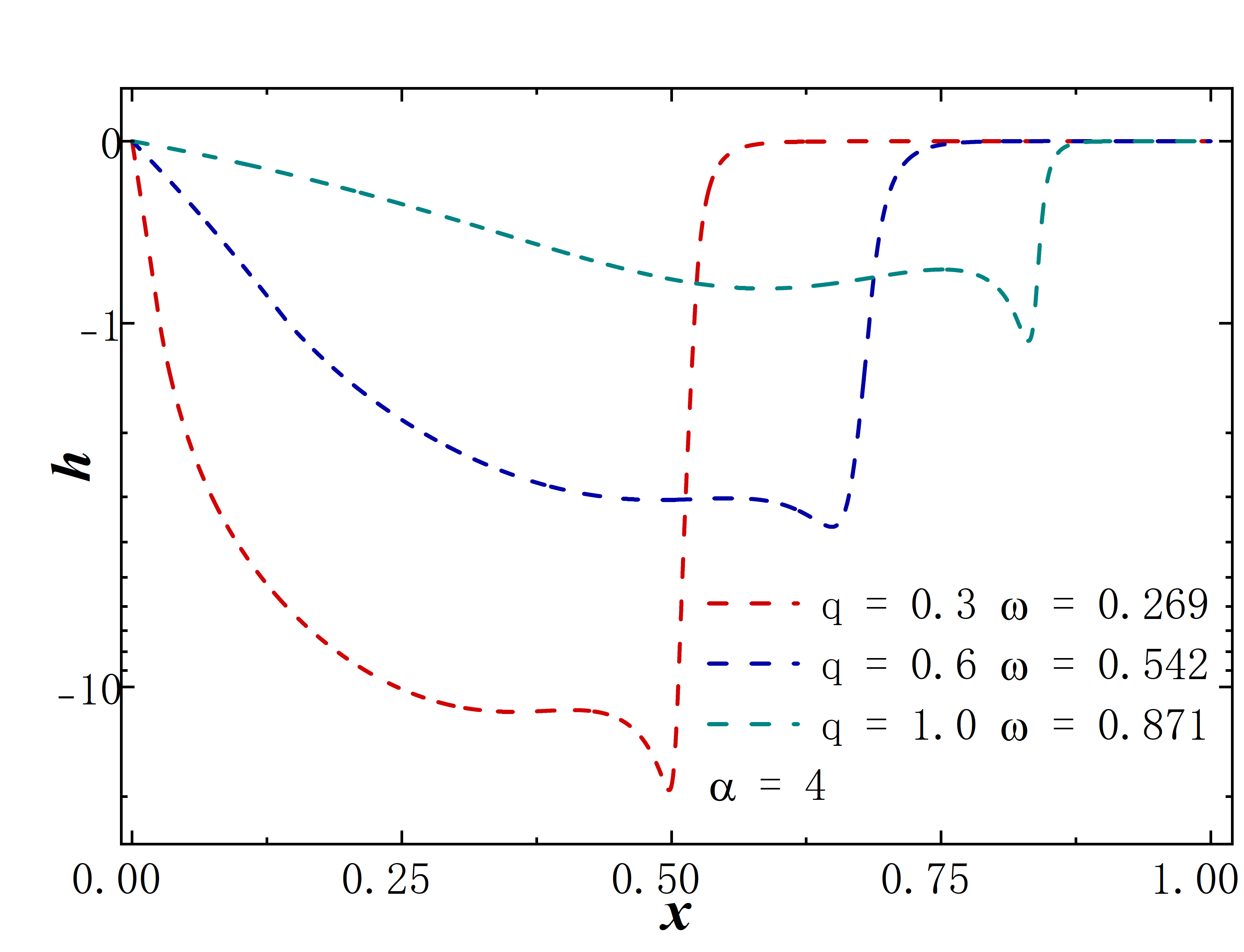}
				\caption{}
			\end{subfigure}
			\begin{subfigure}[b]{0.23\textwidth}
				\includegraphics[width=\textwidth]{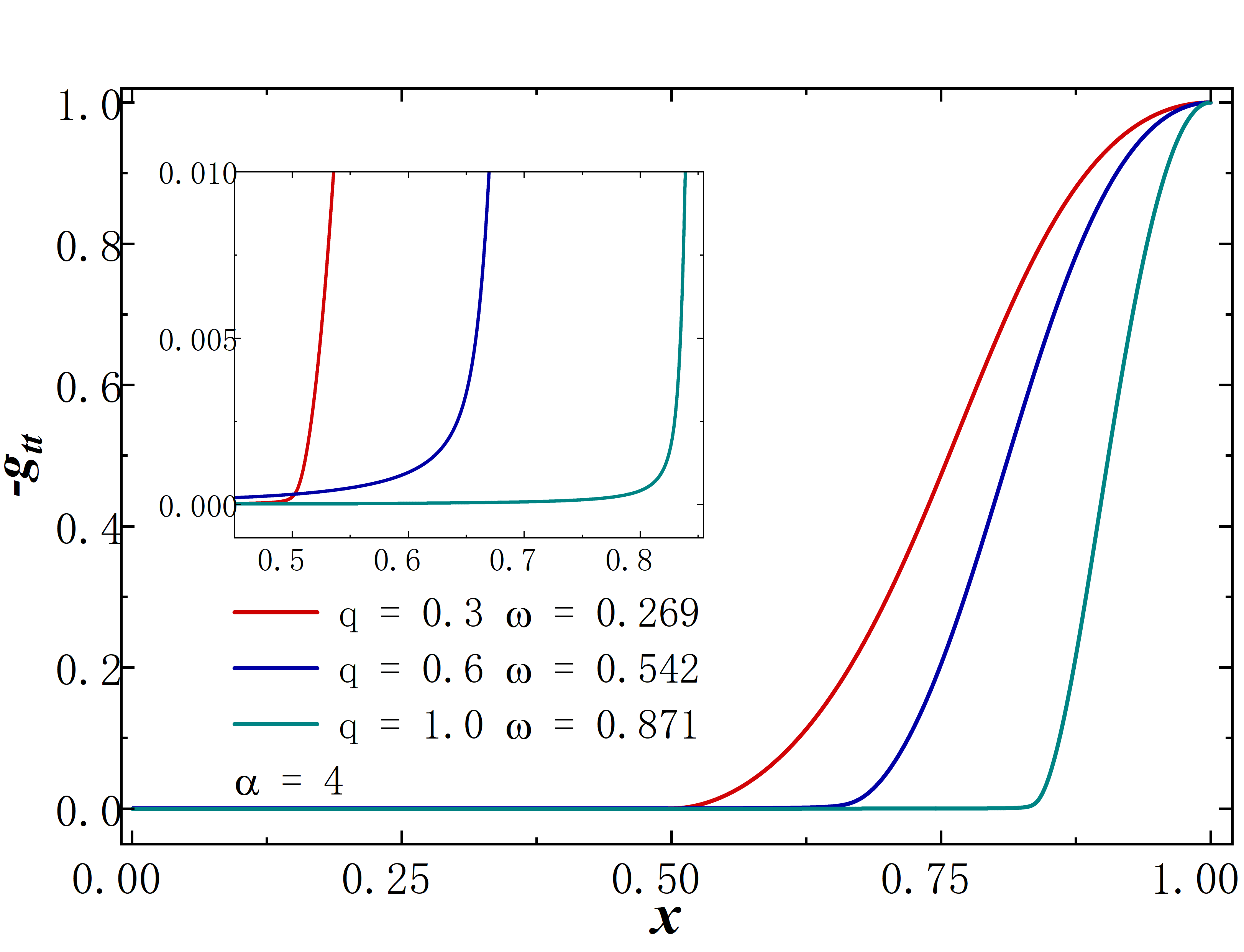}
				\caption{}
			\end{subfigure}
			\begin{subfigure}[b]{0.23\textwidth}
				\includegraphics[width=\textwidth]{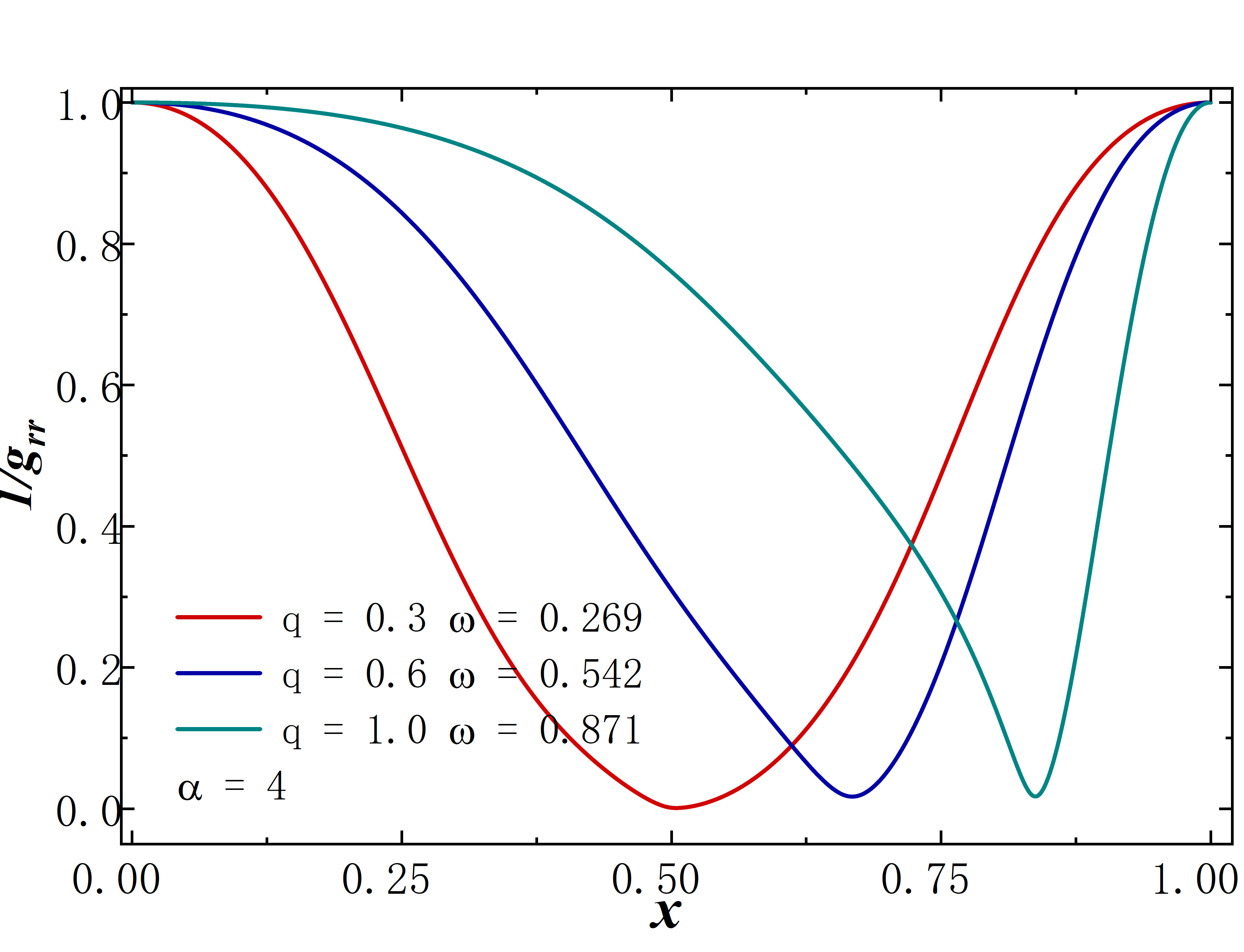}
				\caption{}
			\end{subfigure}
			\begin{subfigure}[b]{0.23\textwidth}
				\includegraphics[width=\textwidth]{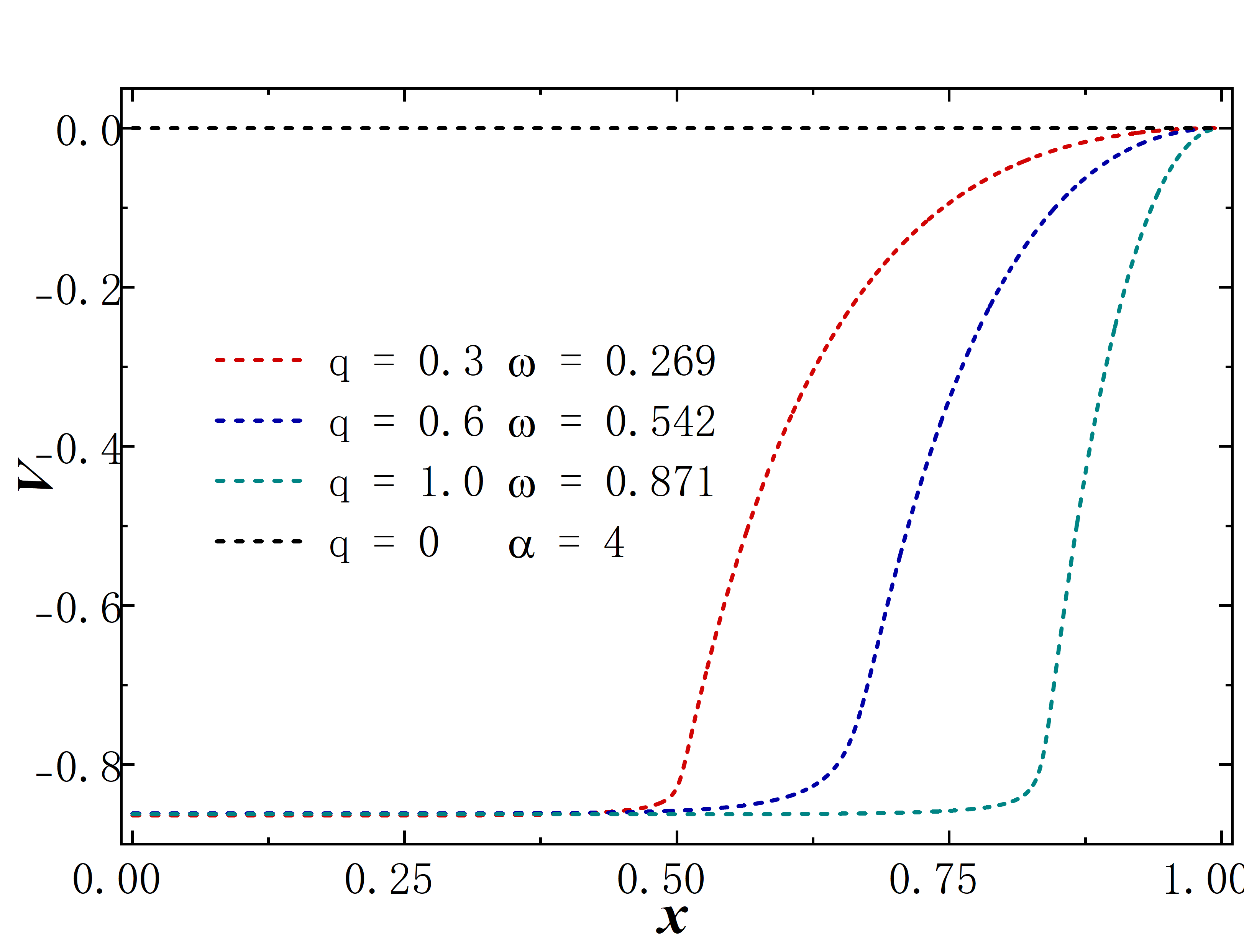}
				\caption{}
			\end{subfigure}
			\hfill
			\begin{subfigure}[b]{0.23\textwidth}
				\includegraphics[width=\textwidth]{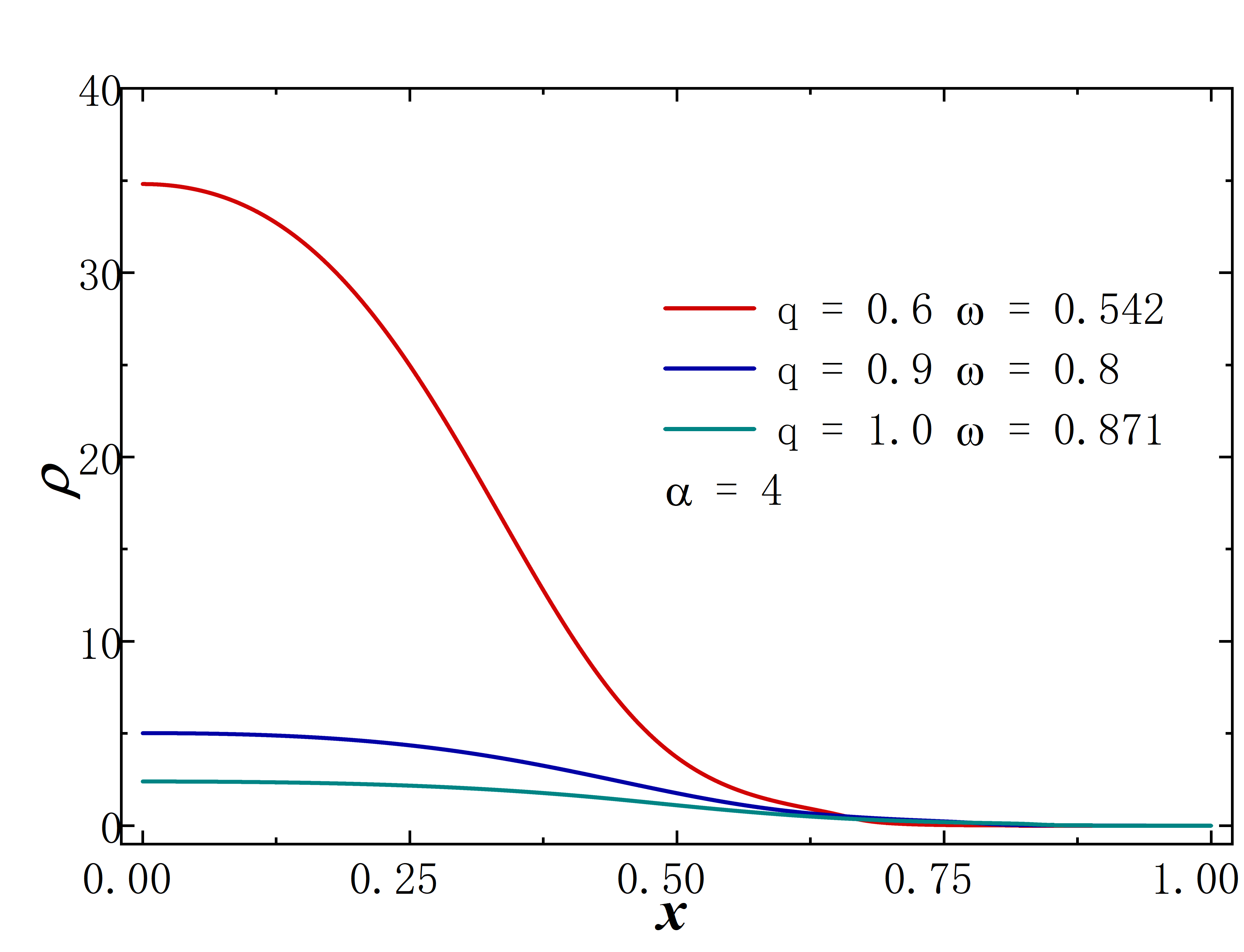}
				\caption{}
			\end{subfigure}
		\end{center}
		\caption{(a) and (b): Proca fields $f$ and $h$ vs. the radial coordinate $x$. (c) and (d): Metric components $-g_{tt}$ and $1/g_{rr}$ vs. the radial coordinate $x$. (e) and (f): Maxwell field $V$ and energy density $\rho$ vs. the radial coordinate $x$. For the three different values of $q$, the frequencies are set to the minimum values allowed for the existence of solutions.}
		\label{p9}
	\end{figure}

	Fig.~\ref{p9} displays the field profiles and metric components at the minimum allowed frequency for various charges. It is worth noting that while the matter fields still concentrate within a characteristic radius $x_c$, it is crucial to emphasize that this behavior does not constitute a true ``frozen state''. The metric components do not vanish sufficiently close to zero (see Table \ref{tab1}). This indicates that low-order Gauss-Bonnet corrections are insufficient to support the fully regular frozen star configuration, which requires the non-perturbative effects of the infinite-derivative tower.
	
	\begin{table*}[ht]
		\centering
		\renewcommand{\arraystretch}{1.5}
		\setlength{\tabcolsep}{15pt}
		\caption{Numerical values of the metric components and critical radius $x_c$ for $n=2$, $\alpha=4$ at selected charges $q$.}
		\begin{tabular}{|c|c|c|c|c|}
			\hline
			\textbf{Charge $q$} & \textbf{Frequency $\omega$} & \textbf{Critical Radius $x_c$} & \boldmath$-g_{tt}(x_c)$ & \boldmath$1/g_{rr}(x_c)$ \\
			\hline
			0.3 & 0.269 & 0.5045 & 0.00046 & 0.00085 \\
			\hline
			0.6 & 0.542 & 0.668  & 0.0089  & 0.0171 \\
			\hline
			1.0 & 0.871 & 0.8378 & 0.0082  & 0.0173 \\
			\hline
		\end{tabular}
		\label{tab1}
	\end{table*}

	\subsection{$n>2$: Higher order correction}
	
	As the correction order $n$ increases, the higher-order curvature terms endow the solutions with entirely new properties. We find that for finite correction orders ($n=3, 4, \dots$) up to the infinite case ($n=\infty$), the systems exhibit remarkably similar physical behaviors. Nevertheless, the $n=\infty$ case also possesses distinct characteristics not found in the finite-order corrections. In this section, we first briefly present the results for $n=3$ to illustrate the general trend, and then focus our detailed discussion on the $n=\infty$ case, with the coupling parameter fixed at $\alpha=4$.
	
\begin{figure}[!htbp]
		\begin{center}
			\begin{subfigure}[b]{0.23\textwidth}
				\includegraphics[width=\textwidth]{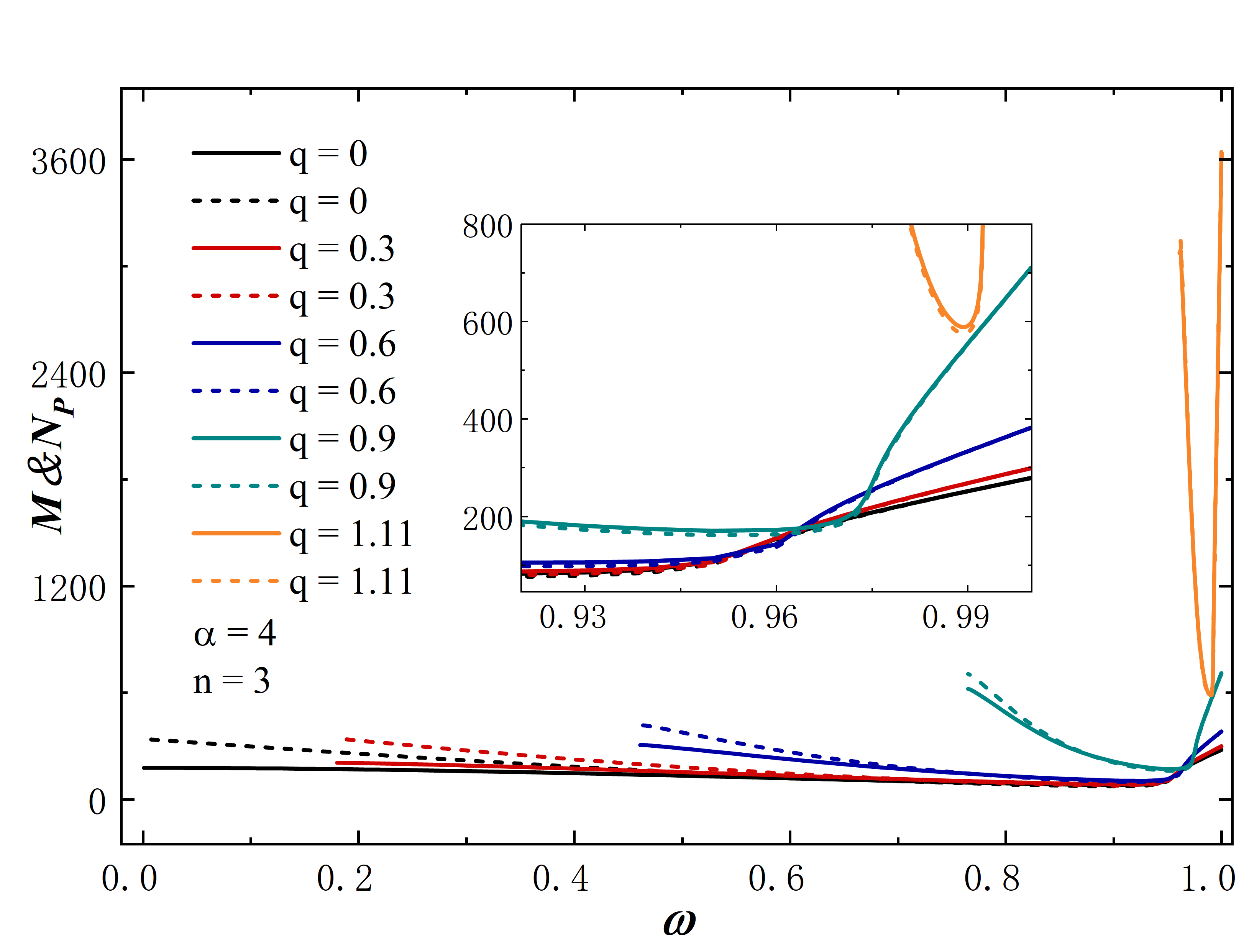}
				\caption{}
			\end{subfigure}
			\begin{subfigure}[b]{0.23\textwidth}
				\includegraphics[width=\textwidth]{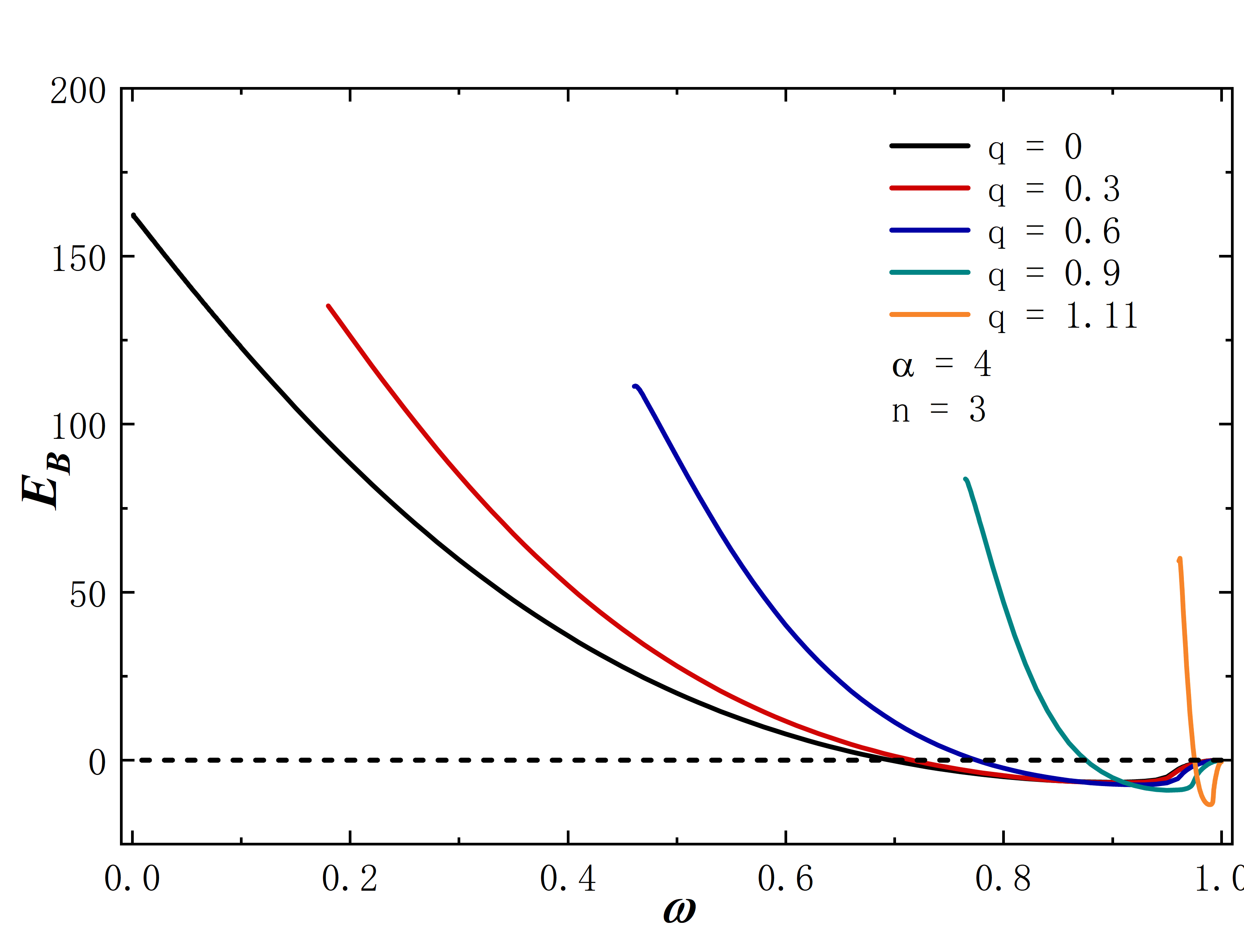}
				\caption{}
			\end{subfigure}
		\end{center}
		\caption{$M$, $N_P$ and the $E_B$ vs. the frequency $\omega$ with different $q$ under $\alpha = 4$ and $n=3$. In (a), the solid lines represent $M$, and dashed lines represent $N_P$.}
		\label{p10}
	\end{figure}

	Fig.~\ref{p10} shows the ADM mass, conserved particle number, and the corresponding binding energy for the $n=3$ case. Similar to the $n=2$ scenario, as the charge $q$ increases from 0, the domain of existence for the solutions gradually narrows. However, a key difference from the infinite-order corrections that will be introduced next is that even when $q$ reaches its maximum allowed value, the upper frequency limit still approaches $\omega \to 1$. Furthermore, unlike the unstable low-order solutions, the $n=3$ solutions remain in a gravitationally bound state ($E_B > 0$) over a wide range of frequencies, indicating enhanced stability.
	
\begin{figure}[!htbp]
		\begin{center}
			\begin{subfigure}[b]{0.23\textwidth}
				\includegraphics[width=\textwidth]{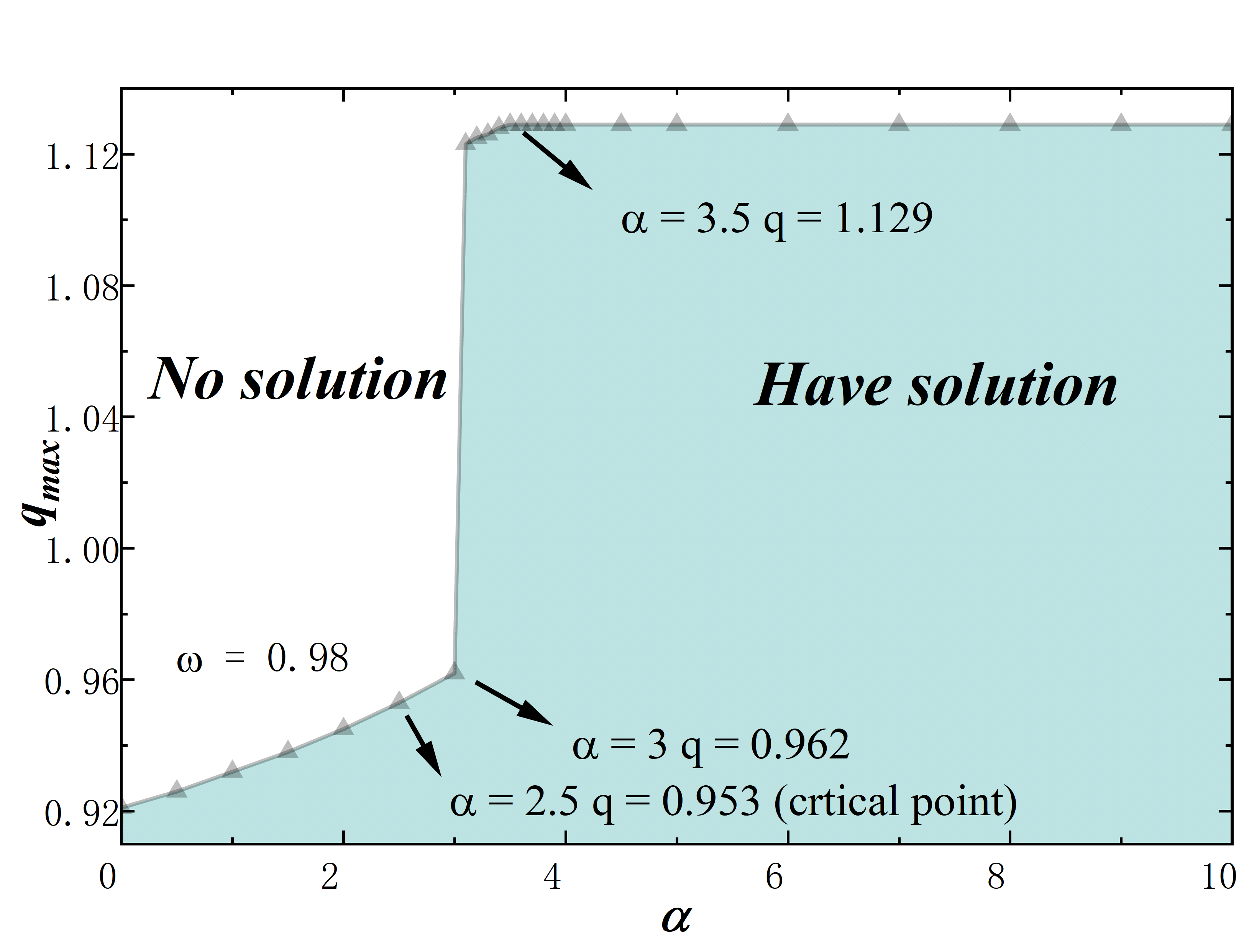}
				\caption{}
			\end{subfigure}
			\begin{subfigure}[b]{0.23\textwidth}
				\includegraphics[width=\textwidth]{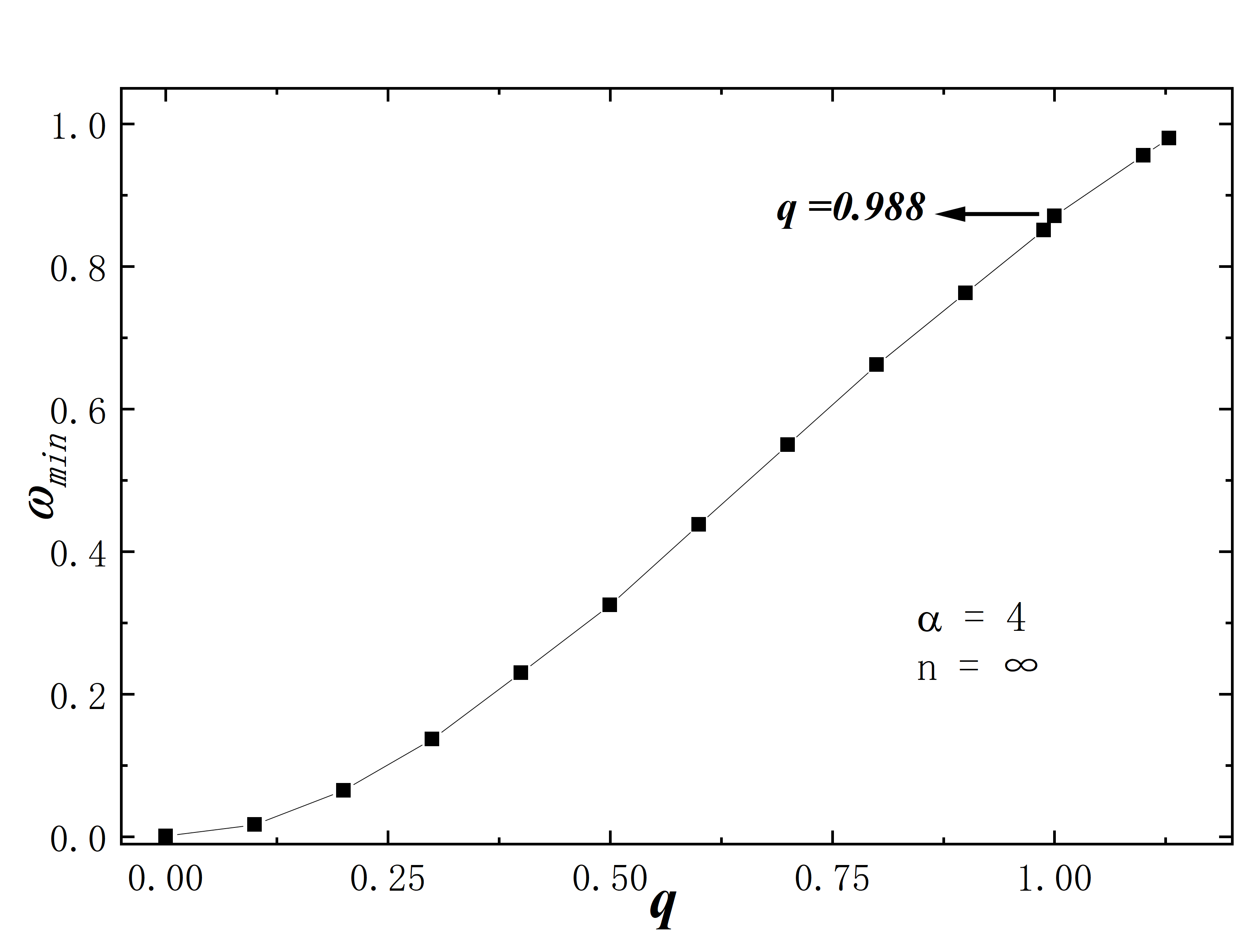}
				\caption{}
			\end{subfigure}
		\end{center}
		\caption{(a): Relationship between the maximum value of the charge parameter $q$ and the coupling parameter $\alpha$ when the fixed frequency $\omega$ is 0.98. (b): Relationship between the minimum value of the $\omega$ and the electric parameter $q$. In both figures, the $n = \infty$.}
		\label{p11}
	\end{figure}

	We strictly analyze the solution space for the infinite-derivative theory ($n=\infty$). In Fig.~\ref{p11} (a), we track the existence boundary $q_{\text{max}}(\alpha)$. The system exhibits a saturation behavior: $q_{\text{max}}$ grows linearly for small $\alpha$ but sharply saturates at $q \approx 1.129$ when $\alpha \ge 3.5$. More importantly, we identify $\alpha \approx 2.5$ as the solution critical point. Beyond this threshold, the multi-branch spiral structure of the fundamental states vanishes, replaced by the monotonic ``frozen'' branch characteristic of non-perturbative regularization. The impact of charge on the frequency spectrum is detailed in Fig.~\ref{p11} (b). Increasing $q$ imposes a lower frequency cutoff $\omega_{\text{min}}$, which rises monotonically. This quantifies the ``unfreezing'' process: electrostatic repulsion prevents the system from accessing the zero-frequency ground state. Furthermore, we report a novel anomaly unique to the $n=\infty$ sector: for hyper-charged states ($q > 0.988$), the upper frequency limit no longer reaches the standard vacuum asymptote ($\omega \to 1$). This implies that sufficiently large charges decouple the solution from the perturbative vacuum.
	
	\begin{figure}[!htbp]
		\begin{center}
			\begin{subfigure}[b]{0.23\textwidth}
				\includegraphics[width=\textwidth]{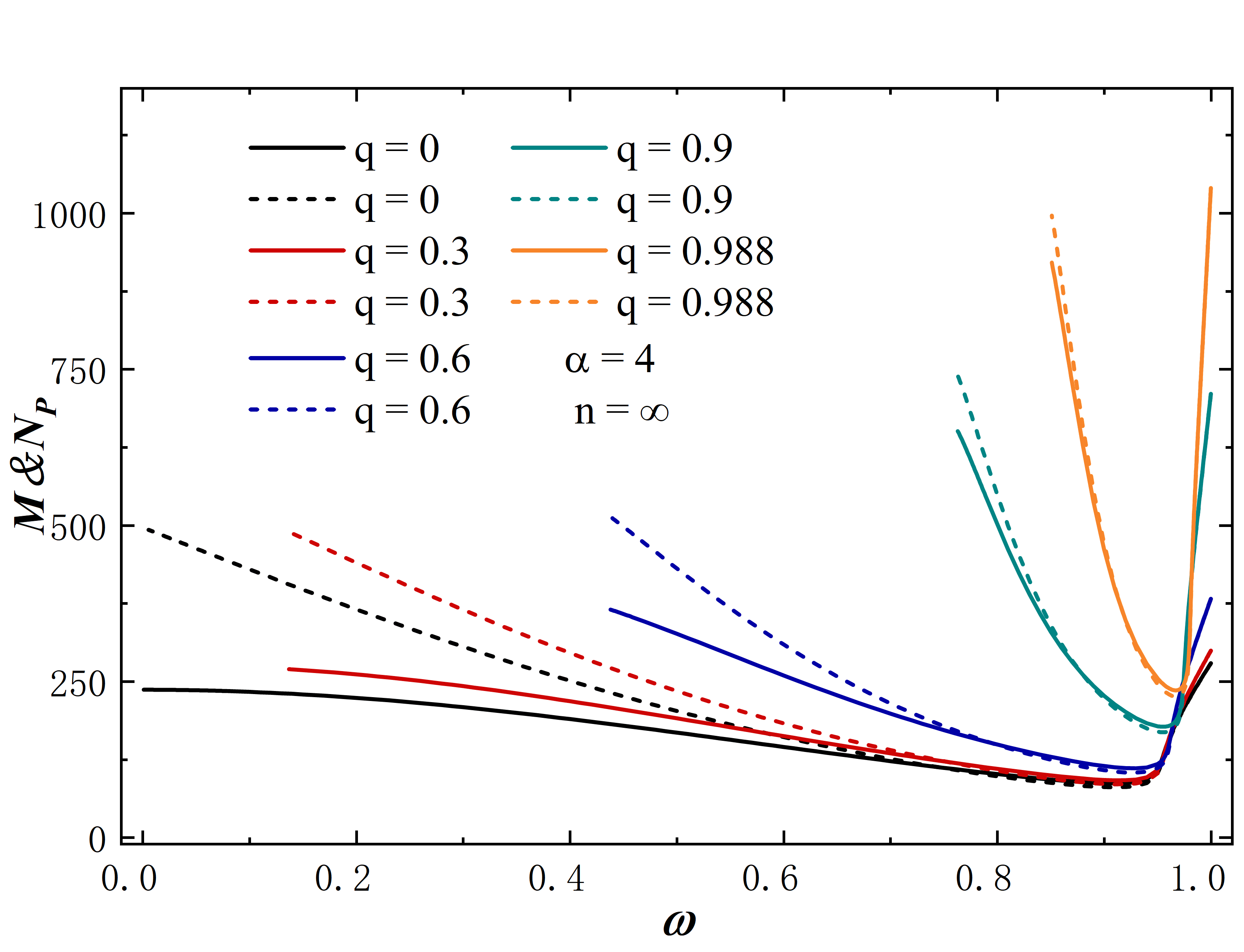}
				\caption{}
			\end{subfigure}
			\begin{subfigure}[b]{0.23\textwidth}
				\includegraphics[width=\textwidth]{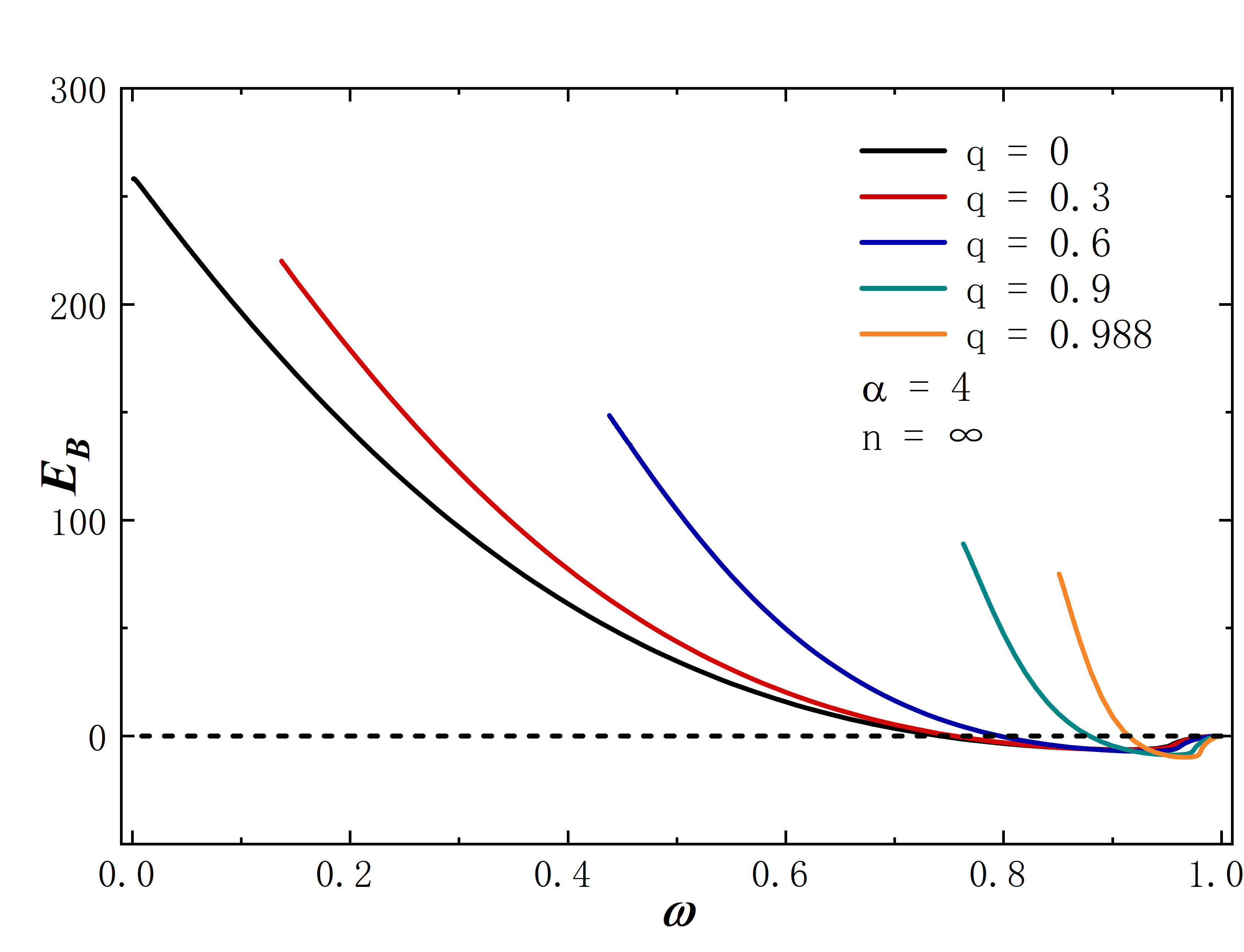}
				\caption{}
			\end{subfigure}
			\begin{subfigure}[b]{0.23\textwidth}
				\includegraphics[width=\textwidth]{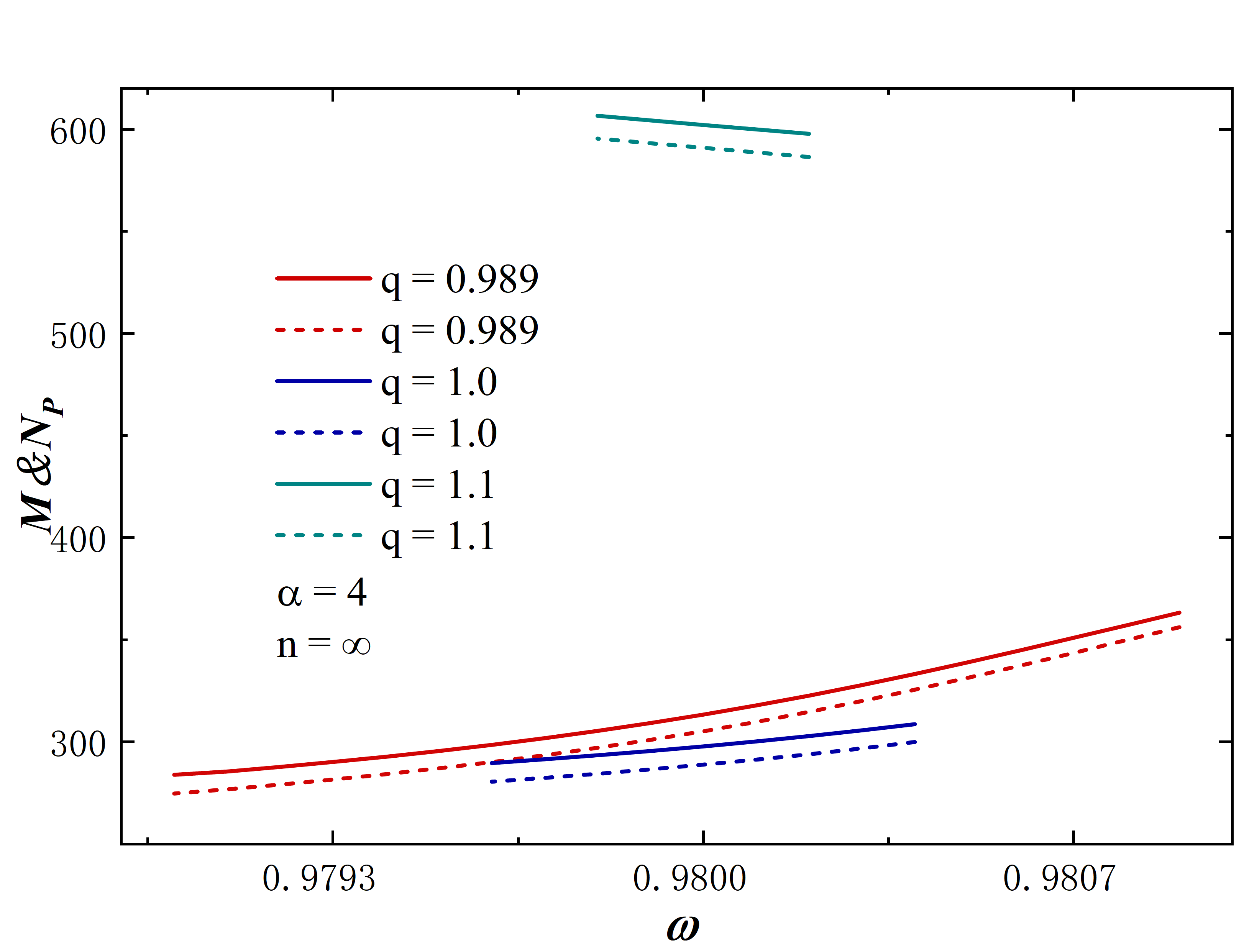}
				\caption{}
			\end{subfigure}
			\begin{subfigure}[b]{0.23\textwidth}
				\includegraphics[width=\textwidth]{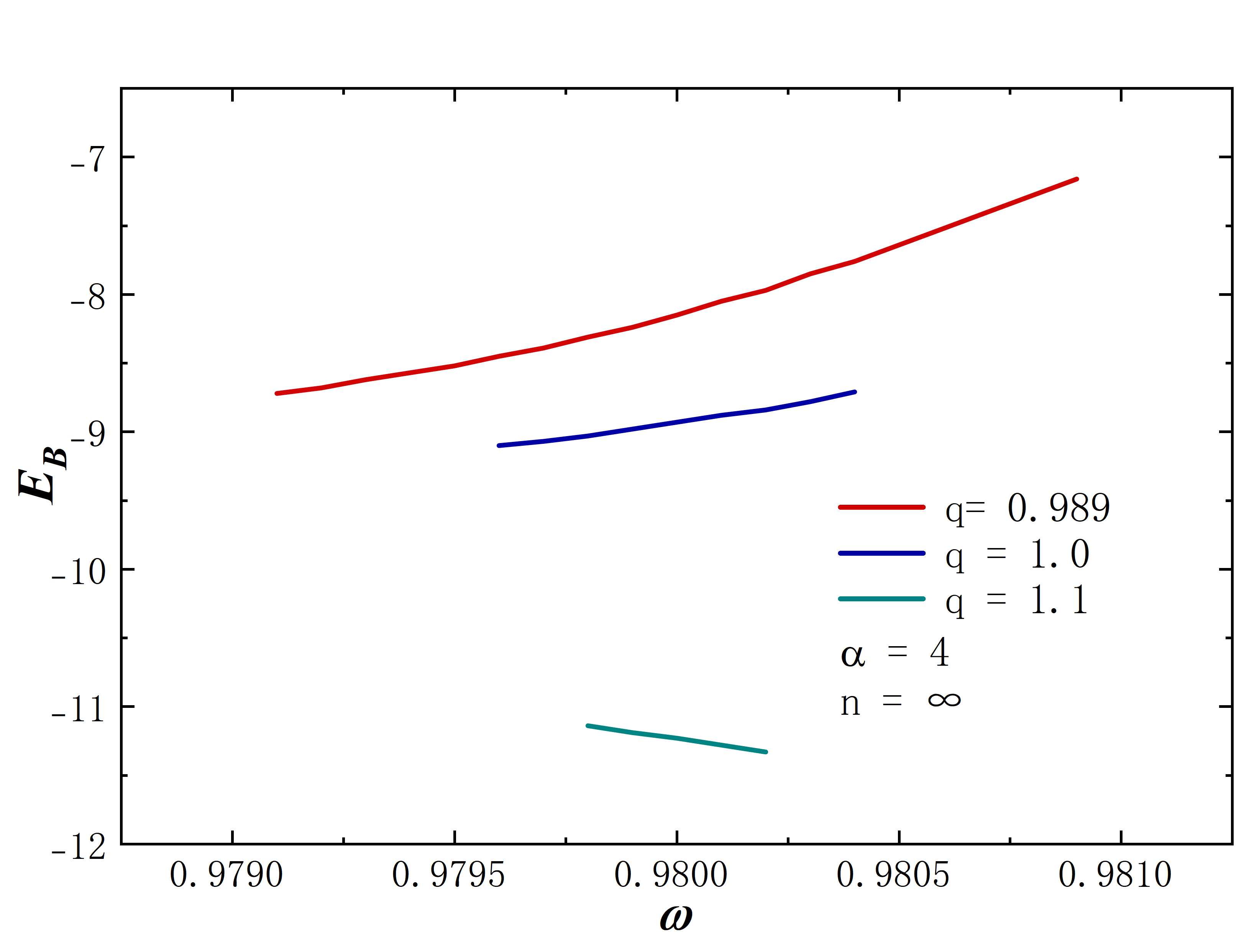}
				\caption{}
			\end{subfigure}
		\end{center}
		\caption{(a) and (b): The ADM mass and conserved particle number, as functions of the field frequency $\omega$ with $q \le 0.988$ and the corresponding binding energy. (c) and (d): The ADM mass and conserved particle number, as functions of the field frequency $\omega$ with $q > 0.988$ and the corresponding binding energy.}
		\label{p12}
	\end{figure}
	
	The ADM mass, particle number, and binding energy are presented in Fig.~\ref{p12}. For charges $q \le 0.988$, the system exhibits positive binding energy over a wide range, indicating stability, while a perturbation analysis is required. However, for the ultra-high charge regime ($q > 0.988$), the binding energy becomes entirely negative within the extremely narrow frequency domain, suggesting that these highly charged configurations are unstable.
	
\begin{figure}[!htbp]
		\begin{center}
			\begin{subfigure}[b]{0.23\textwidth}
				\includegraphics[width=\textwidth]{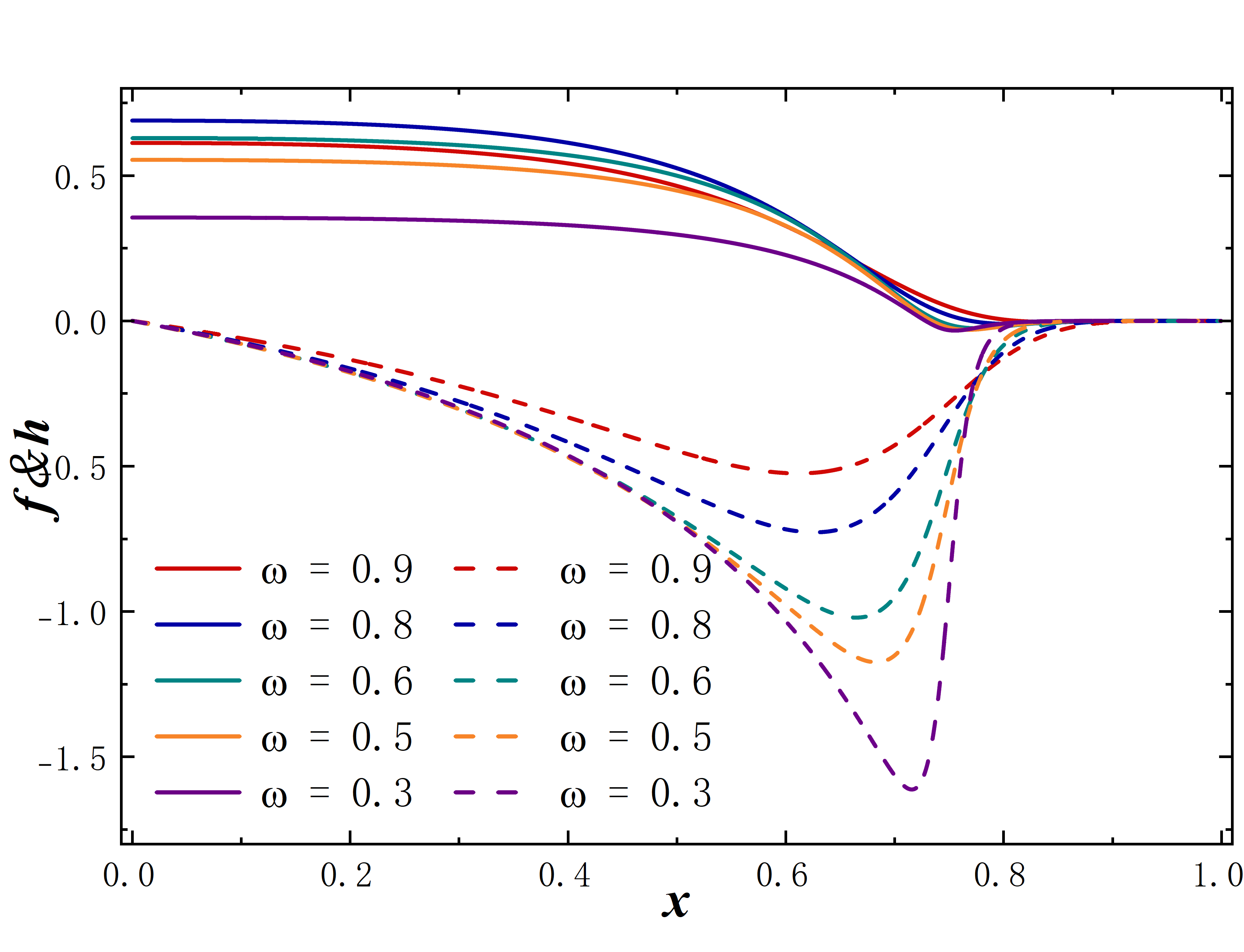}
				\caption{}
			\end{subfigure}
			\begin{subfigure}[b]{0.23\textwidth}
				\includegraphics[width=\textwidth]{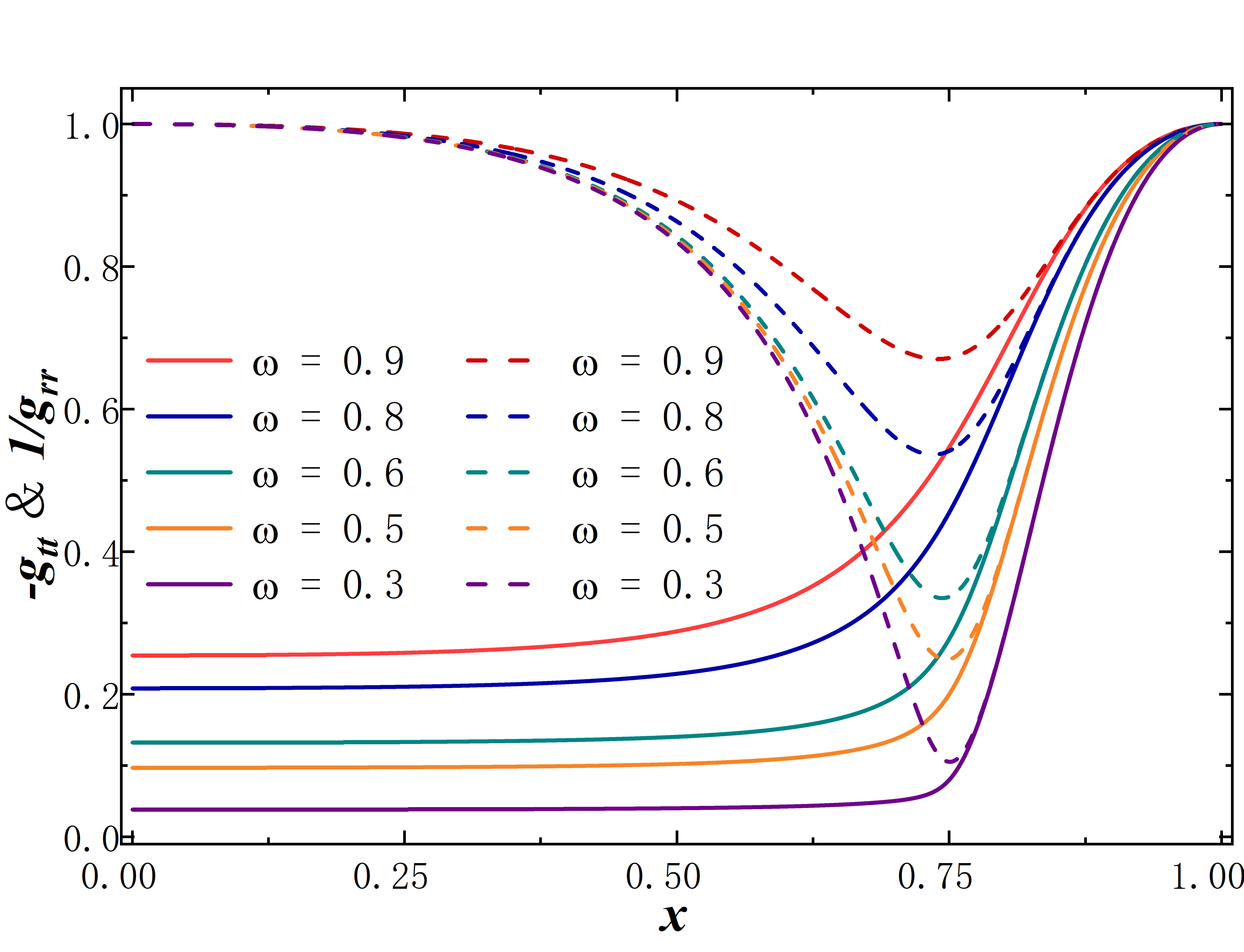}
				\caption{}
			\end{subfigure}
		\end{center}
		\caption{(a): Proca fields $f$ and $h$ vs. the radial coordinate $x$. (b): Metric components vs. the radial coordinate $x$, the solid lines represent $-g_{tt}$, and dashed lines represent $1/g_{rr}$.}
		\label{p13}
	\end{figure}

	First, focusing on the neutral case, we present the Proca field functions and the spacetime metric components for various frequencies $\omega$ in Fig.~\ref{p13}. It is clearly observed that as the $\omega$ decreases, the field becomes increasingly concentrated within a specific radial radius, while the values of $-g_{tt}$ and $1/g_{rr}$ at this location simultaneously tend towards zero. This pattern persists as the $\omega$ approaches the limit $\omega \to 0$, causing the system to enter a ``frozen state''.
	
\begin{figure}[!htbp]
		\begin{center}
			\begin{subfigure}[b]{0.23\textwidth}
				\includegraphics[width=\textwidth]{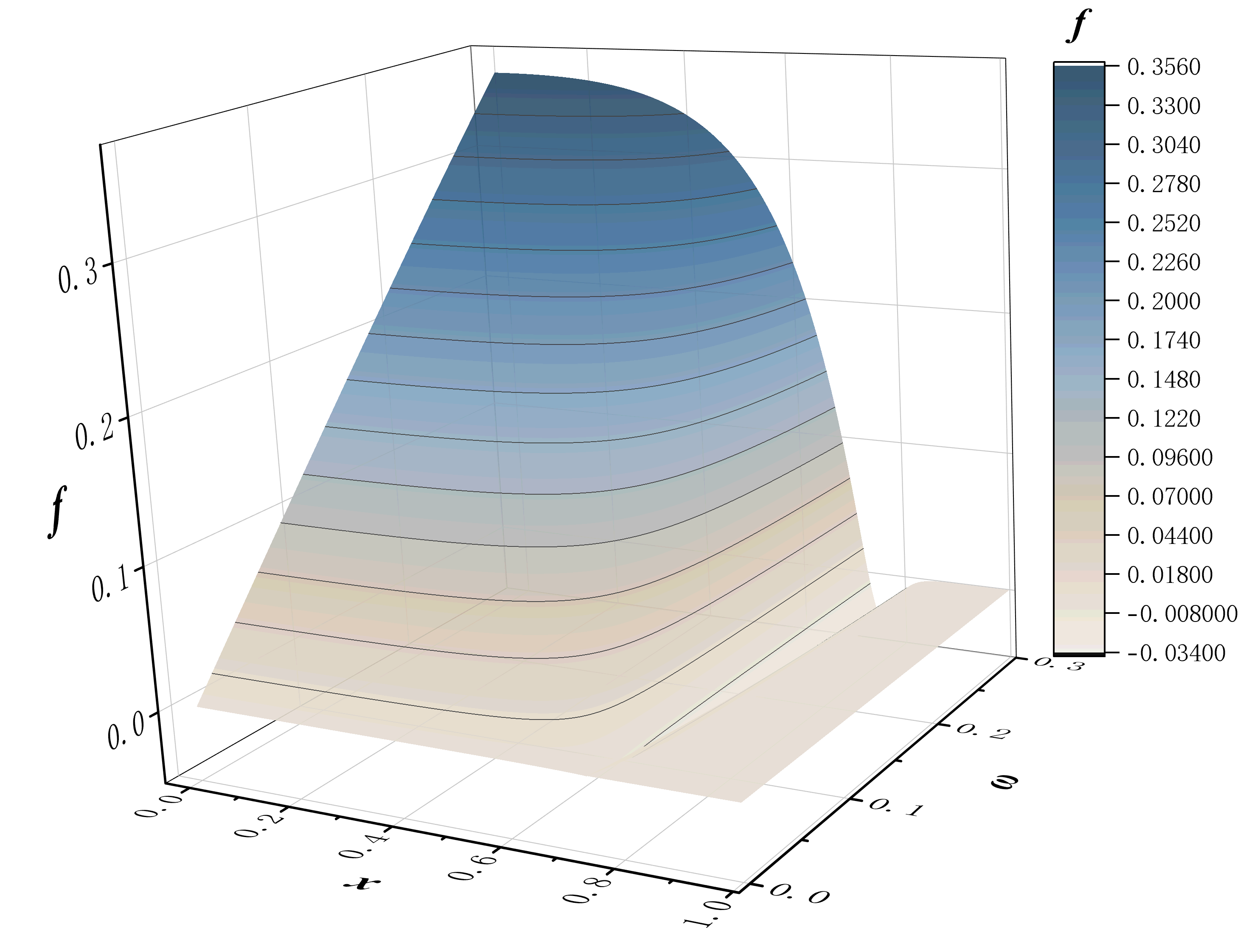}
				\caption{}
			\end{subfigure}
			\begin{subfigure}[b]{0.23\textwidth}
				\includegraphics[width=\textwidth]{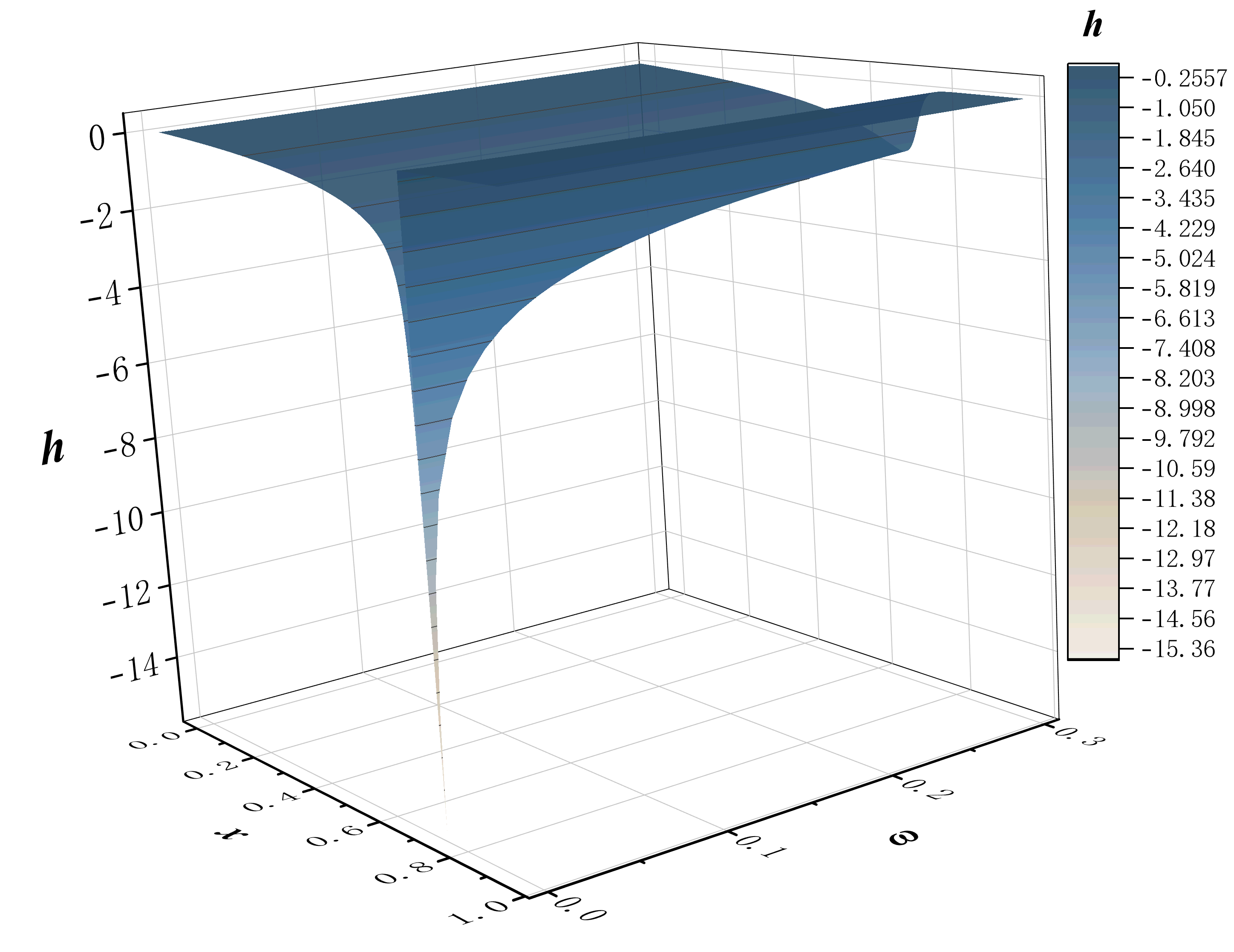}
				\caption{}
			\end{subfigure}
			\begin{subfigure}[b]{0.33\textwidth}
				\includegraphics[width=\textwidth]{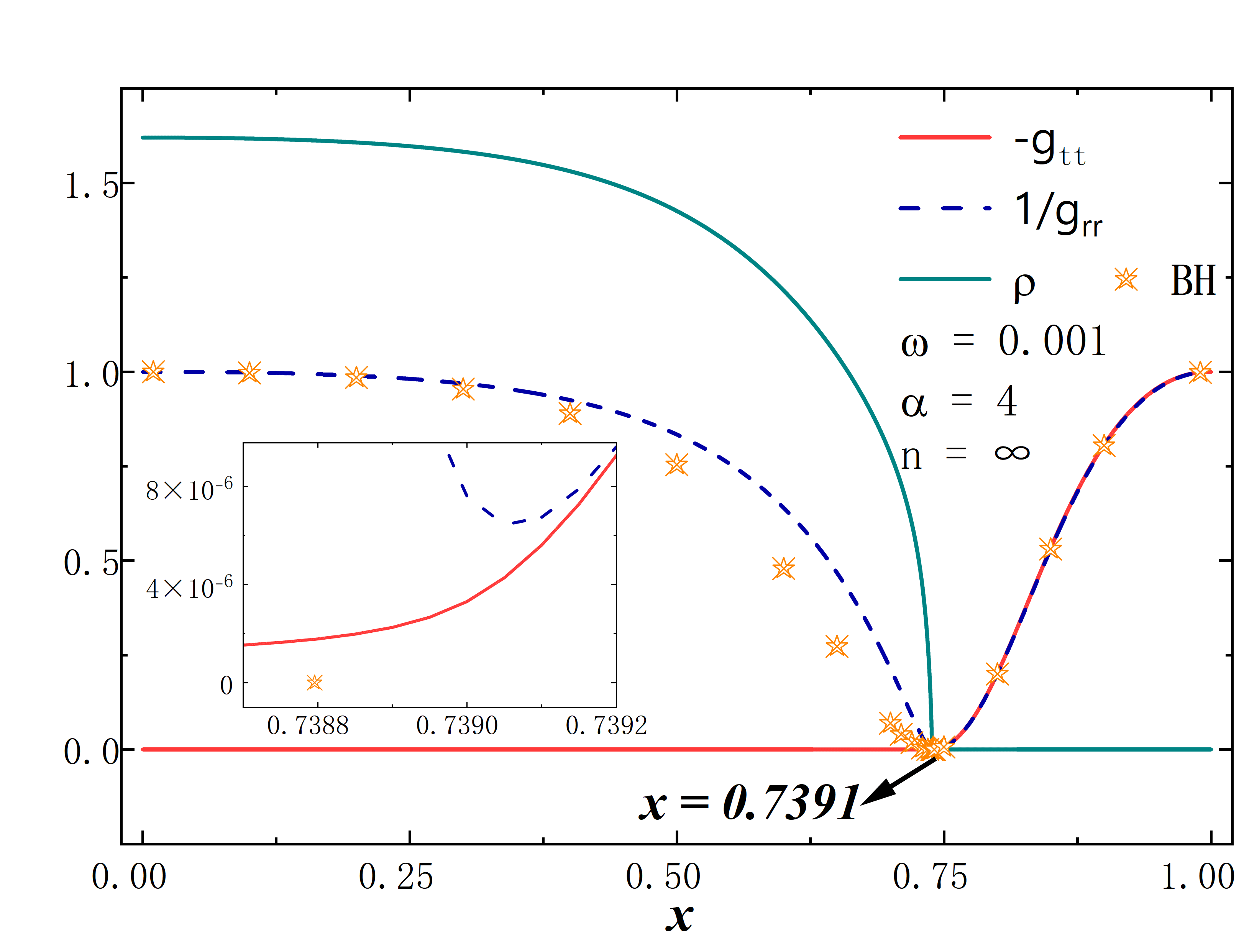}
				\caption{}
			\end{subfigure}
		\end{center}
		\caption{(a) and (b): Three-dimensional surface plots of the Proca field components $f$ and $h$ as functions of the radial coordinate $x$ and the field frequency $\omega$. (c): Metric components and energy density of the numerical solution at $\omega=0.001$, together with the metric functions of the analytic extremal black hole for $n=\infty$.}
		\label{p14}
	\end{figure}

	Furthermore, the defining feature of the $n=\infty$ theory is its ability to support a regular frozen core. In the neutral limit $\omega \to 0$ (Fig.~\ref{p14}, the Proca field retracts entirely into a compact region $x < x_c \approx 0.7391$. Unlike the Gauss-Bonnet case, the metric components here do not diverge; instead, they become vanishingly small at $x_c$ (of order $10^{-6}$, see Table \ref{tab2}), forming a ``quasi-horizon''. This configuration offers a compelling realization of a black hole mimicker. As shown in Fig.~\ref{p14} (c), the energy density is strictly confined within the core. Consequently, the external geometry ($x > x_c$) is identical to that of an extremal black hole (In Appendix \ref{app:vacuum_ninf}, we more clearly present the structural details of this external black hole and the internal frozen star). An external observer would detect the mass of a black hole, yet the interior is horizonless and free of singularities. The ``frozen'' star is thus a regular soliton masquerading as a singular black hole.

	\begin{table*}[ht]
		\centering
		\renewcommand{\arraystretch}{1.5}
		\setlength{\tabcolsep}{15pt} 
		\caption{Numerical values of the metric components and the critical radius $x_c$ for $n=\infty$ ($\alpha=4$). Note the finite but vanishingly small metric values for the neutral case ($q=0$), indicating a quasi-horizon.}
		\label{tab:nin_alpha4_q_values}
		\begin{tabular}{|c|c|c|c|c|}
			\hline
			\textbf{Charge} $q$ & \textbf{Frequency} $\omega$ & \textbf{Critical Radius} $x_c$ & $-g_{tt}(x_c)$ & $1/g_{rr}(x_c)$ \\ 
			\hline
			0 & 0.001 & 0.7391 & $5.9 \times 10^{-6}$ & $6.5 \times 10^{-6}$ \\
			\hline
			0.3 & 0.137 & 0.7470 & $1.2 \times 10^{-4}$ & $2.5 \times 10^{-4}$ \\
			\hline
			0.6 & 0.438 & 0.7717 & $0.0009$ & $0.0015$ \\
			\hline
			0.988 & 0.851 & 0.8488 & $0.0067$ & $0.0133$ \\
			\hline
		\end{tabular}
		\label{tab2}
	\end{table*}

	\begin{figure}
		\begin{center}
			\begin{subfigure}[b]{0.23\textwidth}
				\includegraphics[width=\textwidth]{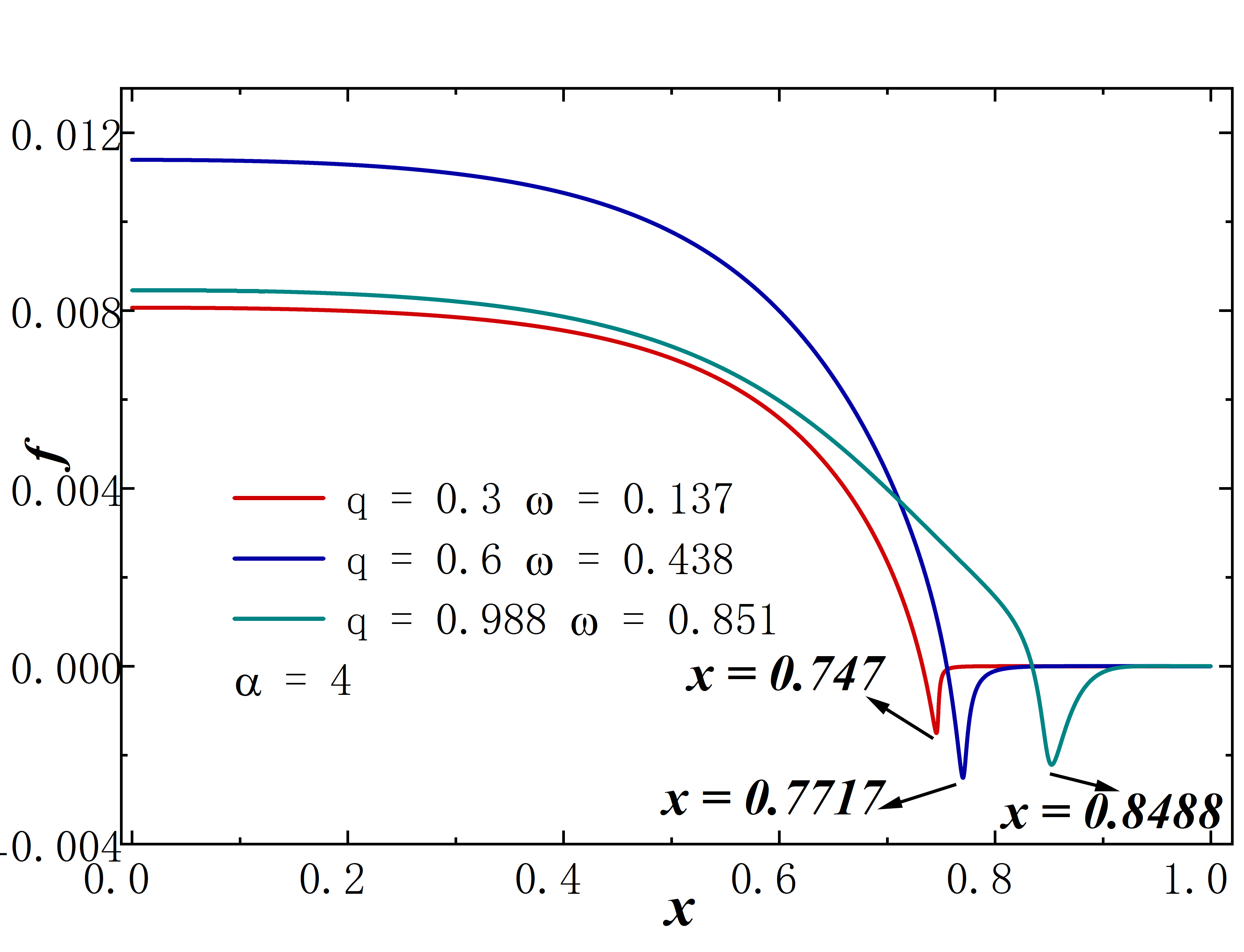}
				\caption{}
			\end{subfigure}
			\begin{subfigure}[b]{0.23\textwidth}
				\includegraphics[width=\textwidth]{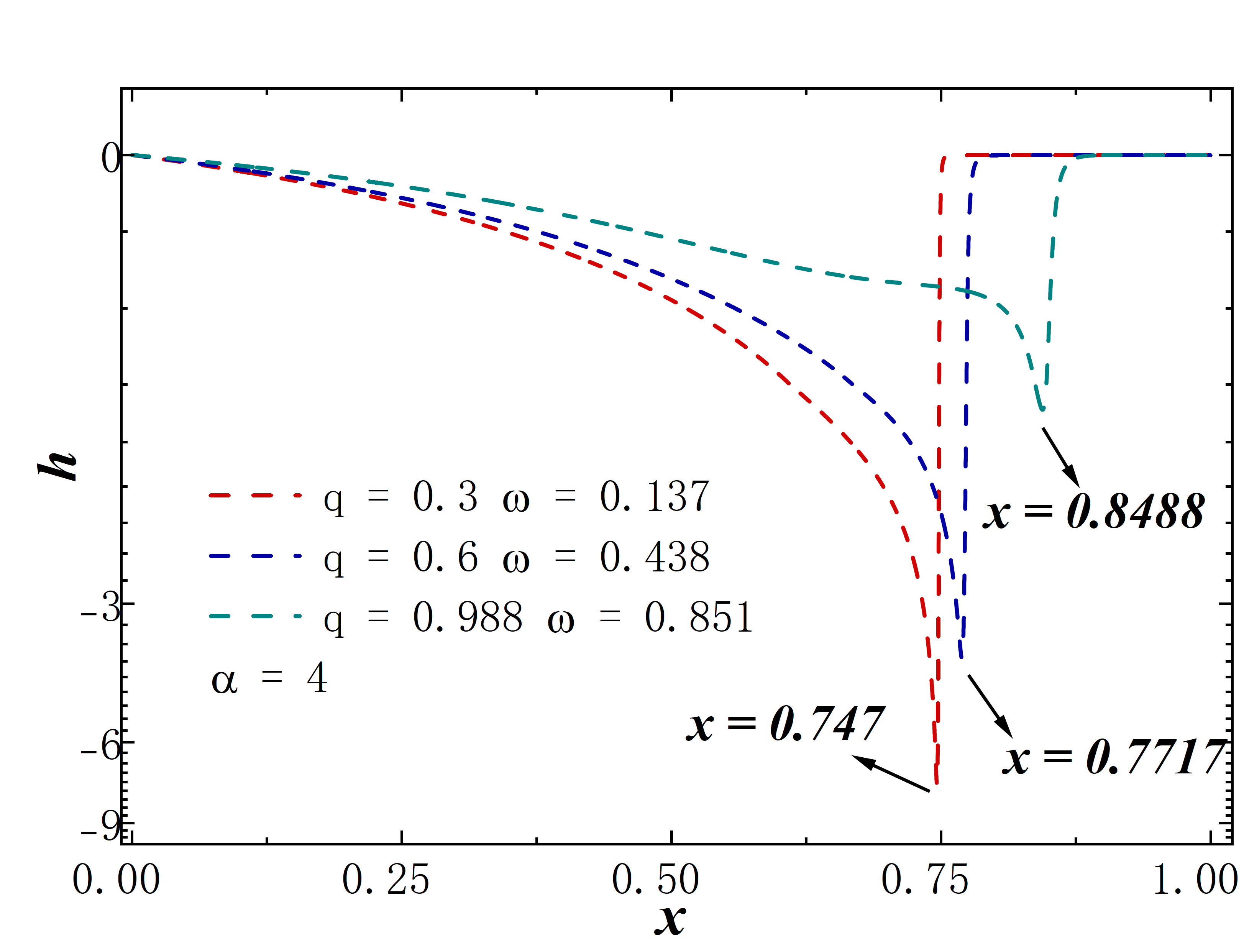}
				\caption{}
			\end{subfigure}
			\begin{subfigure}[b]{0.23\textwidth}
				\includegraphics[width=\textwidth]{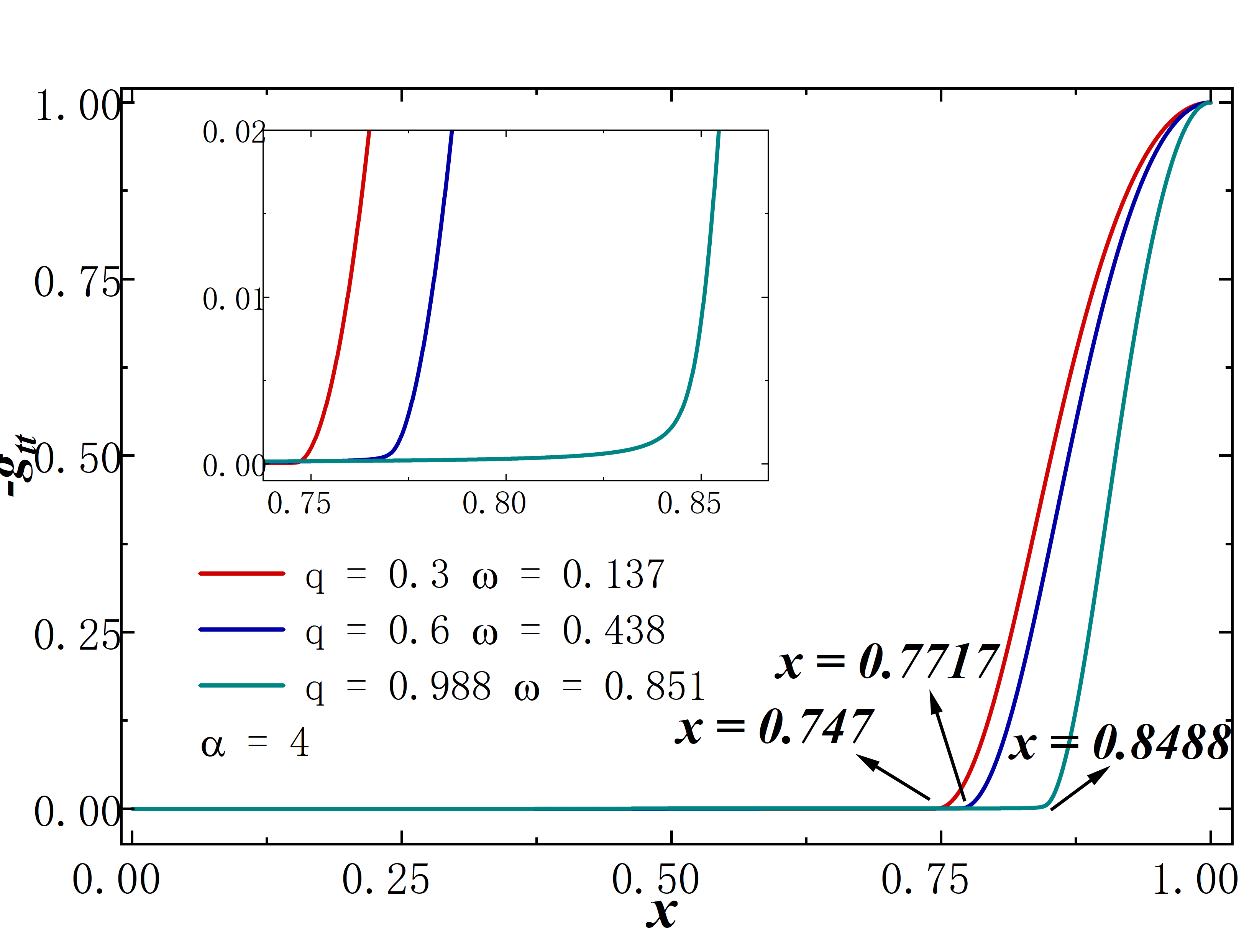}
				\caption{}
			\end{subfigure}
			\begin{subfigure}[b]{0.23\textwidth}
				\includegraphics[width=\textwidth]{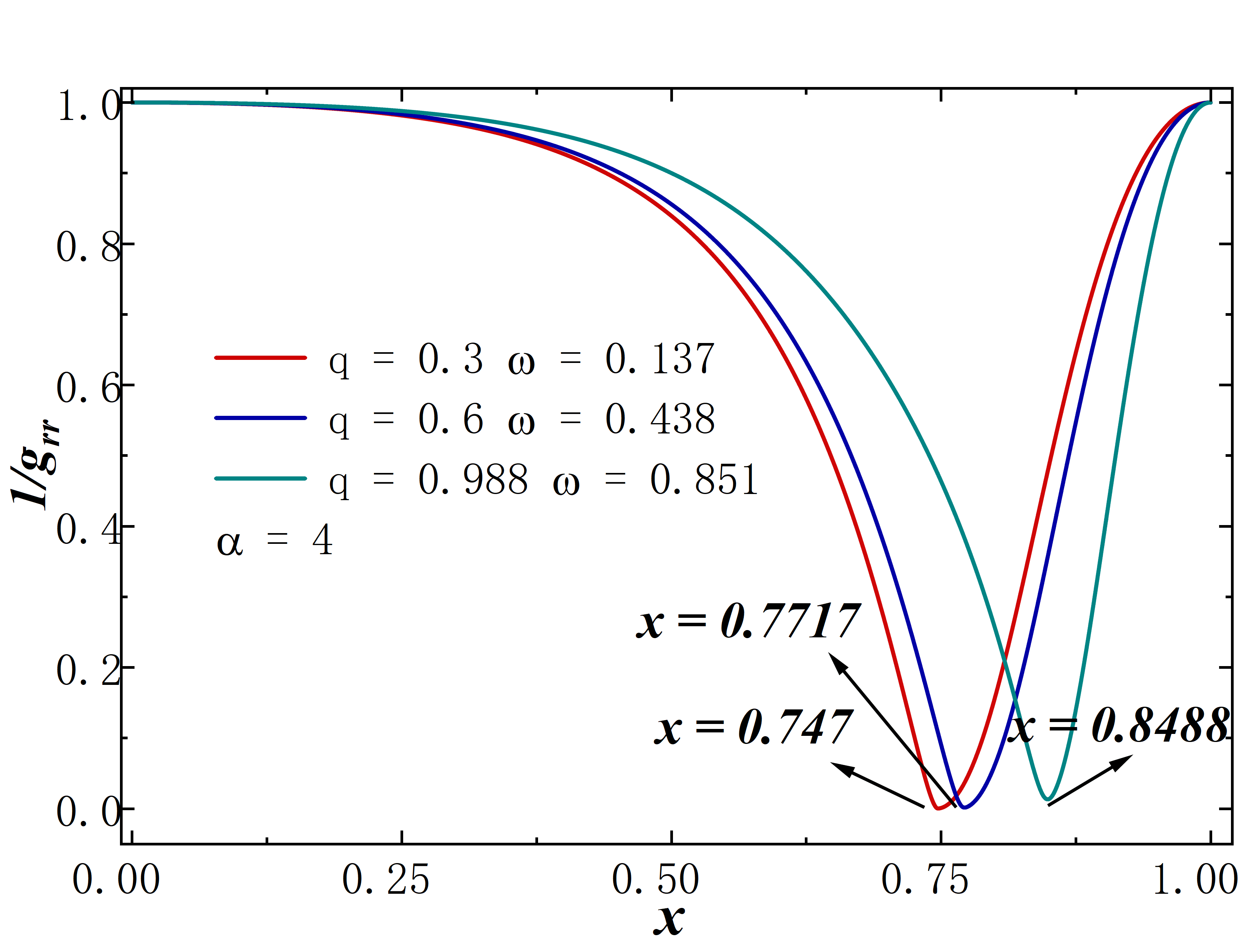}
				\caption{}
			\end{subfigure}
			\begin{subfigure}[b]{0.23\textwidth}
				\includegraphics[width=\textwidth]{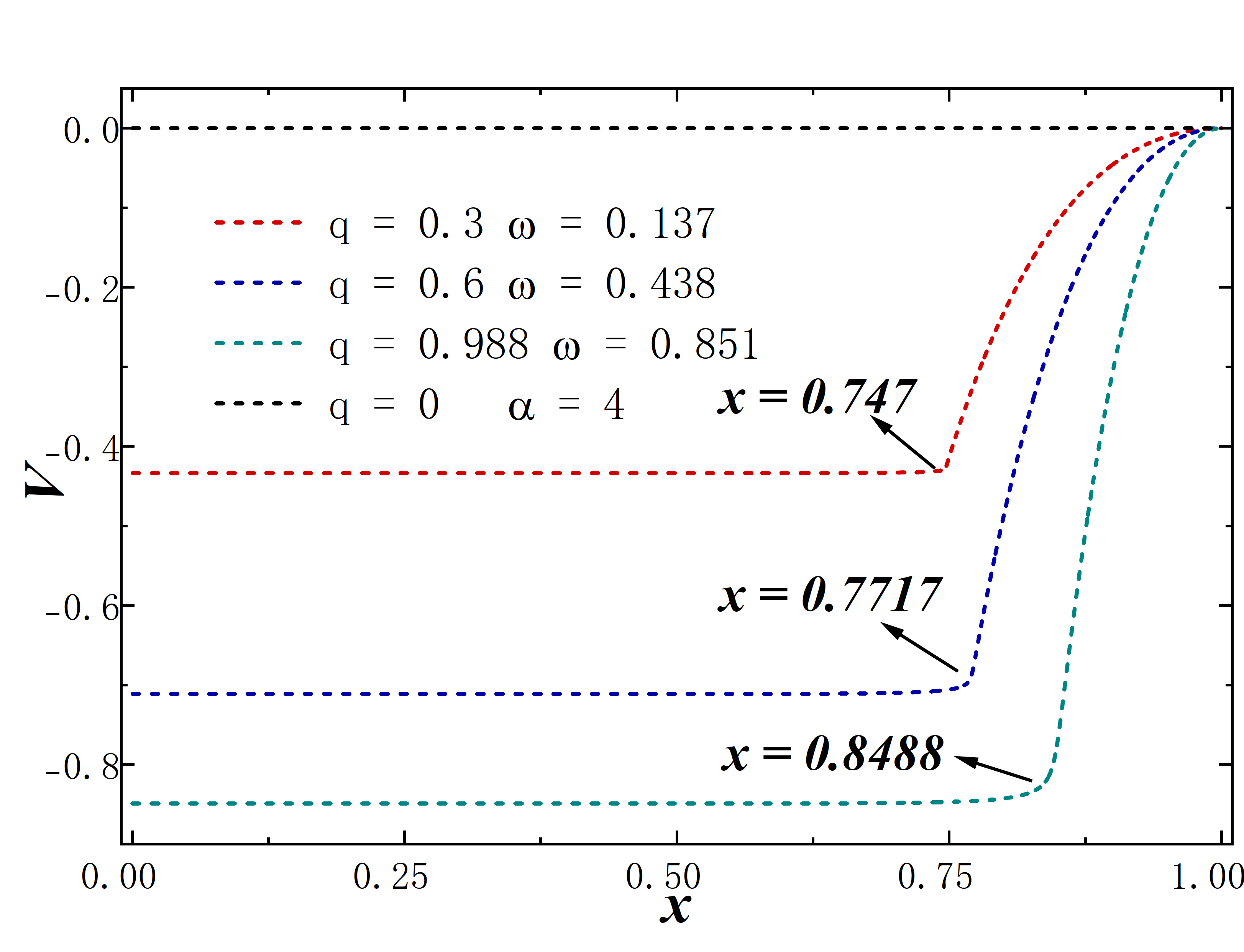}
				\caption{}
			\end{subfigure}
			\begin{subfigure}[b]{0.23\textwidth}
				\includegraphics[width=\textwidth]{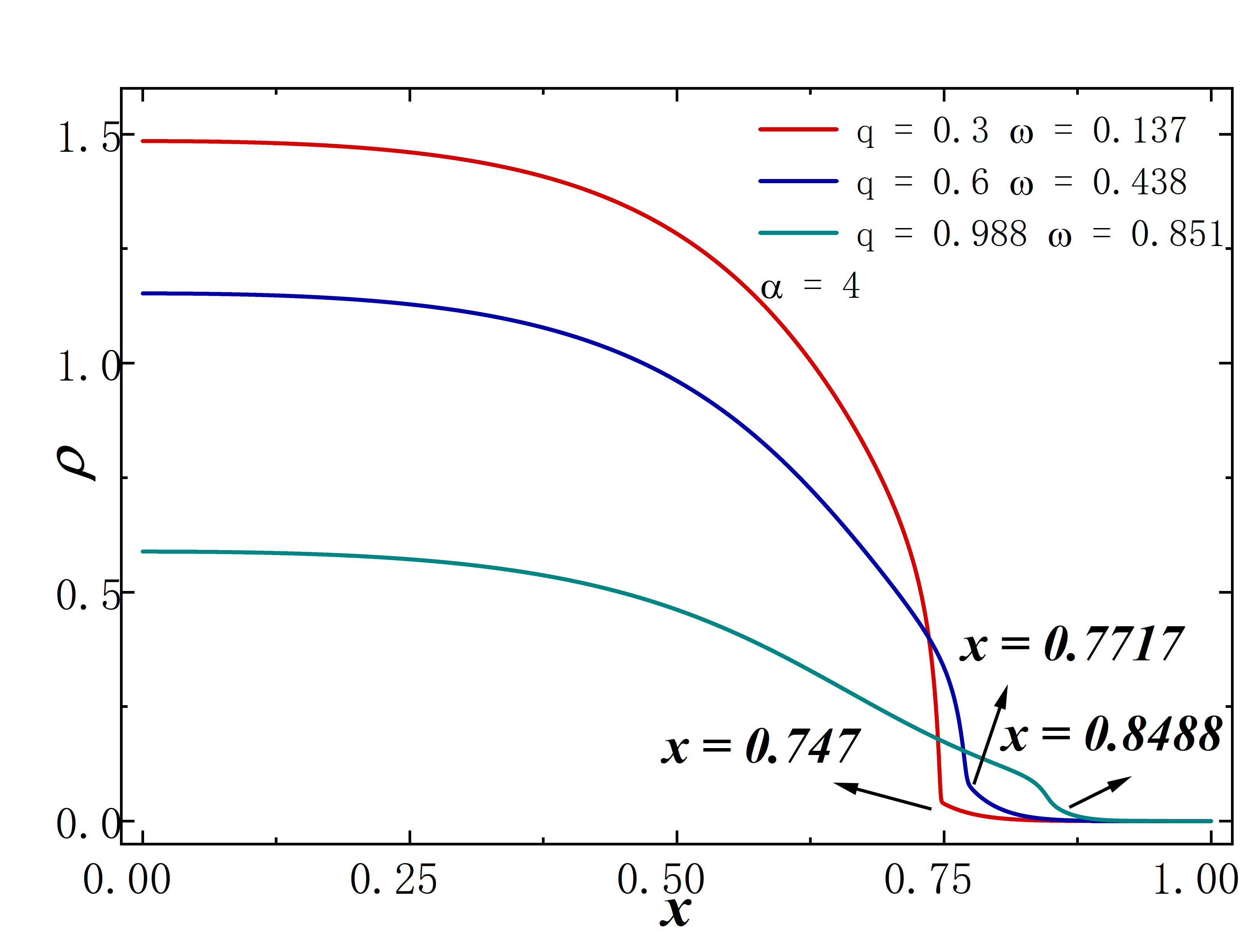}
				\caption{}
			\end{subfigure}
		\end{center}
		\caption{(a) and (b): Proca fields $f$ and $h$ vs. the radial coordinate $x$. (c) and (d): Metric components $-g_{tt}$ and $1/g_{rr}$ vs. the radial coordinate $x$. (e) and (f): Maxwell field $V$ and energy density $\rho$ vs. the radial coordinate $x$. For the three different values of $q$, the frequencies are set to the minimum values allowed for the existence of solutions.}
		\label{p15}
	\end{figure}
	
\begin{figure}
		\begin{center}
			\begin{subfigure}[b]{0.23\textwidth}
				\includegraphics[width=\textwidth]{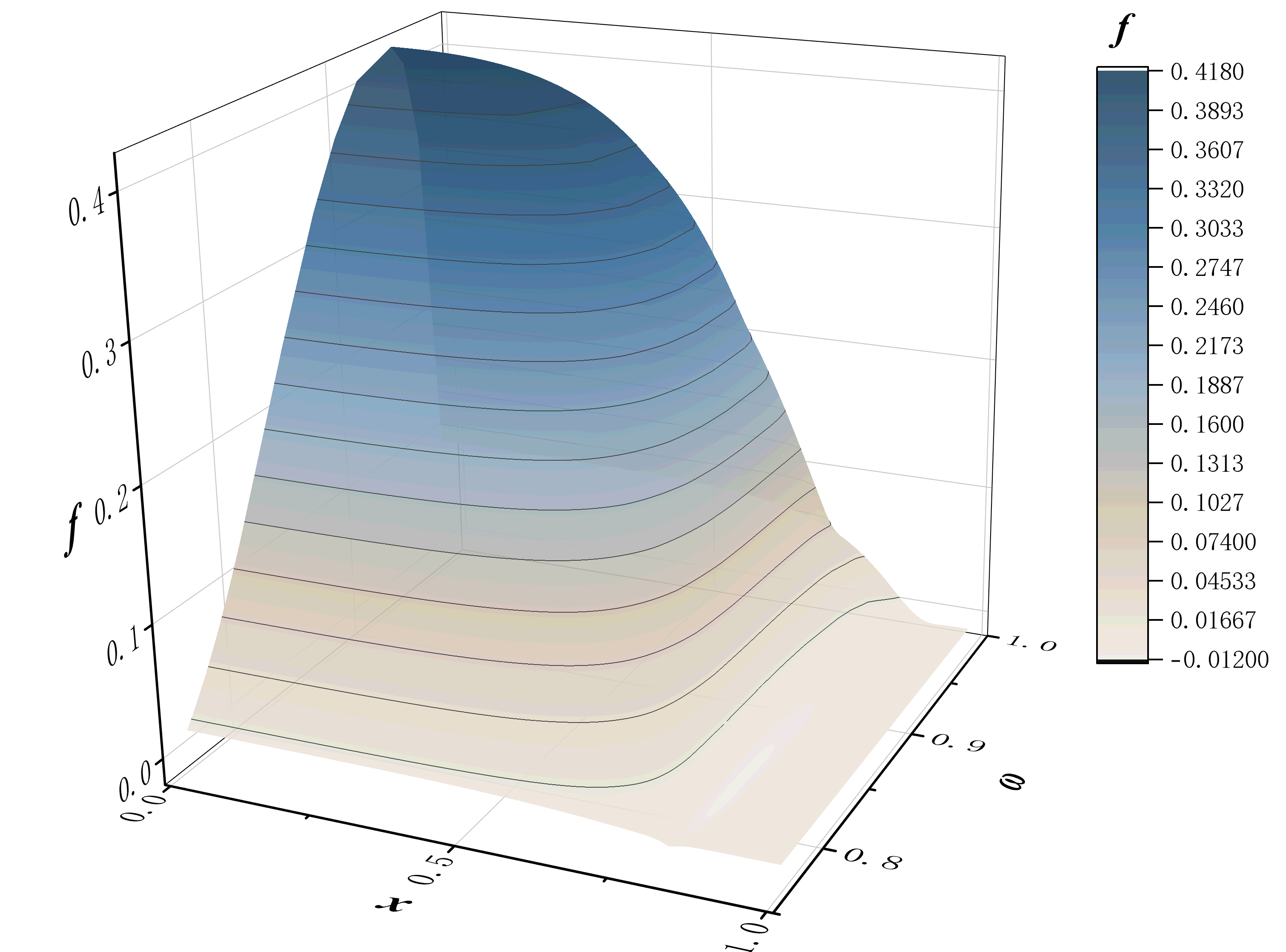}
				\caption{}
			\end{subfigure}
			\begin{subfigure}[b]{0.23\textwidth}
				\includegraphics[width=\textwidth]{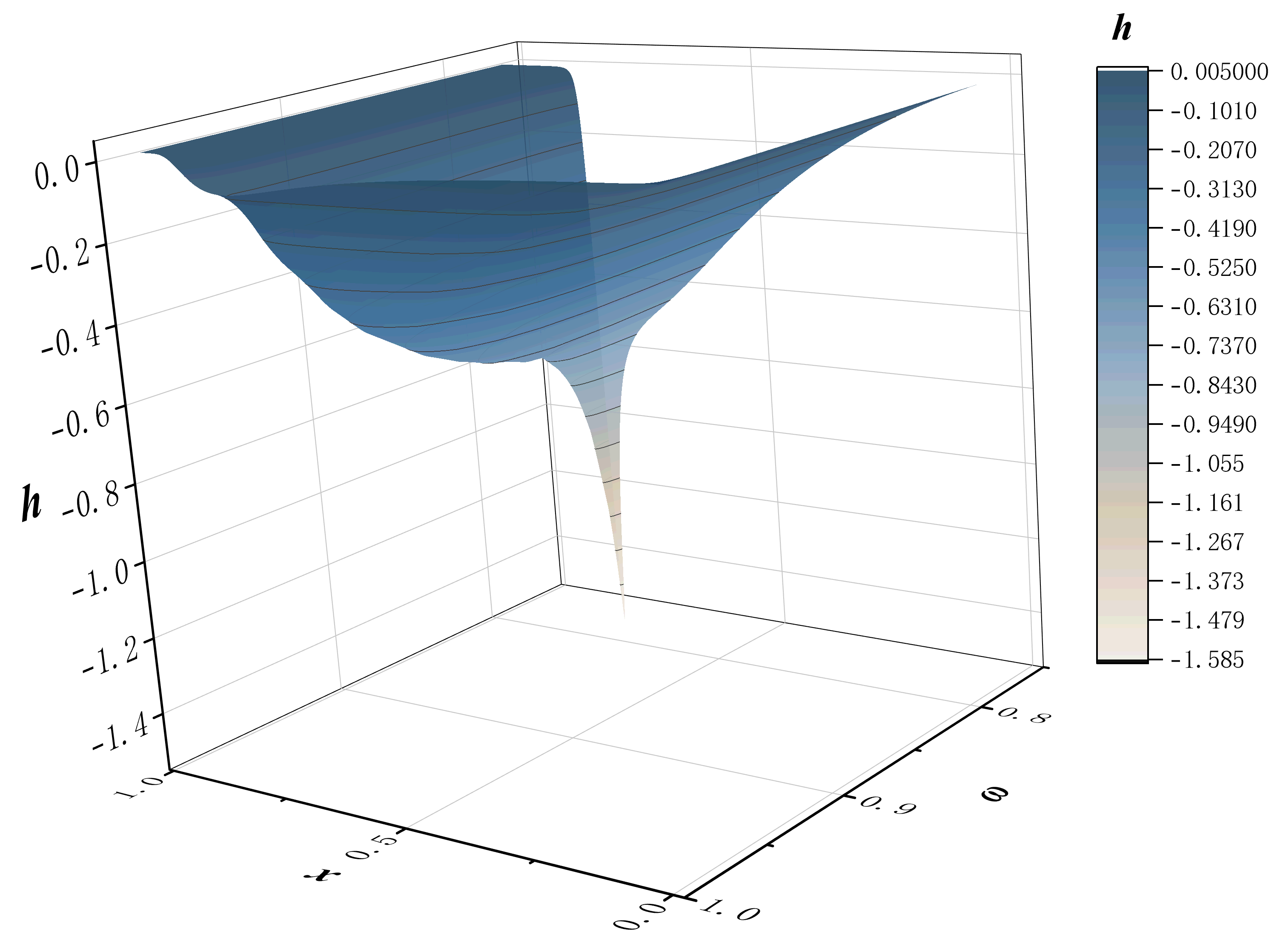}
				\caption{}
			\end{subfigure}
			\begin{subfigure}[b]{0.23\textwidth}
				\includegraphics[width=\textwidth]{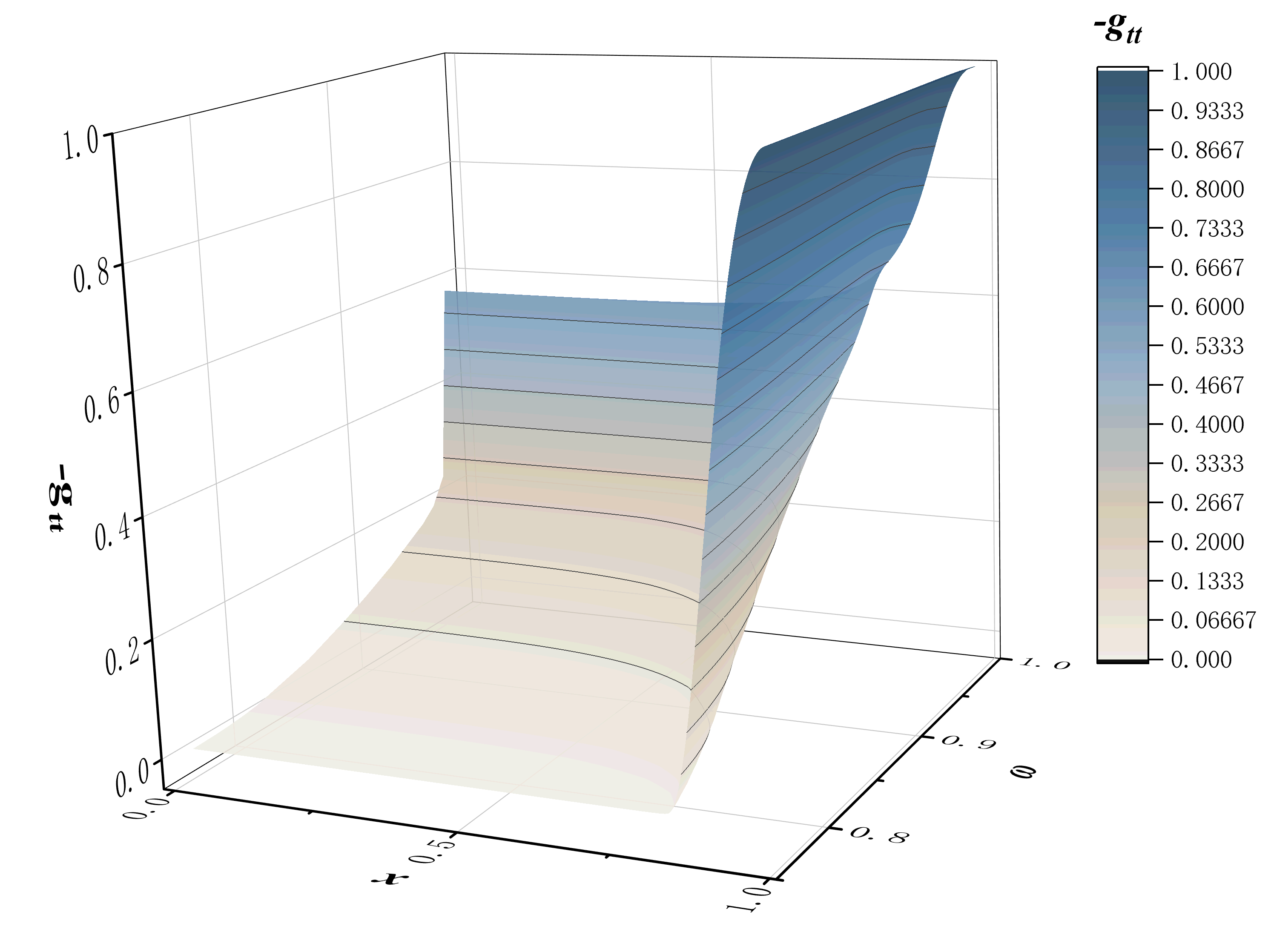}
				\caption{}
			\end{subfigure}
			\begin{subfigure}[b]{0.23\textwidth}
				\includegraphics[width=\textwidth]{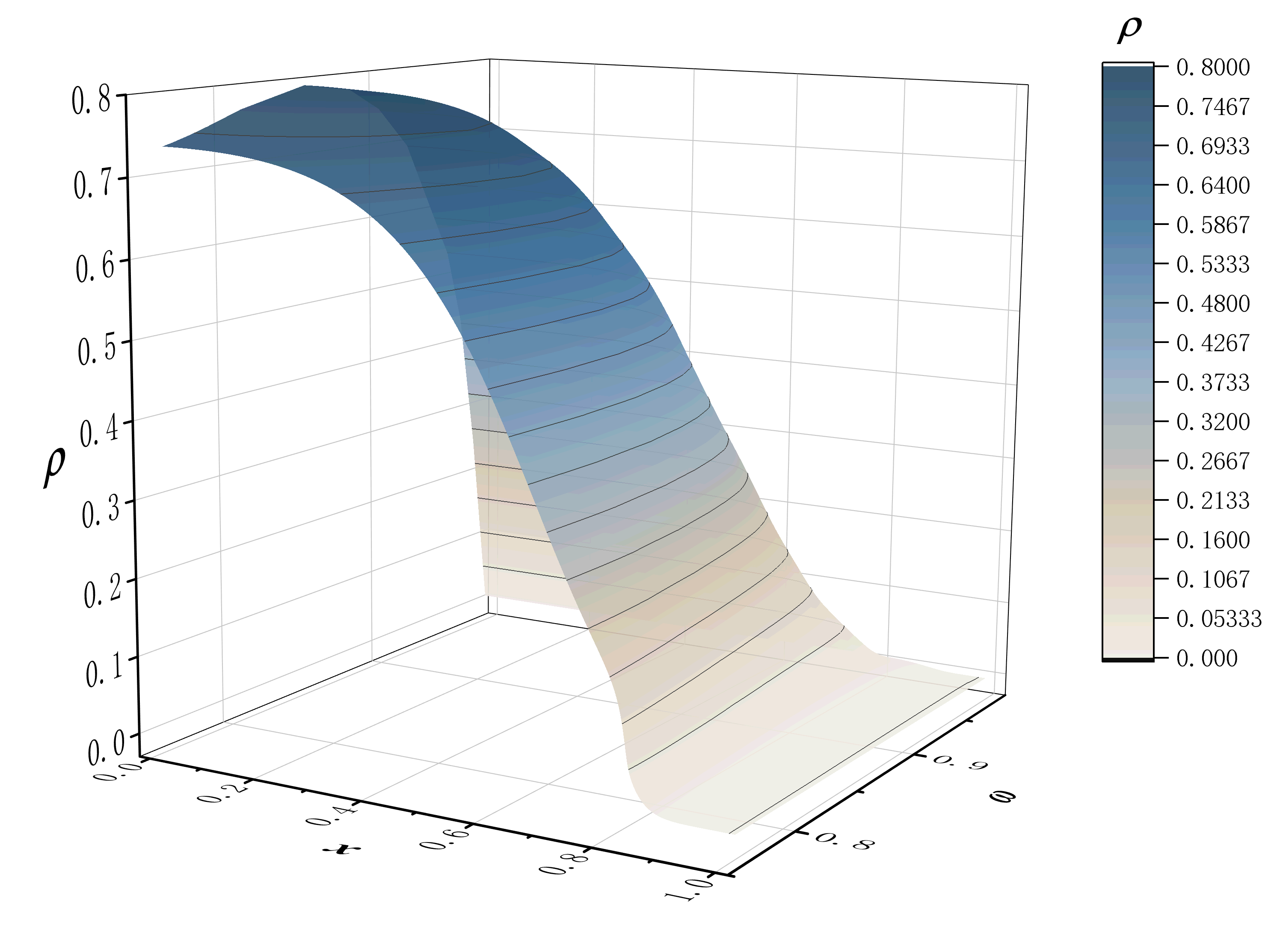}
				\caption{}
			\end{subfigure}
		\end{center}
		\caption{(a) and (b): Three-dimensional surface plots of the Proca field components $f$ and $h$ as functions of the radial coordinate $x$ and the field frequency $\omega$. (c) and (d): Three-dimensional surface plots of the metric $-g_{tt}$ and the energy density $\rho$ as functions of the radial coordinate $x$ and the field frequency $\omega$. All results fixed at $q = 0.9$. }
		\label{p16}
	\end{figure}

	The introduction of charge disrupts this finely tuned ``frozen'' state. As shown in Figs.~\ref{p15} and \ref{p16}, increasing the charge \(q\) forces the matter distribution to delocalize---it effectively ``unfreezes.'' Physically, this reflects a tug-of-war between competing forces. In the neutral case, the frozen state emerges from a delicate balance between gravitational attraction and the short-range repulsion inherent to the infinite-derivative terms. Adding a long-range electrostatic repulsion (the Coulomb force) upsets this balance. It prevents matter from being compressed tightly enough to form a quasi-horizon, thereby raising the minimum possible frequency \(\omega_{\text{min}}\) and restoring a more diffuse, spread-out configuration. In the charged case, the exterior geometry is no longer pure vacuum but electrovac, depending on both the mass \(M\) and the charge \(Q\). As discussed in Appendix~\ref{app:charged_no_universal_matching}, this breaks the one-parameter near-extremal matching that previously led to the remarkable agreement with the neutral extremal geometry.

	\subsection{Energy Condition}
	
	Various energy conditions impose physical viability constraints on matter fields. Here, we list the expressions for several common energy conditions relevant to this model \cite{Bueno:2025tli} and evaluate them for our obtained solutions using numerical methods.
	
	\noindent 1. Weak Energy Condition (WEC). It states that $T_{\mu\nu}t^\mu t^\nu \ge 0$ for any timelike vector $t^\mu$. This implies a non-negative energy density and ensures that the projection of the energy-momentum tensor along null directions is non-negative:

	\begin{equation}\label{wec}
		\rho \geq 0, \quad \rho + P_i \geq 0.
	\end{equation}
	
	\noindent 2. Null Energy Condition (NEC). It requires $T_{\mu\nu}k^\mu k^\nu \ge 0$ for any null vector $k^\mu$:
	
	\begin{equation}\label{nec}
		\rho + P_i \geq 0.
	\end{equation}
	
	\noindent 3. Dominant Energy Condition (DEC). It stipulates that the energy flux measured by any observer must be timelike or null, thereby preserving causality:
	
	\begin{equation}\label{dec}
		\rho \geq |P_i|.
	\end{equation}
	
	\noindent 4. Strong Energy Condition (SEC). It is defined as $R_{\mu\nu}t^\mu t^\nu \ge 0$ for any timelike vector. In $D$ dimensions, this is equivalent to:
	
	\begin{equation}\label{sec}
		\rho + P_i \geq 0, \quad 2 \rho +  P_1 + 3 P_2 \geq 0.
	\end{equation}
	The indices $i=1$ and $i=2$ correspond to the radial and tangential pressures, respectively.
	
\begin{figure}[ht]
		\begin{center}
			\begin{subfigure}[b]{0.23\textwidth}
				\includegraphics[width=\textwidth]{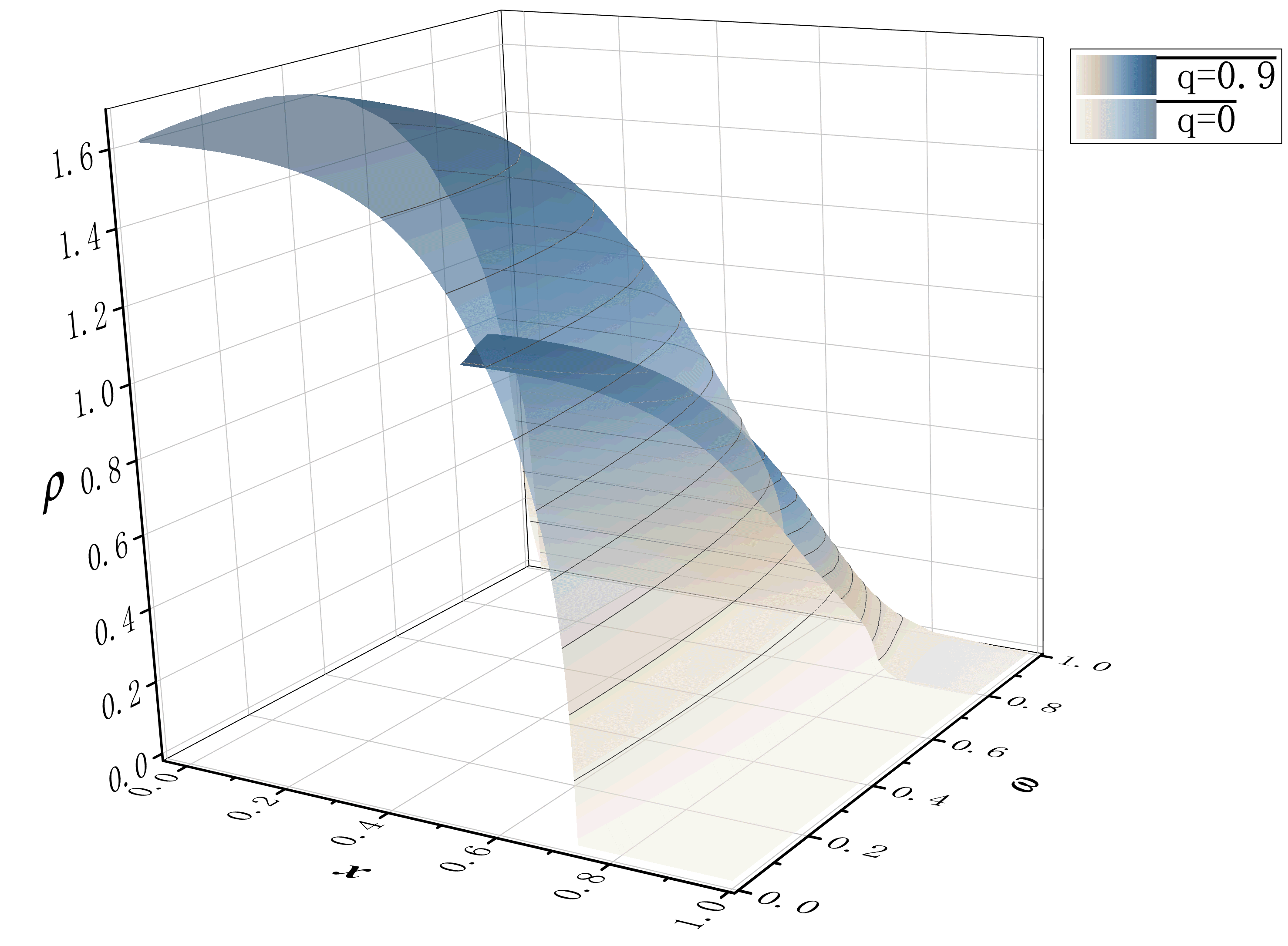}
				\caption{}
			\end{subfigure}
			\begin{subfigure}[b]{0.23\textwidth}
				\includegraphics[width=\textwidth]{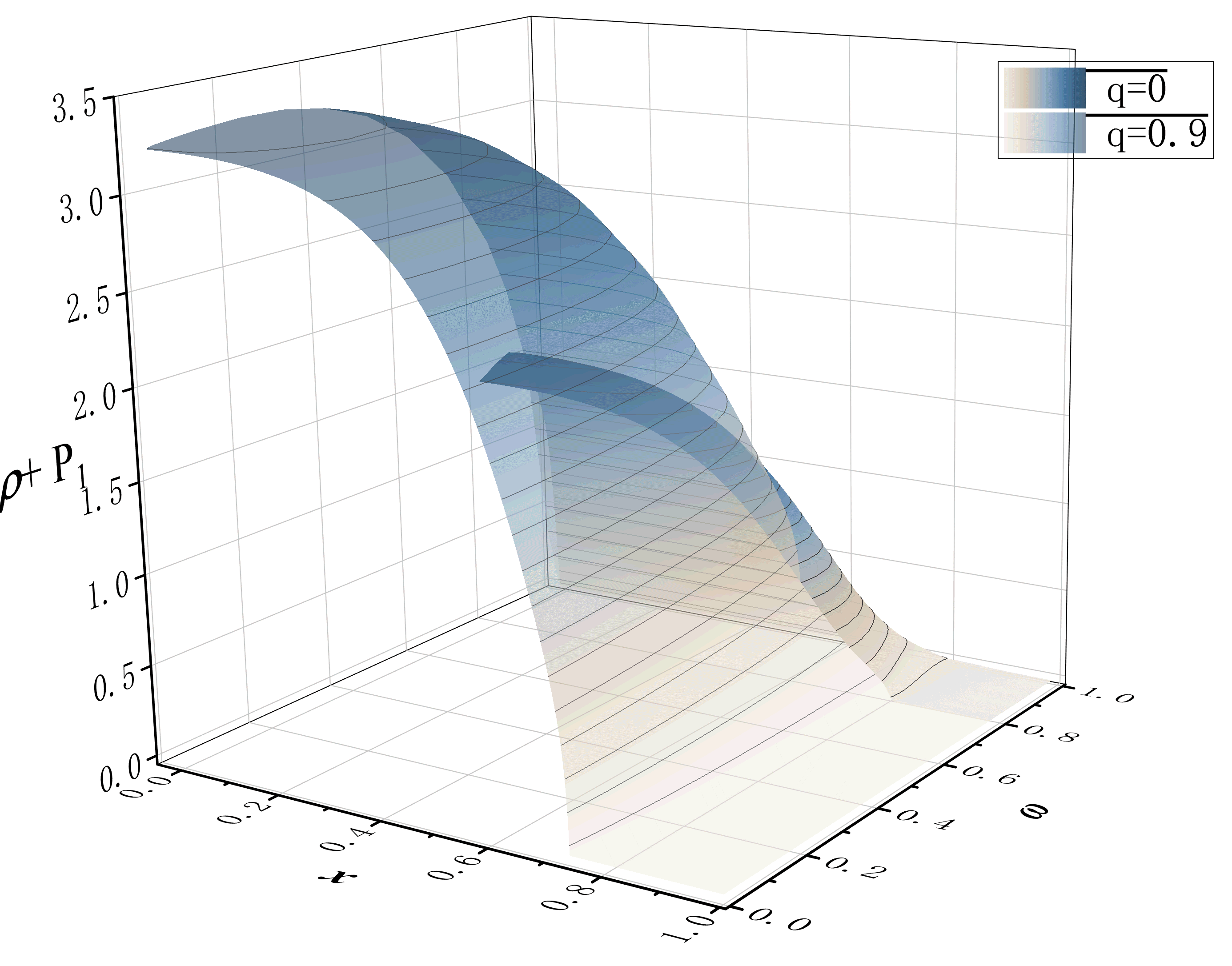}
				\caption{}
			\end{subfigure}
			\begin{subfigure}[b]{0.23\textwidth}
				\includegraphics[width=\textwidth]{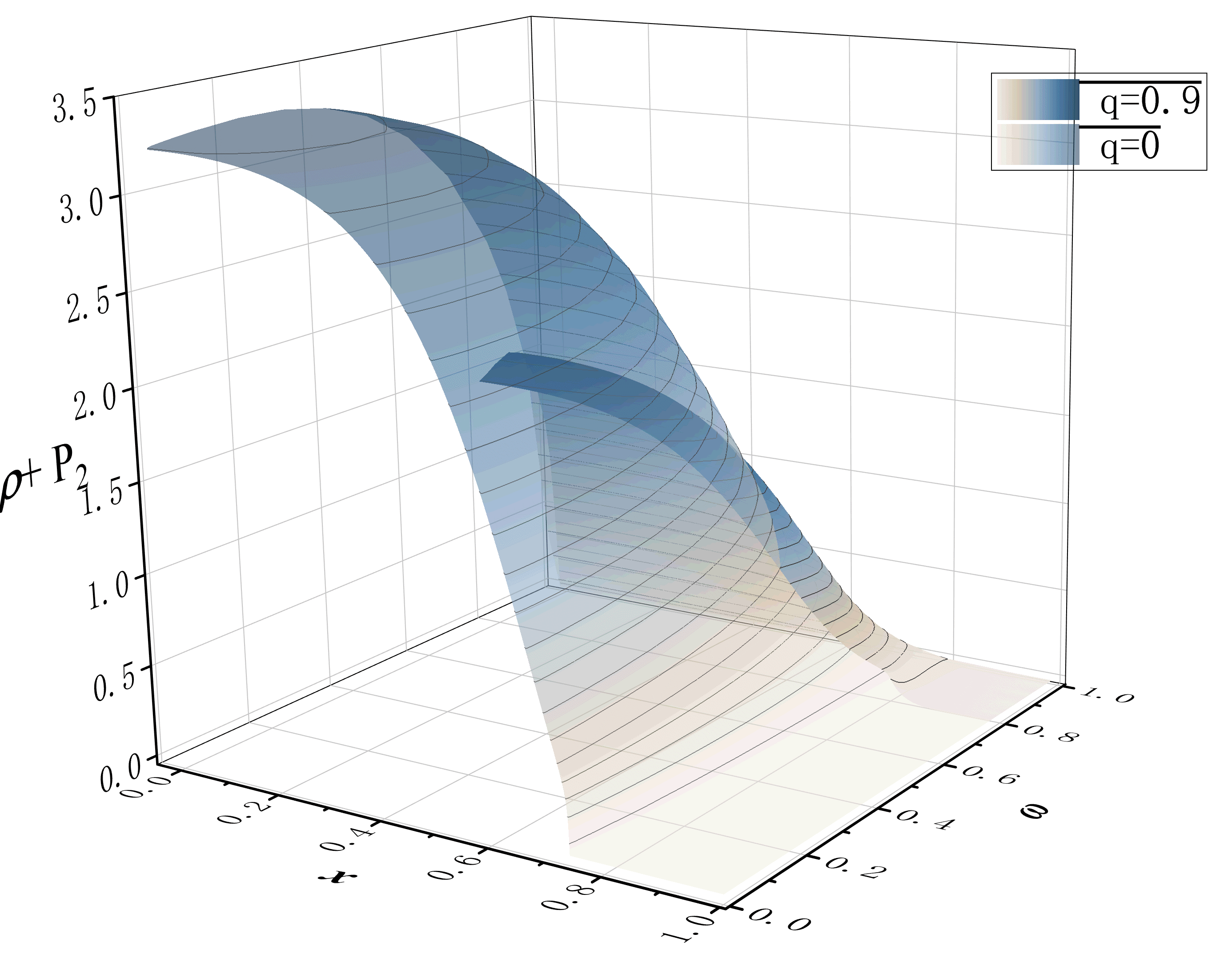}
				\caption{}
			\end{subfigure}
			\begin{subfigure}[b]{0.23\textwidth}
				\includegraphics[width=\textwidth]{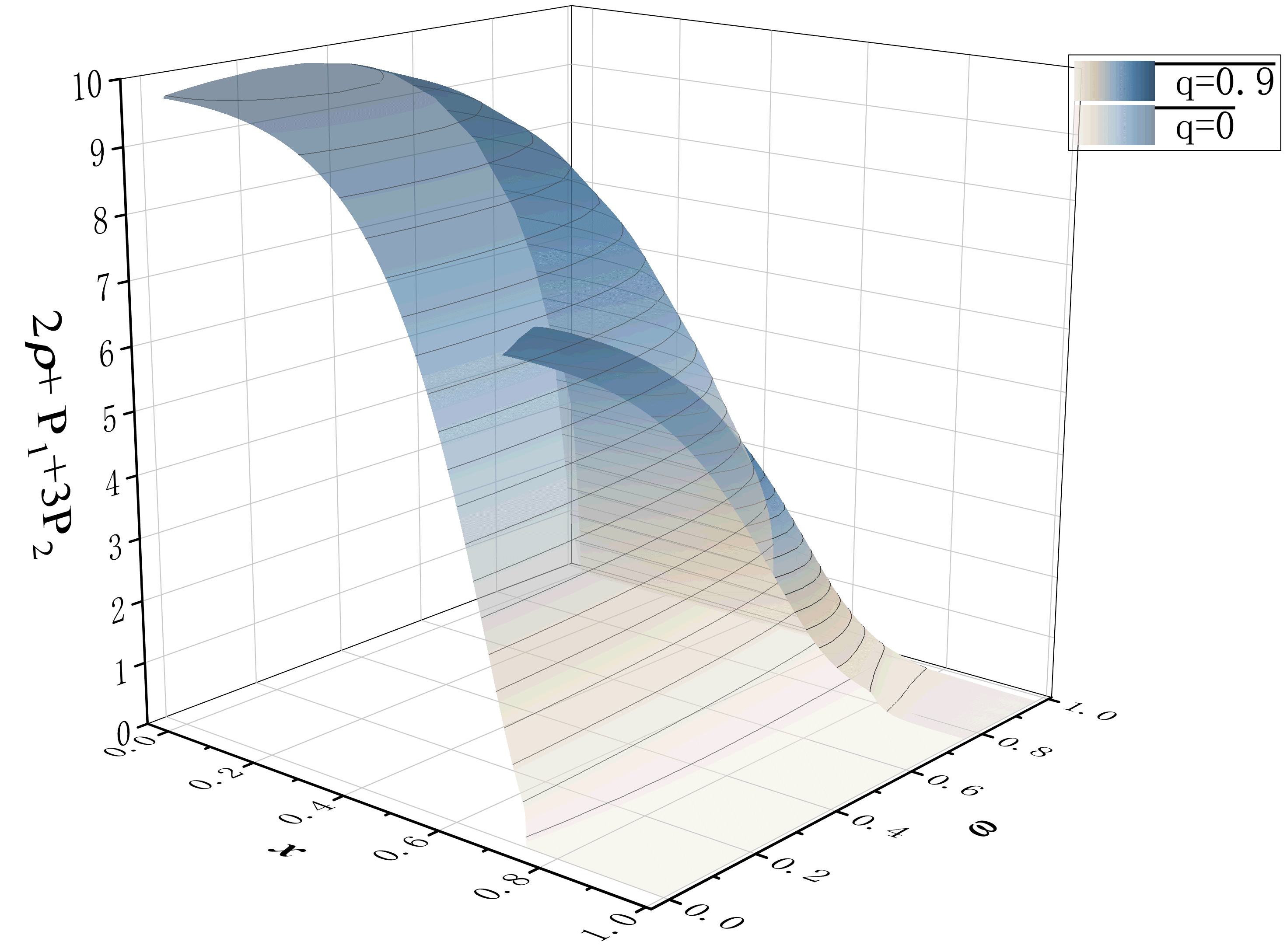}
				\caption{}
			\end{subfigure}
		\end{center}
		\caption{The energy density~$\rho$ and its linear combinations with anisotropic pressures~$P_i$ (e.g., representing energy conditions). 
			Transparent 3D color surfaces represent the solution for~$q=0$, 
			and opaque surfaces correspond to the solution for~$q=0.9$. }
		\label{p17}
	\end{figure}

	We examine these conditions for the representative cases of the ``frozen'' state ($q=0$) and the ``unfrozen'' state ($q=0.9$). As illustrated in Fig.~\ref{p17}, the energy density $\rho$ and radial pressure $P_1$ remain strictly positive globally. Although the tangential pressure $P_2$ exhibits small negative values in a limited frequency band and narrow radial region, its magnitude is insufficient to violate the inequalities.
	
	Consequently, we find that the Proca-Maxwell system coupled to infinite-derivative gravity satisfies all energy conditions everywhere (The conclusion remains the same when $q$ takes other values). This result is significant. It demonstrates a key merit of the pure gravitational regularization scheme: unlike wormhole scenarios or black holes coupled to nonlinear electrodynamics which often require exotic matter \cite{Bronnikov:2000vy,Hao:2025gak,Pattersons:2025bie,Su:2024gxp,Li:2026mam}, regularized gravitational solutions can be obtained while maintaining matter fields that fully respect the standard energy conditions.
	
	\section{Discussion and Conclusion}\label{sec5}
	
	In this work, we have numerically constructed and investigated the Proca-Maxwell system within a five-dimensional quasi-topological gravity theory featuring an infinite tower of higher-derivative terms. We recovered the standard Einstein-Proca-Maxwell system in the $n=1$ limit, which exhibits the characteristic spiral $M$-$\omega$ relation and negative binding energy. For the $n=2$ case (Gauss-Bonnet), we found that solutions develop a central singularity as $\omega \to 0$, although the introduction of electric charge $q$ can mitigate this and support positive binding energy. Crucially, we demonstrated that higher-order corrections ($n \ge 3$ up to $\infty$) are essential for resolving the central singularity, yielding globally regular solutions. In the neutral limit ($q = 0$), the system enters a ``frozen state'' as $\omega \to 0$, where matter fields are confined within a critical radius $x_c$, and the metric components vanish. Externally, this spacetime acts as a novel black hole mimicker. However, increasing $q$ introduces electrostatic repulsion that raises the lower frequency bound, effectively ``unfreezing'' the state.
	
	The physical mechanism underlying these numerical results can be understood as an equilibrium resulting from the competition between ``attractive force'' of gravity and various repulsive supporting forces. In classical gravity, the upper frequency limit for a Proca star is $\omega_{\text{max}} = m$ (where $m=1$ in this paper), corresponds to complete dispersal into Minkowski spacetime. Conversely, the lower frequency limit $\omega_{\text{min}}$ cannot be arbitrarily small and even approach zero, as this would imply an extreme or even singular, gravitational collapse. In the quasi-topological gravity theory, higher-order terms provide an effective ``repulsive force'' in the strong curvature regime that resolves singularities. It is precisely the balance between these competing mechanisms that, under the ``support'' provided by the higher-order corrections, allows the system to ``freeze'' in a certain critical radius as the field frequency approaches zero. Therefore, as $q$ increases from zero, a new repulsive force is introduced into the system. Increasing $q$ introduces electrostatic repulsion that prevents access to this ultra-low frequency regime, raising $\omega_{\text{min}}$ and effectively ``unfreezing'' the configuration.
	
	Looking forward, a natural extension of this work is to move beyond the static limit. While the present study establishes the delicate equilibrium and the ``unfreezing'' mechanism of these Proca-Maxwell configurations, exploring their fully non-linear time evolution remains a critical open frontier. Employing standard numerical relativity techniques to simulate the dynamical stability and time-evolution of these hairy structures will rigorously test this equilibrium under generic perturbations \cite{Choptuik:1992jv,Balakrishna:1997ej,Seidel:1993zk,DiGiovanni:2018bvo,Mio:2026wfh}. Furthermore, as viable black hole mimickers, investigating their phenomenological signatures, particularly their distinctive black hole shadows and gravitational-wave emissions, will provide crucial theoretical templates for future astronomical observations \cite{Fan:2025jow,Arbelaez:2025gwj,Lan:2025brn,Ditta:2024iky,Liu:2024iec,Zeng:2025wlb}. It is expected that the frozen state, owing to its extreme compactness, will generate prominent light rings structures and distinct shadow signatures. Conversely, the charge-induced ``unfreezing'' mechanism reduces the compactness of the configuration, thereby modifying these observational features. Related work is currently in progress.

	\section*{Acknowledgements}
	We are grateful to Shi-Xian Sun, Long-Xing Huang, Jun-Ru Chen, and Tian-Xiang Ma for valuable discussions and suggestions. This work was supported by the National Natural Science Foundation of China under Grants No. 12475051, No. 12375051, and No. 12421005; the science and technology innovation Program of Hunan Province under grant No. 2024RC1050; the Natural Science Fund of Hunan Province under Grant No. 2026JJ20019; the innovative research group of Hunan Province under Grant No. 2024JJ1006.
	
	\appendix
	
	\section{Vacuum exterior solution and quasi-horizon scaling in the $n=\infty$.}
	\label{app:vacuum_ninf}
	%======================================================================
	
	In this appendix, we provide analytic control of the vacuum exterior geometry
	in the $n=\infty$ theory and derive several closed-form relations that are useful
	for diagnosing the quasi-horizon (``frozen'') regime observed numerically.
	
	Throughout we adopt the static spherically symmetric metric
	\begin{equation}
		ds^{2}=-\sigma(r)^{2}N(r)\,dt^{2}+\frac{dr^{2}}{N(r)}+r^{2}d\Omega_{3}^{2},
	\end{equation}
	and the shorthand
	\begin{equation}
		\psi(r)\equiv \frac{1-N(r)}{r^{2}}.
	\end{equation}
	In vacuum (vanishing matter fields), the equations of motion reduce to
	Eq.~(2.13) in the main text:
	\begin{equation}
		\frac{d\sigma}{dr}=0,
		\qquad
		\frac{d}{dr}\!\left[r^{4}\,\mathcal{H}(\psi)\right]=0,
		\label{eq:vacuum_eqs}
	\end{equation}
	where $\mathcal{H}(\psi)$ denotes the (non-polynomial) function built from the
	higher-curvature tower.
	
	\subsection{Regular solution}
	
	With the coupling ansatz $\alpha_n=\alpha^{\,n-1}$, the tower function is
	\begin{equation}
		\mathcal{H}(\psi)=\psi+\sum_{n=2}^{n_{\max}}\alpha^{\,n-1}\psi^{n}.
	\end{equation}
	For $n_{\max}=\infty$ we can resum the series explicitly. Writing $k\equiv n-1$
	so that $k=1,2,\dots$, we obtain
	\begin{align}
		\mathcal{H}(\psi)
		&=\psi+\sum_{n=2}^{\infty}\alpha^{\,n-1}\psi^{n}
		=\psi\left[1+\sum_{k=1}^{\infty}(\alpha\psi)^{k}\right].
	\end{align}
	Using the geometric series $\sum_{k=1}^{\infty}x^{k}=x/(1-x)$ (for $|x|<1$),
	we arrive at the closed form
	\begin{equation}
		\mathcal{H}(\psi)=\frac{\psi}{1-\alpha\psi}
		\qquad (n=\infty).
		\label{eq:H_closed}
	\end{equation}
	
	The first vacuum equation in \eqref{eq:vacuum_eqs} gives $\sigma(r)=\text{const}$.
	Imposing asymptotic normalization $\sigma(\infty)=1$ yields
	\begin{equation}
		\sigma(r)=1.
	\end{equation}
	The second vacuum equation in \eqref{eq:vacuum_eqs} implies the first integral
	\begin{equation}
		r^{4}\,\mathcal{H}(\psi)=\tilde m,
		\label{eq:first_integral}
	\end{equation}
	where $\tilde m$ is an integration constant. As shown in Eq.~(2.15) of the main text,
	it is proportional to the ADM mass $M$:
	\begin{equation}
		\tilde m=\frac{8GM}{3\pi}.
		\label{eq:mtilde_M}
	\end{equation}
	
	Substituting the closed form \eqref{eq:H_closed} into \eqref{eq:first_integral} gives
	\begin{equation}
		r^{4}\,\frac{\psi}{1-\alpha\psi}=\tilde m.
		\label{eq:integral_sub}
	\end{equation}
	Using $\psi=(1-N)/r^{2}$, Eq.~\eqref{eq:integral_sub} becomes
	\begin{equation}
		r^{4}\,\frac{(1-N)/r^{2}}{1-\alpha(1-N)/r^{2}}=\tilde m
		\quad\Longrightarrow\quad
		r^{2}\,\frac{1-N}{1-\alpha(1-N)/r^{2}}=\tilde m.
	\end{equation}
	Defining $y\equiv 1-N$ to simplify the algebra, we have
	\begin{equation}
		r^{2}\,\frac{y}{1-\alpha y/r^{2}}=\tilde m
		\quad\Longrightarrow\quad
		r^{2}y=\tilde m\left(1-\frac{\alpha y}{r^{2}}\right)
		=\tilde m-\tilde m\,\frac{\alpha y}{r^{2}}.
	\end{equation}
	Collecting the $y$-terms and multiplying by $r^{2}$ yields
	\begin{equation}
		r^{4}y+\tilde m\alpha\,y=\tilde m\,r^{2}
		\quad\Longrightarrow\quad
		y=\frac{\tilde m\,r^{2}}{r^{4}+\tilde m\alpha}.
	\end{equation}
	Therefore,
	\begin{equation}
		\boxed{
			N(r)=1-\frac{\tilde m\,r^{2}}{r^{4}+\tilde m\alpha}
		}
		\qquad (n=\infty),
		\label{eq:N_vacuum_ninf}
	\end{equation}
	which is equivalent to Eq.~(2.19) after using \eqref{eq:mtilde_M}.
	
	%----------------------------------------------------------------------
	\subsection{Horizons and the extremal condition}
	%----------------------------------------------------------------------
	
	A horizon radius $r_h$ satisfies $N(r_h)=0$. From \eqref{eq:N_vacuum_ninf} this is
	\begin{equation}
		1-\frac{\tilde m\,r_h^{2}}{r_h^{4}+\tilde m\alpha}=0
		\quad\Longleftrightarrow\quad
		r_h^{4}-\tilde m\,r_h^{2}+\tilde m\alpha=0.
	\end{equation}
	Letting $z\equiv r_h^{2}\ge 0$ gives the quadratic equation
	\begin{equation}
		z^{2}-\tilde m\,z+\tilde m\alpha=0,
	\end{equation}
	with roots
	\begin{equation}
		z_{\pm}=\frac{\tilde m\pm\sqrt{\tilde m^{2}-4\tilde m\alpha}}{2}.
	\end{equation}
	Real horizons exist only when the discriminant is non-negative:
	\begin{equation}
		\tilde m^{2}-4\tilde m\alpha\ge 0
		\quad\Longleftrightarrow\quad
		\tilde m\ge 4\alpha.
		\label{eq:ext_condition}
	\end{equation}
	The extremal case corresponds to a double root, i.e. vanishing discriminant, hence
	\begin{equation}
		\tilde m_{\rm ext}=4\alpha.
	\end{equation}
	The corresponding extremal horizon radius follows from $z_{\rm ext}=\tilde m_{\rm ext}/2$:
	\begin{equation}
		r_{\rm ext}^{2}=2\alpha,
		\qquad
		r_{\rm ext}=\sqrt{2\alpha}.
		\label{eq:rext}
	\end{equation}
	Using \eqref{eq:mtilde_M}, the extremal mass is
	\begin{equation}
		\boxed{
			M_{\rm ext}=\frac{3\pi}{2G}\,\alpha.
		}
		\label{eq:Mext}
	\end{equation}
	In the numerical units $4\pi G=1$ (i.e. $G=1/4\pi$), Eq.~\eqref{eq:Mext} reduces to
	$M_{\rm ext}=6\pi^{2}\alpha$.
	
	%----------------------------------------------------------------------
	\subsection{Horizonless quasi-horizon and useful scaling relations}
	%----------------------------------------------------------------------
	
	For non-vacuum case, the geometry is horizonless but $N(r)$
	develops a global minimum, which provides an analytic proxy for the ``quasi-horizon''
	observed numerically in the frozen regime.
	
	Differentiating \eqref{eq:N_vacuum_ninf} gives
	\begin{equation}
		N'(r)=
		-\frac{2\tilde m r(\tilde m\alpha-r^{4})}{(r^{4}+\tilde m\alpha)^{2}}.
	\end{equation}
	Besides the trivial root $r=0$, the non-trivial stationary point satisfies
	\begin{equation}
		\tilde m\alpha-r^{4}=0
		\quad\Longrightarrow\quad r_{c}=(\tilde m\alpha)^{1/4}.
		\label{eq:rc}
	\end{equation}
	One easily checks that $N'(r)<0$ for $0<r<r_c$ and $N'(r)>0$ for $r>r_c$, hence $r_c$
	is the global minimum of $N(r)$.
	
	Evaluating \eqref{eq:N_vacuum_ninf} at $r=r_c$ and using $r_c^{4}=\tilde m\alpha$ yields
	\begin{align}
		N_{\min}\equiv N(r_c)
		&=1-\frac{\tilde m r_c^{2}}{r_c^{4}+\tilde m\alpha}
		=1-\frac{\tilde m r_c^{2}}{2\tilde m\alpha}
		=1-\frac{1}{2}\sqrt{\frac{\tilde m}{\alpha}}.
	\end{align}
	Therefore,
	\begin{equation}
		\boxed{
			N_{\min}=1-\frac{1}{2}\sqrt{\frac{\tilde m}{\alpha}}.
		}
		\label{eq:Nmin}
	\end{equation}
	The extremal limit $\tilde m\to 4\alpha$ indeed gives $N_{\min}\to 0$.
	
	It is convenient to introduce the dimensionless ratio
	\begin{equation}
		\delta\equiv \frac{\tilde m}{\tilde m_{\rm ext}}=\frac{\tilde m}{4\alpha}
		=\frac{M}{M_{\rm ext}}.
	\end{equation}
	In terms of $\delta$, Eqs.~\eqref{eq:rc}--\eqref{eq:Nmin} become
	\begin{equation}
		N_{\min}=1-\sqrt{\delta},
		\qquad
		r_c=\sqrt{2\alpha}\,\delta^{1/4},
		\qquad
		\frac{r_c}{r_{\rm ext}}=\delta^{1/4}.
		\label{eq:delta_relations}
	\end{equation}
	Eliminating $\delta$ between $r_c$ and $N_{\min}$ gives a particularly useful
	self-consistency relation:
	\begin{equation}
		r_c^{2}=2\alpha\,(1-N_{\min}).
		\label{eq:rc_Nmin_relation}
	\end{equation}
	
	Finally, in the compactified numerical coordinate $x=r/(1+r)$ used in Eq.~(4.2),
	the extremal radius \eqref{eq:rext} corresponds to
	\begin{equation}
		\boxed{
			x_{\rm ext}=\frac{r_{\rm ext}}{1+r_{\rm ext}}
			=\frac{\sqrt{2\alpha}}{1+\sqrt{2\alpha}}.
		}
	\end{equation}
	Similarly, $r_c=x_c/(1-x_c)$, and Eq.~\eqref{eq:rc_Nmin_relation} can be rewritten as
	\begin{equation}
		x_c=\frac{\sqrt{2\alpha(1-N_{\min})}}{1+\sqrt{2\alpha(1-N_{\min})}}.
	\end{equation}
	
	\paragraph{Near-extremal expansion.}
	Let $M=M_{\rm ext}-\Delta M$ with $\Delta M\ll M_{\rm ext}$.
	From $N_{\min}=1-\sqrt{M/M_{\rm ext}}$ we obtain
	\begin{equation}
		N_{\min}=1-\sqrt{1-\frac{\Delta M}{M_{\rm ext}}}
		\simeq \frac{1}{2}\frac{\Delta M}{M_{\rm ext}}
		+\mathcal{O}\!\left(\frac{\Delta M}{M_{\rm ext}}\right)^{2},
	\end{equation}
	i.e.
	\begin{equation}
		\boxed{
			\Delta M\simeq 2M_{\rm ext}\,N_{\min}
			\qquad (\text{near extremality}).
		}
		\label{eq:near_ext}
	\end{equation}
	This relation is useful for converting an empirical scaling of $\Delta M(\omega)$ into
	the corresponding scaling of $N_{\min}(\omega)$ in the frozen limit. In this work we set $\alpha=4$, for which the extremal mass is found to be $M=24\pi^{2}\approx 236.871$. The event horizon is located at $x=\frac{2\sqrt{2}}{1+2\sqrt{2}}\approx 0.7388$.
	In the frozen configuration, for a frequency $\omega=0.001$, the minimum of $N$
	occurs at $x=0.7391$, where the ADM mass of the system is
	$M_{\mathrm{ADM}}=236.879$.
	Substituting these values into Eq.~\eqref{eq:near_ext}, we find that the numerical result is
	in good qualitative agreement with the analytic estimate.

	\section{Electrovacuum solution for $n=\infty$ (vanishing Proca field)}
	\label{app:electrovac_ninf}
	%----------------------------------------------------------------------
	
	In the electrovacuum limit the Proca field vanishes,
	\begin{equation}
		f(r)=0,\qquad h(r)=0,
	\end{equation}
	while the Maxwell potential is kept as $A=V(r)\,dt$.
	In this case the system reduces to a purely gravitational sector sourced by a
	radial electric field.
	
	\paragraph{Maxwell equation.}
	Setting $f=h=0$ in the Maxwell equation (Eq.~(2.11) in the main text) and using the
	spherically symmetric ansatz yields
	\begin{equation}
		r\,\sigma'(r)\,V'(r)-\sigma(r)\left(3V'(r)+rV''(r)\right)=0.
		\label{eq:Maxwell_reduced}
	\end{equation}
	In electrovacuum we still have $\sigma'(r)=0$ (see below), hence $\sigma(r)=1$ after
	imposing the asymptotic normalization.
	Equation \eqref{eq:Maxwell_reduced} then integrates to
	\begin{equation}
		V'(r)=\frac{\widehat Q}{r^{3}},
		\qquad
		V(r)=V_{\infty}-\frac{\widehat Q}{2r^{2}}.
		\label{eq:V_solution_hatQ}
	\end{equation}
	We fix the residual gauge freedom by choosing $V_{\infty}=0$.
	
	\paragraph{First integral of the gravitational equation.}
	With $f=h=0$, the metric equation (Eq.~(2.7) in the main text) reduces to
	\begin{equation}
		\frac{d}{dr}\!\left[r^{4}\mathcal{H}(\psi)\right]
		=\frac{2}{3}\,\frac{r^{3}}{\sigma^{2}}\,V'(r)^{2}.
		\label{eq:grav_reduced_electrovac}
	\end{equation}
	The $\sigma$-equation (Eq.~(2.8) in the main text) has a vanishing right-hand side when
	$f=h=0$, hence $\sigma' = 0$ and $\sigma=1$.
	Using \eqref{eq:V_solution_hatQ} in \eqref{eq:grav_reduced_electrovac} gives
	\begin{equation}
		\frac{d}{dr}\!\left[r^{4}\mathcal{H}(\psi)\right]
		=\frac{2}{3}\,r^{3}\left(\frac{\widehat Q^{2}}{r^{6}}\right)
		=\frac{2\widehat Q^{2}}{3}\,r^{-3}.
	\end{equation}
	Integrating once we obtain
	\begin{equation}
		r^{4}\mathcal{H}(\psi)=\tilde m-\frac{\widehat Q^{2}}{3r^{2}},
		\label{eq:first_integral_electrovac_hatQ}
	\end{equation}
	where $\tilde m$ is the integration constant related to the ADM mass,
	\begin{equation}
		\tilde m=\frac{8GM}{3\pi},
	\end{equation}
	as in Eq.~(2.15) of the main text.
	To match the standard $1/r^{4}$ Coulomb falloff in five dimensions and reproduce
	Eq.~(2.21) in the main text, it is convenient to define the physical charge parameter $Q$
	through
	\begin{equation}
		\frac{\widehat Q^{2}}{3}\equiv \frac{GQ^{2}}{3\pi^{3}}
		\qquad\Longleftrightarrow\qquad
		\widehat Q^{2}=\frac{GQ^{2}}{\pi^{3}}.
		\label{eq:Qhat_to_Q}
	\end{equation}
	Then \eqref{eq:first_integral_electrovac_hatQ} becomes
	\begin{equation}
		\boxed{
			r^{4}\mathcal{H}(\psi)=\tilde m-\frac{GQ^{2}}{3\pi^{3}r^{2}}.
		}
		\label{eq:first_integral_electrovac}
	\end{equation}
	
	\paragraph{Solving for $N(r)$ in the $n=\infty$ theory.}
	For $n_{\max}=\infty$ we have
	\begin{equation}
		\mathcal{H}(\psi)=\frac{\psi}{1-\alpha\psi}.
	\end{equation}
	Substituting into \eqref{eq:first_integral_electrovac} yields
	\begin{equation}
		r^{4}\,\frac{\psi}{1-\alpha\psi}=\tilde m-\frac{GQ^{2}}{3\pi^{3}r^{2}}.
		\label{eq:psi_equation_electrovac}
	\end{equation}
	Dividing by $r^{4}$ we write the right-hand side as
	\begin{equation}
		\frac{\psi}{1-\alpha\psi}=
		\frac{\tilde m}{r^{4}}-\frac{GQ^{2}}{3\pi^{3}r^{6}}.
	\end{equation}
	Solving for $\psi$ gives
	\begin{align}
		\psi
		&=\left(1-\alpha\psi\right)\left(\frac{\tilde m}{r^{4}}-\frac{GQ^{2}}{3\pi^{3}r^{6}}\right)
		\nonumber\\
		&=\left(\frac{\tilde m}{r^{4}}-\frac{GQ^{2}}{3\pi^{3}r^{6}}\right)
		-\alpha\psi\left(\frac{\tilde m}{r^{4}}-\frac{GQ^{2}}{3\pi^{3}r^{6}}\right).
	\end{align}
	Bringing all $\psi$-terms to the left-hand side, we obtain
	\begin{equation}
		\psi\left[
		1+\alpha\left(\frac{\tilde m}{r^{4}}-\frac{GQ^{2}}{3\pi^{3}r^{6}}\right)
		\right]
		=
		\left(\frac{\tilde m}{r^{4}}-\frac{GQ^{2}}{3\pi^{3}r^{6}}\right).
	\end{equation}
	Multiplying numerator and denominator by $r^{6}$ yields
	\begin{equation}
		\psi
		=
		\frac{\tilde m r^{2}-\dfrac{GQ^{2}}{3\pi^{3}}}
		{r^{6}+\alpha\left(\tilde m r^{2}-\dfrac{GQ^{2}}{3\pi^{3}}\right)}.
		\label{eq:psi_solution_electrovac}
	\end{equation}
	Using $\psi=(1-N)/r^{2}$ we obtain
	\begin{equation}
		N(r)
		=
		1-\frac{r^{2}\left(\tilde m r^{2}-\dfrac{GQ^{2}}{3\pi^{3}}\right)}
		{r^{6}+\alpha\tilde m r^{2}-\alpha\dfrac{GQ^{2}}{3\pi^{3}}}
		=
		1+\frac{r^{2}\left(\dfrac{GQ^{2}}{3\pi^{3}}-\tilde m r^{2}\right)}
		{r^{6}+\alpha\tilde m r^{2}-\alpha\dfrac{GQ^{2}}{3\pi^{3}}}.
		\label{eq:N_electrovac_compact}
	\end{equation}
	Finally, inserting $\tilde m=\frac{8GM}{3\pi}$ and multiplying numerator and denominator
	by $3\pi^{3}$, we arrive at
	\begin{equation}
		\boxed{
			N(r)=
			1+\frac{G r^{2}\left(Q^{2}-8M\pi^{2}r^{2}\right)}
			{3\pi^{3}r^{6}-GQ^{2}\alpha+8GM\pi^{2}\alpha r^{2}}.
		}
		\label{eq:N_electrovac_ninf_final}
	\end{equation}
	This reproduces Eq.~(2.21) of the main text. Setting $Q=0$ reduces the solution back to
	Eq.~(2.19).
	
	\paragraph{Asymptotic expansion.}
	For completeness, expanding \eqref{eq:N_electrovac_compact} at large $r$ gives
	\begin{equation}
		N(r)=1-\frac{\tilde m}{r^{2}}+\frac{GQ^{2}}{3\pi^{3}r^{4}}+\mathcal{O}(r^{-6})
		=
		1-\frac{8GM}{3\pi r^{2}}+\frac{GQ^{2}}{3\pi^{3}r^{4}}+\mathcal{O}(r^{-6}),
	\end{equation}
	which confirms the interpretation of $M$ and $Q$ as the ADM mass and electric charge
	parameters in five dimensions.
	
	%----------------------------------------------------------------------
	\subsection{Why the charged case does not exhibit the same ``universal matching'' as the neutral frozen state}
	\label{app:charged_no_universal_matching}
	%----------------------------------------------------------------------
	
	The neutral frozen state studied in the main text admits an exceptionally sharp geometric
	interpretation: as $\omega\to 0$ (for $q=0$) the matter distribution becomes confined inside
	a critical radius $r_c$ and the exterior region becomes vacuum. Consequently, the exterior
	metric is forced onto the \emph{one-parameter} vacuum family \eqref{eq:N_vacuum_ninf},
	and in the frozen limit the ADM mass approaches the \emph{unique} extremal value
	$M\to M_{\rm ext}(\alpha)$, yielding a universal near-extremal matching.
	
	Once electric charge is introduced, this tight correspondence is generically lost for three
	independent reasons.
	
	\paragraph{(i) The exterior is never vacuum.}
	For $q\neq 0$, even if the Proca field becomes very localized, the Maxwell field remains
	long-ranged and does not vanish outside the star. Hence the exterior geometry is governed
	by the \emph{two-parameter} electrovac family \eqref{eq:N_electrovac_ninf_final}, labelled
	by $(M,Q)$, rather than the one-parameter vacuum family labelled by $M$ only.
	
	\paragraph{(ii) No universal extremal mass.}
	In the neutral case the extremality condition fixes a unique value $\tilde m_{\rm ext}=4\alpha$
	(or $M_{\rm ext}\propto \alpha$), see Eq.~\eqref{eq:ext_condition}.
	In contrast, for the electrovac solution the horizon equation $N(r_h)=0$ becomes a cubic
	polynomial in $z\equiv r_h^{2}$,
	\begin{equation}
		P(z)\equiv z^{3}-\tilde m z^{2}+(\alpha\tilde m+\beta)z-\alpha\beta=0,
		\qquad
		\beta\equiv \frac{GQ^{2}}{3\pi^{3}},
		\label{eq:horizon_cubic}
	\end{equation}
	and extremality requires a \emph{double root}, i.e.\ $P(z_\star)=0$ and $P'(z_\star)=0$.
	Solving these two conditions yields a \emph{curve} in the $(\tilde m,\beta)$ plane, which can
	be written parametrically as
	\begin{equation}
		\boxed{
			\beta_\star(z)=\frac{z^{3}(z-2\alpha)}{(z-\alpha)^{2}},
			\qquad
			\tilde m_\star(z)=\frac{3z^{2}+\beta_\star(z)}{2z-\alpha}.
		}
		\label{eq:ext_curve_charged}
	\end{equation}
	Therefore, unlike the neutral case, there is \emph{no} unique $M_{\rm ext}(\alpha)$: the
	extremal mass depends on the charge parameter $Q$ (or $\beta$).
	A charged Proca-star branch would have to dynamically approach the extremal curve
	\eqref{eq:ext_curve_charged} in the $(M,Q)$ plane in order to mimic an extremal charged
	black hole, which is not enforced by the field equations.
	
	\paragraph{(iii) The ``quasi-horizon'' diagnostic becomes intrinsically two-dimensional.}
	Even in the horizonless regime, the location of the global minimum of $N(r)$ is no longer
	fixed by a simple closed form (contrast Eq.~\eqref{eq:rc} in the neutral case).
	Writing \eqref{eq:N_electrovac_compact} in terms of $z=r^{2}$,
	\begin{equation}
		N(z)=1+\frac{\beta z-\tilde m z^{2}}{z^{3}+\alpha\tilde m z-\alpha\beta},
	\end{equation}
	one finds that the stationary condition $dN/dz=0$ leads to the quartic equation
	\begin{equation}
		\boxed{
			\tilde m z^{4}-2\beta z^{3}-\alpha\tilde m^{2}z^{2}+2\alpha\beta\tilde m z-\alpha\beta^{2}=0,
			\qquad (\beta\neq 0).
		}
		\label{eq:rc_quartic_charged}
	\end{equation}
	For $\beta=0$ this collapses to $z^{2}=\alpha\tilde m$ and reproduces the neutral result
	$r_c^{4}=\alpha\tilde m$, but for $\beta\neq 0$ the ``critical radius'' depends on both
	$\tilde m$ and $\beta$, hence on both $(M,Q)$.
	As a result, there is no one-dimensional universal relation such as
	$N_{\min}=1-\sqrt{M/M_{\rm ext}}$ in the charged case.
	
	\paragraph{Implications for charged Proca stars.}
	In the Proca-Maxwell solitons studied in the main text, the global charge is not an
	independent parameter but is tied to the particle number, $Q=qN_P$.
	Consequently, along a given numerical branch both $M(\omega)$ and $Q(\omega)$ vary with
	the frequency, so the exterior metric cannot approach a single fixed electrovac template in the
	same rigid way as in the neutral frozen limit. Moreover, the long-range Coulomb repulsion
	introduces a lower-frequency cutoff $\omega_{\min}(q)$, which dynamically prevents the
	$\omega\to 0$ approach that was essential for the neutral near-extremal matching.
	Finally, for the coupling choice
	$\alpha_n=\alpha^{n-1}$ the charged black-hole solution \eqref{eq:N_electrovac_ninf_final}
	develops a singularity near $r=0$, indicating that even the electrovac background itself does not
	admit the same ``regular core + extremal exterior'' interpretation as the neutral case.

\end{document}